\documentclass[10pt]{article}
\usepackage{eurosym}
\usepackage{epigraph}
\usepackage{amsfonts}
\usepackage{soul}
\usepackage{bbm}
\usepackage{amsmath}
\usepackage{multirow}
\usepackage{amssymb}
\usepackage{natbib}
\usepackage{color}
\usepackage{setspace}
\usepackage{epsf}
\usepackage{bm}
\usepackage{latexsym}
\usepackage{amssymb}
\usepackage{geometry}
\usepackage{amsmath}
\usepackage{graphicx}
\usepackage{graphics}
\usepackage{epsf}
\usepackage{float}
\usepackage{enumerate}
\usepackage{hyperref}
\usepackage{url}
\hypersetup{linkcolor=blue,citecolor=black,urlcolor=black,colorlinks=true}
\RequirePackage{amsopn} \RequirePackage{amsfonts}
\RequirePackage{amsthm}
\newtheorem{prop}{Proposition}

\newtheorem{lemma}{Lemma}

\theoremstyle{definition}

\newtheorem{ex}{Example}

\newcommand{\field}[1]{\mathbb{#1}}
\newcommand{\R}{\field{R}}

\renewcommand{\Pr}{\field{P}}
\newcommand{\Ind}[1]{\field{I}_{\{#1\}}}

\def\EXP{{\bf E}}

\definecolor{light-gray}{gray}{0.80}
\definecolor{yyellow}{gray}{.50}
\definecolor{red-orange}{rgb}{1,0.1,0}
\definecolor{blue}{rgb}{0,0,1}
\definecolor{dark-blue}{rgb}{0,0, 0.55}
\definecolor{dark-red}{rgb}{0.55, 0, 0}

\begin{titlepage}
\end{titlepage}
\begin{document}


\title{Crowding out the truth? A simple model of misinformation, \\ polarization and meaningful social interactions\thanks{We thank Roland B\'{e}nabou, Alexander Frug, Ga\"{e}l Le Mens, and audiences at the 8th International Conference on Computational Social Science in Chicago (IC$^2$S$^2$ 2022) and at the 5th Economics of Media Bias Workshop (Berlin 2022) for helpful comments. Fabrizio Germano acknowledges financial support from Grant PID2020-115044GB-I00//AEI/10.13039/501100011033 and
from the Spanish Agencia Estatal de Investigaci\'{o}n (AEI), through the Severo Ochoa Programme for Centres of Excellence in R\&D (Barcelona School of Economics CEX2019-000915-S). Francesco Sobbrio is grateful to Nando Pagnoncelli and IPSOS for allowing access to the data of the Polimetro.}}
\author{Fabrizio Germano\thanks{Department of Economics and Business, Universitat
Pompeu Fabra, and BSE,
{\tt fabrizio.germano@upf.edu}} 
\hspace{.15in} Vicen\c{c} G\'{o}mez\vspace{15pt}\thanks{Department of Information and Communications Technologies,
Universitat Pompeu Fabra, Barcelona, {\tt vicen.gomez@upf.edu}} 
\hspace{.15in} Francesco
Sobbrio\vspace{15pt}\thanks{Department of Economics and Finance, Tor Vergata University of Rome and CESifo, {\tt francesco.sobbrio@uniroma2.it}}}

\date{\today \vspace{20pt}}

\maketitle

\begin{abstract}
This paper provides a simple theoretical framework to evaluate the effect of key parameters of ranking algorithms, namely popularity and personalization parameters, on measures of platform engagement, misinformation and polarization. The results show that an increase in the weight assigned to online social interactions (e.g., likes and shares) and to personalized content may increase engagement on the social media platform, while at the same time increasing misinformation and/or polarization. By exploiting Facebook's 2018 ``Meaningful Social Interactions'' algorithmic ranking update, we also provide direct empirical support for some of the main predictions of the model.

\bigskip

\end{abstract}


\thispagestyle{empty}
\begingroup
\setstretch{1.2}


\maketitle
\newpage
\section{Introduction}
Recent revelations by whistle-blowers at Facebook have once again brought to the attention of the public the risks and dangers associated with the algorithms of digital platforms to manage their informational content.\footnote{See, for example, \url{https://www.wsj.com/articles/the-facebook-files-11631713039}} The algorithms used by social media like Facebook, Twitter or Instagram or by search engines like Google and Bing decide what information to show to users and, importantly, also in what order to show it. Indirectly, they determine what information is more or less relevant for any given user. A rapidly growing body of empirical research has documented how social media platforms may foster polarization and misinformation \citep{allcott_2020,ditella_2021,levy2021} sometimes associated with a tangible impact \citep{bursztyn_2019,muller_2020,muller_2021,amnesty_2022}. In particular, there are journalistic and academic claims suggesting that such adverse effects may be a consequence of the way profit-maximizing social media platforms design their algorithms \citep{cnn_2021,lauer_2021}, namely with the objective of ensuring a high level of engagement \citep{hao_2017}.

In this paper, we provide a theoretical framework---and related empirical evidence---to assess whether it is indeed the case that algorithmic rules that tend to be desirable from the perspective of social media platforms may instead lead to detrimental effects for their users and, more broadly, for the health of democracies. We build on and extend our previous work \citep{germano_2019,germano_2020} to develop a model where a platform ranks news items (e.g., posts, tweets, etc.), while individuals sequentially access the platform to decide which news items to click and possibly  ``highlight" (e.g., like, share, comment or retweet). At the center of our model, there is an endogenous ranking algorithm that decides the order of news items to be displayed to a given user. In particular, the ranking evolves according to the \textit{popularity} of news items, which is a weighted combination of the clicks and highlights received by that news item. Simply put, the more people click and the more people highlight a news item, the higher the probability that the news item will go up in the ranking and will be then displayed in a higher-order position. The model also allows for assessing the role of \emph{personalization}, that is, when the platform provides a different ranking of the news items to different individuals. 

To preserve tractability, the choices of an individual over which news items to click and highlight are modeled as driven by behavioral traits that are rooted in ample empirical evidence. In terms of clicking choices, we assume that with some positive probability individuals have some preference for choosing confirmatory news \citep{gentzkow_2010,yom_et_al,white2015belief,flaxman_et_al} and, at the same time, also for news items that are higher ranked \citep{pan_et_al,NovareseWilson2013,glick_et_al,epstein2015search}. In terms of highlights, we assume that with some probability individuals highlight a news item, provided it is sufficiently close to their prior beliefs \citep{garz_2020} and the more so the more extreme their prior beliefs are \citep{bakshy2015exposure,grinberg_2019,pew_2019,hopp_2020}.

Armed with this theoretical framework, we then proceed to assess the impact of popularity-driven and personalized rankings on $(i)$ platform engagement (defined in terms of the overall number of clicks and highlights); $(ii)$ misinformation (defined as the average distance between the information content present in the news items chosen by individuals and the true state of the world) and $(iii)$ polarization (defined as the average distance between the information content present in the news items chosen by individuals belonging to different groups).

The paper provides insights on whether and when ranking algorithms may lead to a trade-off between platform and user welfare. First, we show that increasing the weight given to highlights in the popularity ranking might be desirable from the platform's perspective as it increases engagement. Yet, it is detrimental from a public policy perspective as it also leads to higher levels of misinformation---\textit{crowding-out the truth}---and polarization. For completeness, we also show that such trade-off would not be present if the propensity to highlight a ``like-minded" news item was not higher for people with more extreme priors. This difference is relevant as previous research \citep{bakshy2015exposure} has shown that in the case of ``hard" (e.g, national, political) news, the propensity to highlight contents is indeed higher for individuals with more extreme prior whereas the same does not apply to ``soft" news (e.g., entertainment). Accordingly, our results suggest that the trade-off between engagement and misinformation/polarization is not much of a concern in the case of ``soft" news while it might instead be particularly relevant in the case of political news. For what concerns personalization, the results show that a trade-off between engagement and polarization is always present regardless of whether the propensity to highlight content is correlated with extreme priors or not. That is, increasing the degree of personalization in the ranking algorithm is conducive to a higher level of engagement yet also to a higher degree of polarization.

In terms of the empirical relevance of our theoretical insights, first, we point out how the detrimental impact of personalization on political polarization implied by our model is very much in line with the empirical literature on this issue (e.g., \citealt{levy2021,dujgar2022,huszar_2022}). Most importantly, we also provide direct evidence on the impact of increasing the weight given by platforms to highlighted content. In particular, we leverage a rich survey dataset from Italy and exploit Facebook's ``Meaningful Social Interaction'' (MSI) algorithmic ranking update implemented in January 2018, which significantly boosted the weight given to comments and shares in the Facebook's ranking algorithm.\footnote{See \url{https://www.facebook.com/business/news/news-feed-fyi-bringing-people-closer-together}.} We estimate a Differences-in-Differences empirical model comparing the ideological extremism and affective polarization of people interviewed after the Meaningful Social Interaction (MSI) algorithm was introduced (i.e., January-June 2018) and that use internet to form an opinion relative to those of people also using internet to form an opinion who were interviewed before such a change (i.e., June-December 2018) and at the same time relative to people interviewed after such change in the algorithm who were not using internet as one of the main sources to form an opinion. The results confirm some of the key theoretical predictions of the model: namely Facebook's 2018 MSI update led to an increase in ideological extremism and affective polarization in Italy.\footnote{The theoretical predictions of the model pointing out the role of social media algorithms in fostering misinformation, are also consistent with \cite{vosoughi2018spread} providing evidence that false stories spread faster than true ones on Twitter. Similarly, \cite{mosleh_2020} points out the presence of a negative correlation between the veracity of a news item and its probability of being shared on Twitter.} 

 To the best of our knowledge, this is the first paper to explore both theoretically and empirically how an algorithmic boost given to highlighted content may affect platform engagement and social welfare. 
 The model generalizes and extends the ones of \cite{germano_2019} and \cite{germano_2020}. The present setting differs in several key aspects. First, signals are drawn from a continuous distribution: individuals observe whether an item reports like-minded news, but then need to actually click on the item in order to learn the actual signal (and update their beliefs accordingly). Second, we allow for a broader set of clicking behavior by individuals than just confirmatory or ranking-driven types. Third, most importantly, the present model also allows individuals to {\em highlight} news items and explores how such action might impact platform engagement, misinformation and polarization. Fourth, we implicitly endogenize the ranking weights assigned by digital platforms when considering which highlight and personalization weights would maximize engagement. Last but not least, we evaluate the impact of ranking algorithms along different metrics meant to be informative for social welfare (including measures of platforms and consumers' welfare).

 Our model is complementary to the one of \cite{acemoglu_2022} who focus on endogenous social networks and fact-checking.\footnote{See also \cite{azzimonti_2022} for a model of diffusion of misinformation on social media via internet bots.} In particular, as in our model, \cite{acemoglu_2022} show that platforms have an incentive to increase personalization (more homophilic communication patterns) as this increases platform engagement. In their setting, this is detrimental in terms of social welfare as it increases the level of misinformation. Instead, in our case, more personalization increases polarization yet it does not affect the overall level of misinformation, since our model does not embed the issue of fact-checking and cannot therefore capture such an effect. At the same time, because we explicitly model the endogenous dynamic ranking used by social media platforms, we are instead able to provide insights on the incentives---and possible perverse effects on social welfare---of such platforms to boost the weight given to content highlighting in their ranking algorithm. More generally, our paper relates to the literature analyzing the effects of ranking algorithms on democratic outcomes. This literature encompasses communication 
scholars \citep{hargittai2004changing,granka,napoli2015social}, computer scientists \citep{choo_et_al,menczer2006googlearchy,pan_et_al,glick_et_al,flaxman_et_al,bakshy2015exposure,hao_2017,tabibian_2020},  economists \citep{levy2019echo,germano_2020,van_gils_2020,acemoglu_2022}, legal scholars \citep{goldman2006search,grimmelmann,sunstein2009}, 
media activists \citep{pariser}, psychologists \citep{epstein2015search}, political scientists \citep{putnam2001,hindman,lazer2015rise,tucker_2018}, and
sociologists \citep{tufekci2015algorithmic,tufekci_2018}.

\section{The Model}
\label{section:model}

 At the center of the model is a digital platform characterized by its ranking algorithm, which ranks and directs individuals to different news items (e.g., websites, Facebook posts, tweets), based on the popularity of individuals' choices. Such news items may be used by individuals to obtain information on an unknown cardinal state of the world $\theta \in \R$ (e.g., net benefits of vaccines, consequences of inaction on global warming, optimal foreign policy intervention, etc.). The ranking of each news item is inversely related to its  popularity, where the popularity is determined by the number of clicks and the number of ``highlights'' received by a given item (e.g., likes received by a Facebook post/number of shares, like/retweets of a tweet, etc.). Each click has a weight of one, and each ``highlight''  has an additional weight of $\eta \ge 0$. In the following subsections we provide a formal and detailed description of the different elements of our model.

\subsection{News items and Individuals}
There are $M>2$ news items, each of which carries an informative signal on the state of the world $y_m \in \R$, and which is drawn randomly and independently from $N(\theta, \sigma_{y}^2)$ (we use $g(y)$ to denote the corresponding density function). There are $N$ individuals, each of whom receives a private informative signal on the state of the world $x_n \in \R$, which is drawn randomly and independently from $N(\theta, \sigma_{x}^2)$ (we use $f(x)$ to denote the corresponding density function).

To model individuals' clicking behavior, we further assume there is a benchmark $\widehat{\theta} \in \R$---non-informative with respect to $\theta$---which allows individuals to sort news items into ``like-minded" or not. That is, we assume that, leaving aside the order of news items provided by the ranking algorithm, individuals are able to see whether a news item is reporting a ``like-minded" information or not. Yet they need to click on the news item in order to see the actual signal $y_m$.  This assumption is meant to capture a rather typical situation, where individuals observe the ``coarse'' information provided in the landing page by the platform (e.g., infer the basic stance of a news item, whether Left or Right, pro or anti something, from the website title, Facebook post intro, first tweet in a thread, etc.), yet, in order to learn the actual content of the news (i.e., the cardinal signal  $y_m$) and update her beliefs, the individual has to click on the news item.

We formally translate this setting into assuming that an individual is able to observe whether her own signal $x_n$ and the news items' signals $y_m$ are above or below $\widehat{\theta}$. 
Accordingly, for each individual, the signal $x_n$ has an associated binary signal indicating whether such signal is above or below $\widehat{\theta}$: sgn$(x_n) \in \{-1, 1\}$, where sgn$(x_n) =-1$ if $x_n < \widehat{\theta}$ and sgn$(x_n) = 1$ if $x_n \ge \widehat{\theta}$. Similarly, for each news item, the signal $y_m$ has an associated binary signal sgn$(y_m) \in \{-1, 1\}$, where sgn$(y_m) =-1$ if $y_m < \widehat{\theta}$ and sgn$(y_m) = 1$ if $y_m \ge \widehat{\theta}$. 
 
From this we can compute $M_{-}$ and $M_{+}$ as the set of news items with binary signal respectively $-1$ and $1$ (by slight abuse of notation, we also use $M_{-}$ and $M_{+}$ to denote the number of news items in $M_{-}$ and $M_{+}$ respectively). 

That is, given the individual's signal $x_n$, the benchmark $\widehat{\theta}$ allows the individual to sort news items into ``like-minded" or not before actually clicking or reading. For most of the paper, we focus on the case where the benchmark separates signals in roughly symmetrical groups, i.e. $\widehat{\theta} \approx\theta$.\footnote{Say, $| \widehat{\theta} - \theta | < \min \left\{ \frac{\sigma_x}{4}, \frac{\sigma_y}{4} \right\}$. In \ref{section:NCaverageop}, we discuss the case, where $\widehat{\theta}$ and $\theta$ are far apart. \ref{section:hetaverageop} discusses the case where individuals have heterogeneous benchmarks $\widehat{\theta}_n$.}

At the same time, individuals have to actually click on news item $m$ in order to learn its cardinal signal $y_m$. In particular, we assume that, absent ranking effects, the individual's choice about which news item to click on depends on her ``clicking type'' ($\tau^c_n$). To encompass all possible clicking behaviour, we consider three {\em clicking types}: 
\begin{itemize}
\item {\em confirmatory type} ($\tau^c_n = \tau_C$): clicks with propensity  $\gamma_C$ on a news item with the same sign as her own signal sgn$(x_n)$, with $1/2 < \gamma_C < 1$, and with propensity $1-\gamma_C$ on one of opposite sign; 
\item {\em exploratory type} ($\tau^c_n = \tau_E$): clicks with propensity $\gamma_E$ on a news item with the same sign as her own signal sgn$(x_n)$, with $0 < \gamma_E < 1/2$, and with propensity $1-\gamma_E$ on one of opposite sign;
\item {\em indifferent (purely ranking-driven) type} ($\tau^c_n = \tau_I$): clicks with equal propensity $\gamma_I = 1/2 = 1-\gamma_I$ on an outlet of either sign.
\end{itemize}
The three types occur with probabilities, respectively, $p_C\geq0, p_E\geq0$ and $p_I\geq0$, such that $p_C+p_E+p_I=1$. Similar to the literature on political economy that parametrizes the fraction of different types of voters (e.g., \citealt{krasa_2009,krishna_2011,galasso_2011}), the model does not micro-found the individuals' clicking choices. At the same time, it is easy to see that the confirmatory type might be driven by a preference for like-minded news \citep{mullainathan_2005, bernhardt_2008,gentzkow_2010,sobbrio_2014,gentzkow_2015}.\footnote{See \citet{yom_et_al,flaxman_et_al,white2015belief} for empirical evidence on confirmation bias by users of digital platforms.} Similarly, the exploratory type might be the by-product of incentives to cross-check different information sources \citep{rudiger_2013,athey_2018}. Finally, the indifferent type allows us to consider the role of individuals with a high attention bias or search cost \citep{pan_et_al,glick_et_al,NovareseWilson2013}.\footnote{We assume clicking types to be independent from the individual's prior $x_n$. Nonetheless, when formalizing the individual's choice over which news items to ``highlight'', we assess how polarization is impacted when the choice to ``highlight'' is correlated with the individual's prior beliefs.} Notice that we specify these three different types to encompass all possible clicking behaviour (confirmatory, exploratory, ranking-driven), yet the key insights of the model will hold true even if we were to focus only on one or two of such types.

The binary signal sgn$(x_n)$ together with the individual's clicking type $\tau^c_n \in  T^c \equiv \{ \tau_C, \tau_E, \tau_I \}$ determine the {\em propensity with which individual $n$ will click on an item $m$, absent ranking}:
\begin{equation}
\hspace{-.06in}
\varphi_{n,m} = 
\left\{
\begin{array}
[c]{cl}
\frac{\gamma_k}{[m]} & \text{ if } \tau^c_n = \tau_k, \text{sgn}(x_n) = \text{sgn}(y_m), k = C, E, I 
\\  \\
\frac{1-\gamma_k}{[m]} & \text{ if } \tau^c_n = \tau_k,  \text{sgn}(x_n) \ne  \text{sgn}(y_m),k = C, E, I,
\end{array}
\right. 
\end{equation}
where $[m] = M_{-}$ if sgn$(y_m)=-1$ and $[m] = M_{+}$ if sgn$(y_m)=1$.

\subsubsection{Individual choice over which news items to click on}
We now generalize the individual choice function to take into account the fact that individuals see the news items presented in a given order, following the ranking $r_n = \left( r_{n,m} \right)_{m \in M}$, where $r_{n,m}$ is the rank of news item $m$ as seen by individual $n$. We assume that individuals have an {\em attention bias} calibrated by the parameter $\beta > 1$, with the interpretation that, a news item of equal sign but placed one position higher in the ranking has a likelihood $\beta$ times larger to be clicked on than the one in the lower position. Together with the propensity to click absent ranking, these jointly determine the probability with which individuals click on news items. We define the {\em probability of individual $n$ clicking on news item $m$} as: 
\begin{equation} \label{eq:clickprob}
\rho_{n,m} = \frac{\beta^{(M-r_{n,m})}\varphi_{n,m}}{\sum_{m' \in M} \beta^{(M-r_{n,m'})} \varphi_{n,m'}} .
\end{equation}

\subsubsection{Individual choice over which news items to highlight}

After clicking on a given news item $m$, the individual sees the actual signal $y_m \in \R$ and then decides whether or not to {\em highlight} such a news item (e.g, like, share, comment, retweet, etc.). This depends on the individual's ``highlighting type'' ($\tau^h_n$). We consider two {\em highlighting  types}: 
\begin{itemize}
\item {\em passive type} ($\tau^h_n = \tau_P$): never highlights a news item regardless of her signal;
\item {\em active type} ($\tau^h_n = \tau_A$): highlights a news item if and only if the news item's signal is sufficiently close to her own signal, $y_m \in H (x_n)$,
\end{itemize}
where $H (x_n) \equiv [ x_n - \sigma_{x}/2 , x_n + \sigma_{x}/2 ]$ and $\sigma_{x}$ is the standard deviation of the individual's signal $x_n$. 
We assume the highlighting types, $\tau_n^h \in T^h \equiv \{ \tau_P, \tau _A \}$, occur with probabilities $p_P$ and $p_A$ respectively, where $p_P+p_A=1$. 
Fixing the probability of being a {\em passive type} $\tau_P$ given by $p_P = 1- p_A$, we consider two alternative cases for the probability of being an {\em active type} $p_A$:

\begin{itemize}
\item {\em flat case}: $p_A$ is a constant in $(0, 1)$;
\item {\em non-flat case:} $p_A$ is a function of the signal received by the individual given by:
\begin{equation} \label{eq:BMA}
p_A(x_n)=1-e^{-\frac{1}{2\alpha}\left(\frac{x_n - \widehat{\theta}}{\sigma_x} \right)^{2\alpha}}  , \, \,\, \, \alpha \ge 1 \, .
\end{equation}
\end{itemize}

In words, we always assume that individuals highlight only if the news item reports a signal sufficiently close to their prior \citep{garz_2020}. Moreover, in the flat case the probability of highlighting a news item ($p_A$) is independent of the individual's signal $x_n$ (again, provided that the news item reports a signal sufficiently close to individual's prior). By contrast, in the non-flat case, the highlighting probability is correlated with the individual's signal. Specifically, $p_A$ increases with the (square of the) deviation of $x_n$ from the benchmark $\widehat{\theta}$, normalized by $\sigma_x$.\footnote{ Results are robust to replacing $\theta$ for $\widehat{\theta}$ in Eq.~(\ref{eq:BMA}).} However, while the specific functional form assumed in Eq.~(\ref{eq:BMA}) is not crucial for our results, what matters is that individuals with more extreme prior beliefs are more likely to be active and hence to highlight a given news item $y_m$ (provided it is within $H(x_n)$ and hence sufficiently close to the individual's signal $x_n$).

As with the individuals' clicking choice, the highlighting choice is not derived from a maximization problem of the individual. Nevertheless, the correlation between extreme beliefs and propensity to highlight present in the non-flat case is reminiscent of the link between overconfidence and ideological extremism modeled by \cite{ortoleva_2015}. Importantly, the non-flat case is also rooted in observed empirical regularities on how individuals with more extreme ideological beliefs tend to be more actively engaged and also more likely to highlight items on social media platforms \citep{bakshy2015exposure,grinberg_2019,pew_2019,hopp_2020}.
In particular, by using data on over 10 million Facebook users in the US, \cite{bakshy2015exposure} provide evidence on the ideological distribution of shared contents.\footnote{More specifically, they measure the ideological alignment of content shared on Facebook as the average affiliation of sharers weighted by the total number of shares. Similar evidence is presented by the authors when weighting by the total number of distinct URL shared.} The data clearly show a bimodal distribution with large mass on the tails of the distribution, i.e., individuals with more extreme preferences account for a larger proportion of the overall shared contents on Facebook. Remarkably, such a bimodal distribution is present only when looking at the distribution of shares related to contents defined as ``hard information'' (e.g., national news, politics, world affairs). By contrast, no such bimodality is present when looking at ``soft information'' content (e.g., sport, entertainment, travel). This suggests that the bimodal distribution observed in the shares of hard information is unlikely to be driven by large tails in the ideological distribution of Facebook's users (i.e., a bimodal distribution of Facebook users' ideology) or by a larger density of the network in such tails. Rather, such a bimodal distribution of shares is likely to be driven by users with more extreme ideological preferences having a higher propensity of sharing hard information contents.\footnote{See also \cite{grinberg_2019,pew_2019,hopp_2020} for additional empirical evidence on the positive correlation between extreme political preference and the propensity to share contents in social media.}

\begin{figure}[t]
  \centering
  \includegraphics[width=15.5cm]{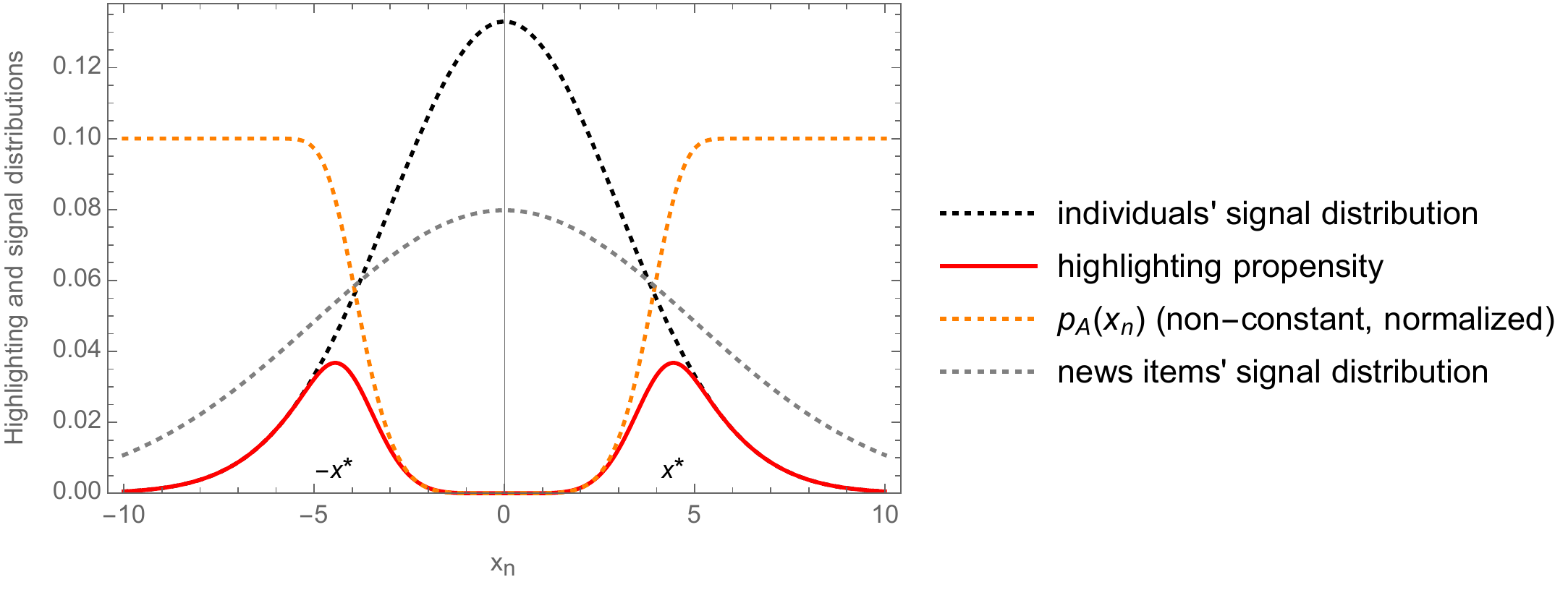}
    \caption{Individuals’ signal distribution and highlighting propensity in the non-flat case, and items' signal distribution for $\widehat{\theta}=\theta=0$; $-x^*, x^*$ denote the values of $x_n$ where the highlighting propensity is locally maximal.}
  \label{fig_highlightdistr}
\end{figure}

Figure  \ref{fig_highlightdistr} provides a graphical representation of the assumed signal distributions of individuals (black dashed line) and of news items (gray dashed line), in the case where $\widehat{\theta}=\theta=0$ and $\sigma_{x}^2<\sigma_{y}^2$. The Figure also plots the distribution of the probability of being an active type (i.e., of highlighting news items provided that its signal is sufficiently close to the individual prior) in the non-flat case (orange dashed line). The red solid line shows the resulting distribution of the highlights in the non-flat case.

\subsection{Platform and Ranking Algorithm}

We consider {\em popularity-based rankings} that evolve as a function of the clicking and highlighting behavior of the individuals, as well as {\em personalized rankings} that may further depend on the identity of the individuals doing the search.

\subsubsection{Popularity Ranking}  \label{sect:popranking}

After each individual makes her choices, the algorithm updates the popularity of each news item such that a click has a weight of 1 and a highlight has a weight of $\eta \in \R_{+}$. That is, starting from $\kappa_{0,m} \in \R_{+}$, the {\em popularity} of each news $m$, $\kappa_{n,m}$, for $n \ge 1$, is updated according to:
\begin{equation} \label{eq:clickpop}
\kappa_{n,m} = \kappa_{n-1, m} +
\left\{
\begin{array}
[c]{cl}
0 & \text{ if } m \text{ is not clicked on by } n \\ 
1 &  \text{ if } m \text{ is clicked on and not highlighted by } n \\ 
1 + \eta & \text{ if } m \text{ is clicked on and highlighted by } n .
\end{array}
\right. 
\end{equation}
The {\em ranking} of the news that individual $n$ sees  $\left( r_{n,m} \right)_{m \in M}$ is inversely related to the popularity before she clicks: 
\begin{equation}
 r_{n,m} < r_{n,m'} \iff \kappa_{n-1,m} < \kappa_{n-1,m'} . 
 \end{equation}
For convenience we also keep track of the traffic a news item receives without counting the highlights. Hence,  starting again from $\widehat{\kappa}_{0,m} = \kappa_{0,m} \in \R_{+}$, the {\em number of clicks} on news item $m$, $\widehat{\kappa}_{n,m}$, for $n \ge 1$, is updated according to:
\begin{equation} \label{eq:clicks}
\widehat{\kappa}_{n,m} = \widehat{\kappa}_{n-1, m} +
\left\{
\begin{array}
[c]{cl}
0 & \text{ if } m \text{ is not clicked on by } n \\ 
1 &  \text{ if } m \text{ is clicked on by } n .
\end{array}
\right. 
\end{equation}

Similarly, we track keep of the highlight engagement of the individuals, that is based solely on the number of highlights
a news item receives, without counting the clicks. Thus again, defining $\widetilde{\kappa}_{0,m} = \kappa_{0,m} \in \R_{+}$, the {\em number of highlights} of news item $m$, $\widetilde{\kappa}_{n,m}$, for $n \ge 1$, is updated according to:
\begin{equation}
\widetilde{\kappa}_{n,m} = \widetilde{\kappa}_{n-1, m} +
\left\{
\begin{array}
[c]{cl}
0 & \text{ if } m \text{ is not highlighted by } n \\ 
1 &  \text{ if } m \text{ is highlighted by } n .
\end{array}
\right. 
\end{equation}

\subsubsection{Popularity Ranking with Personalization} 
\label{model:personalization}

Consider now the case where the ranking, while still being based on popularity, weights differently the clicks from different groups. This can be done in various ways. Because in this simple model individuals only enter once in the platform, the model does not allow the personalization to be based strictly speaking on previous clicks and highlights. Nevertheless, we assume the algorithm can somehow deduce the sign of the individuals' signals (e.g., using the location of her IP address,  cookies from past browsing history, etc.) so that individuals are naturally divided into two groups, say $x_n \in L$ if their signal satisfies sgn$(x_n) = -1$, and $x_n \in R$ if sgn$(x_n) = 1$. Choices and highlights are determined as above, but the difference is that now there are two rankings, $r_{n,m}^L$ and $r_{n,m}^R$, whereby individuals in $L$ see $r_{n,m}^L$ when doing their search, while individuals in $R$ see $r_{n,m}^R$. Moreover, the ranking of group $g \in \{ L, R \}$ depends only in part on the clicks and highlights of individuals from the opposite group.\footnote{The symbol $g$ denoting the group is not to be confused with the function $g(y)$ describing density the news items' signals.} Specifically, starting from $\kappa^g_{0,m} \in \R_{+}$, the {\em popularity for group} $g$ of each news item $m$, $\kappa^g_{n,m}$, for $n \ge 1$, is updated according to:
\begin{equation} \label{eq:clickpers}
\hspace{-.15in}
\kappa^g_{n,m} = \kappa^g_{n-1, m} +
\left\{
\begin{array}
[c]{cl}
0 & \text{ if } m \text{ is not clicked on by } n \\ 
1 &  \text{ if } m \text{ is clicked on by } n \in g \text{ and not highlighted by } n \\ 1 + \eta & \text{ if } m \text{ is clicked on by } n \in g \text{ and highlighted by } n \\
\lambda &  \text{ if } m \text{ is clicked on by } n \notin g \text{ and not highlighted by } n \\
\lambda(1 + \eta) & \text{ if } m \text{ is clicked on by } n \notin g \text{ and highlighted by } n ,
\end{array}
\right. 
\end{equation}
where the parameter $\lambda$, $0 \le \lambda \le 1$, determines how much clicks and highlights from the opposite group $g' \ne g$ count for the ranking seen by group $g$. When $\lambda=0$ each group sees a fully personalized ranking, independent of the clicks and highlights of the other group. When $\lambda =1$ clicks from both groups count the same, so that the two rankings are identical, and we get back the case of a single ranking as in the previous subsection. 

As before, the {\em ranking} of news item $m$ that individual $n \in g$ sees  $\left( r_{n,m}^g \right)_{m \in M}$ is inversely related to the popularity of $m$ before $n$ clicks: 
\begin{equation}
 r_{n,m}^g < r_{n,m'}^g \iff \kappa_{n-1,m}^g < \kappa_{n-1,m'}^g , \hspace{.1in} g \in \{ L, R \}. 
\end{equation} 
We also keep track of traffic and engagement of news items separately for each group. 
Starting again from $\widehat{\kappa}_{0,m}^g = \kappa_{0,m} \in \R_{+}$, the {\em number of clicks by group } $g$ on website $m$, $\widehat{\kappa}_{n,m}^g$, for $n \ge 1$ and $g \in \{ L, R \}$, is updated according to:
\begin{equation}
\widehat{\kappa}_{n,m}^g = \widehat{\kappa}_{n-1, m}^g +
\left\{
\begin{array}
[c]{cl}
0 & \text{ if } m \text{ is not clicked on by } n   \\ 
1 &  \text{ if } m \text{ is clicked on by } n , n \in g \\
0 &  \text{ if } m \text{ is clicked on by } n , n \notin g ,
\end{array}
\right. 
\end{equation}
The same can be done by counting the highlights without counting the clicks, which yields the measure of highlight engagement of group $g$, $\widetilde{\kappa}_{n,m}^g$, for $g\in \{ L, R \}$.

\subsection{Key Parameters and Evaluation Indices}\label{sec:indexes}

In summary, the model is described by the following items: 

\begin{itemize}
\item Information structure: $(\theta, \widehat{\theta}; \sigma_x^2, \sigma_y^2)$
\item News items: $M; M_{-}, M_{+}$
\item Individuals' clicking and highlighting behavior: 
$N; ( (p_C, p_E, p_I)$, $(\gamma_C, \gamma_E, \gamma_I), \beta)$; $((p_P, p_A), \alpha)$
\item Platform and ranking algorithm: $(\eta, \lambda)$
\end{itemize}
It describes a process, where $N$ individuals sequentially access a digital platform to click and potentially highlight one of the $M$ endogenously ranked items. Parameters and distributions for the signals and preferences are fixed.

To evaluate the effect of the highlighting and personalization parameters of the ranking algorithm on various aspects of social welfare, we formally define a few indices. Let $y(n) \in M$ denote the signal of the news item clicked on by individual $n$, and let $L$ ($R$) denote the individuals with signals $x_n$ with sign$(x_n)=-1$ ($=+1$). 
Then we can define the following indices:

\begin{itemize}
\item {\em Engagement} on item $m$ by group $g$: $ENG_m^g = \widehat{\kappa}^g_{N,m} +
\widetilde{\kappa}^g_{N,m}$ (clicking and highlighting by grop $g$);
\item {\em Total Engagement}: $ENG= \sum_{m \in M} \left( ENG_m^L + ENG_m^R \right)$ (total clicking and highlighting);
\item {\em Misinformation}: $MIS =\frac{1}{N}\sum_{n \in N} 
\left| y(n) - \theta \right|$;
\item {\em Polarization}: $POL = \frac{1}{N} \left| \sum_{n\in R} y(n) - \sum_{n' \in L} y(n') \right|$;
\end{itemize}

The first two indices are meant to capture one the key dimension digital platforms care about:  user engagement, or the expected amount of activity generated by the individuals.\footnote{In particular, the willingness to increase engagement was behind the boost in $\eta$ implemented in 2018 by Facebook with the stated objective of increasing {\em meaningful social interactions} (see Section \ref{sec:empirical} for a related discussion).}  The third index is a straightforward measure of misinformation capturing the average distance between the information carried by the news items chosen by individuals and the true state of the world. The forth index measures polarization as the average distance between the information provided by the news items chosen by individuals in group $R$ with the respect to the one provided by the news items chosen by individuals in group $L$.\footnote{Notice that we abstract from the specific belief updating of each individual. Our focus is on the comparative static effect of changes in the algorithm parameters ($\eta$ and $\lambda$) on misinformation and polarization. Accordingly, the proposed misinformation and polarization indices will be informative on such effects as long as individuals update their beliefs in the direction of the signal carried by the news item they click on, $y(n)$.}

Finally, we aim to asses the impact of highlighting and personalization weights on the overall social welfare. In principle, such weights might have opposite effects on the welfare of the platform versus the one of the individuals. Accordingly, we consider two sources of social welfare, namely, one based on what concerns the platform: generating high levels of engagement ($ENG$); and another based on what may concern the users of the platform: guaranteeing low levels of misinformation ($MIS$) and polarization ($POL$). For convenience, we capture all these aspects in a single measure of {\em welfare} of the form:
\begin{equation}\label{eq:welfare}
    W_{\psi}(\eta, \lambda) = \psi \cdot ENG ( \eta, \lambda) - (1-\psi) \cdot MIS( \eta, \lambda) \cdot POL( \eta, \lambda),
\end{equation}
where $0 \le \psi \le 1$ is a weight for the relative importance of the platform's welfare (high $ENG$) relative to the users' welfare (low $MIS$ and $POL$).

\section{Engagement, Misinformation, Polarization and Social Welfare: Analytical Results}

\label{section:results}

In this section, we study the mechanics behind the dynamic interplay between individuals’ clicking and highlighting behavior and the platform’s ranking algorithm, based on popularity and personalization. We start by discussing what happens to key variables when the highlighting behavior affects the ranking of the news items. In our model, it matters greatly whether individuals' highlighting behavior is what we call {\em flat} or {\em non-flat}. To simplify the discussion, in this section we assume that $\widehat{\theta} = \theta$. That is, we focus on the symmetric case where signals are symmetrically distributed around the benchmark.\footnote{Allowing for heterogeneous benchmarks across individuals $\widehat{\theta}_n$ does not affect the key insights of the model, see \ref{section:hetaverageop}. Similarly, allowing for small asymmetries in the distribution of signals with respect to the benchmark ($| \widehat{\theta} - \theta | < \min \left\{ \frac{\sigma_x}{4}, \frac{\sigma_y}{4} \right\}$) does not change the results qualitatively. Instead, allowing for large asymmetries may change the results. \ref{section:NCaverageop} discusses the case where $\widehat{\theta}$ and $\theta$ are far apart.}

We first present a preliminary discussion of the mechanism linking $\eta$ and the dynamics of clicking and highlighting in the flat and non-flat case. We then provide analytical results characterizing the impact of $\eta$ and $\lambda$ on platform engagement, misinformation and polarization.

\subsection{Preliminaries: Increasing Meaningful Social Interactions in the Flat and Non-Flat Cases}

A key objective of our model is to help understand the effect of the weight of a highlight ($\eta$) on engagement, misinformation, and polarization.
As we will show, especially on misinformation and polarization, the effect crucially depends on the propensity with which individuals decide to highlight news items. In particular, what matters in our model is whether the function $p_A$ describing the probability of an individual being active is flat or whether it is non-flat. The reason is that, as the weight of a highlight increases, more items close to the truth ($y$'s $\approx \theta$) are highlighted in the flat case and consequently also clicked on, whereas, in the non-flat case, it is items farther from the truth ($y$'s $\approx -x^*, x^*$; see Fig.~\ref{fig_highlightdistr}) that are highlighted and consequently also clicked on more frequently. 

\begin{figure*}[!t]
  \centering
\includegraphics[width=.45\linewidth]{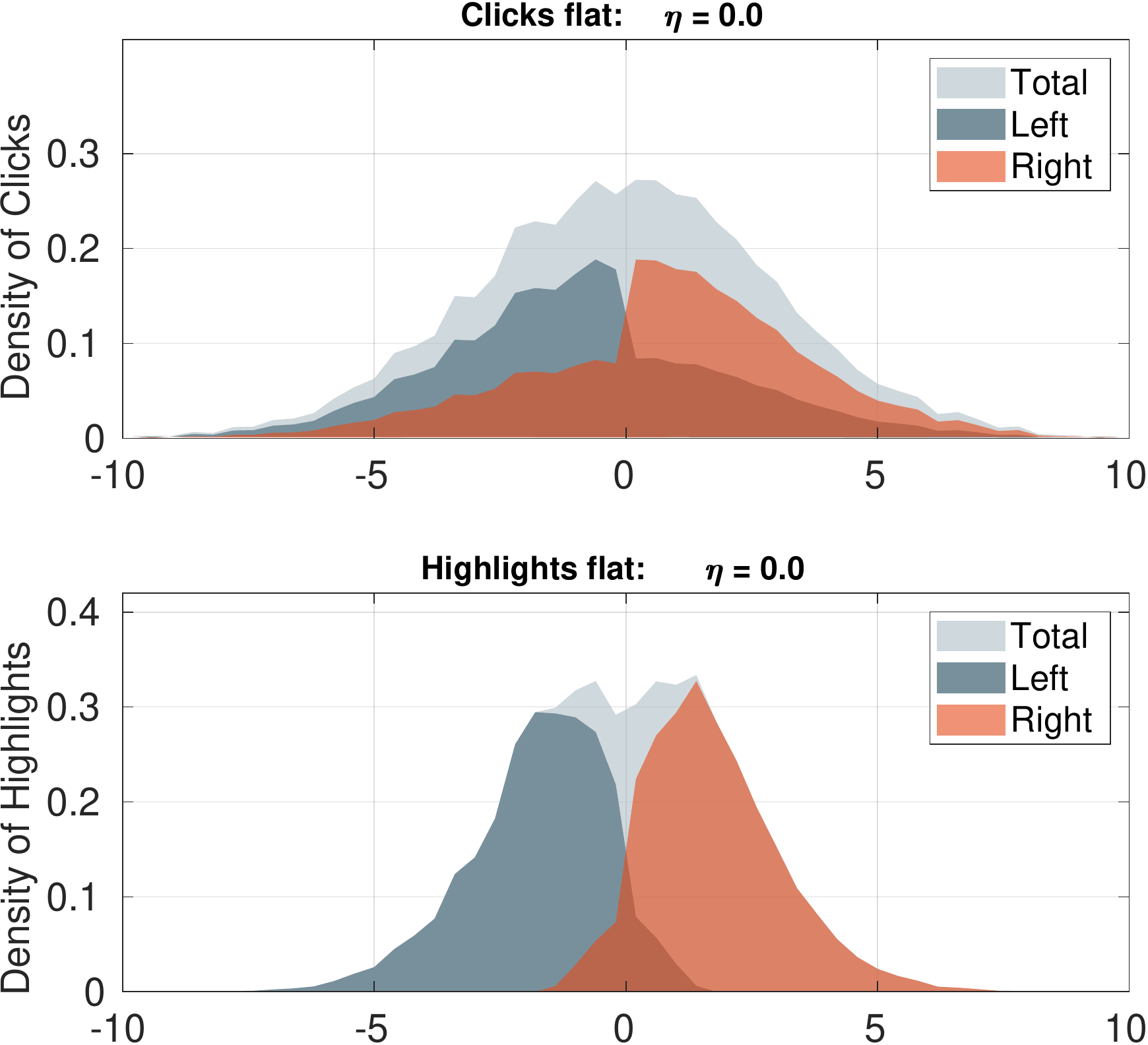}
\includegraphics[width=.45\linewidth]{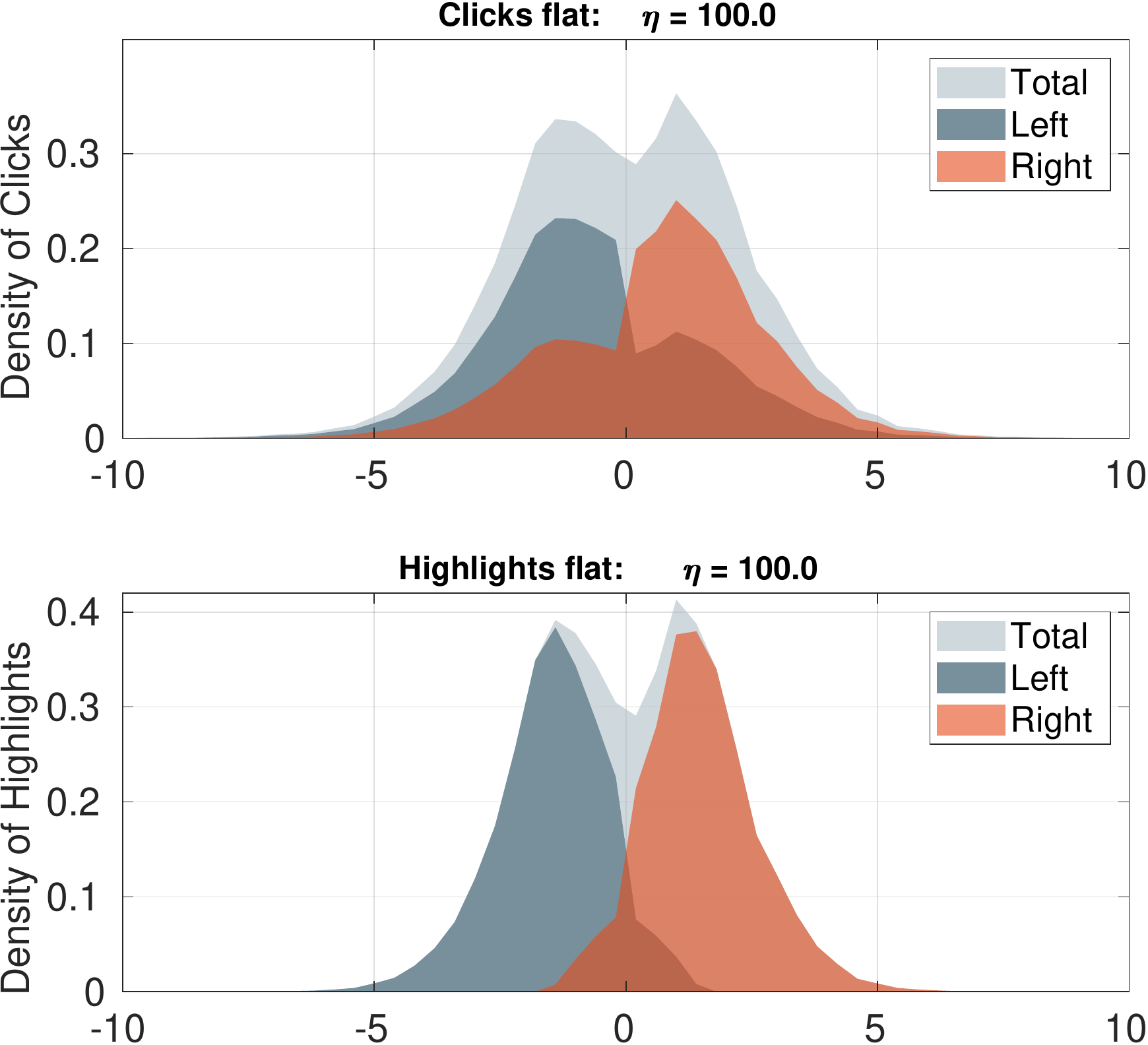}
     \caption{Users’ clicking (top) and highlighting (bottom) behavior for $\eta=0$~(left) and for $\eta=100$~(right) in the \emph{constant (flat)} highlighting case.
     For increasing values of $\eta$, both polarization and misinformation \emph{decrease}. Polarization decreases from an average value of $1.8$ (stev $0.5$) to $1.3$ (stdev $0.3$), and the misinformation also decreases
     from an average value of $2.4$ (stev $0.6$) to $1.7$ (stdev $0.3$). See Section~\ref{section:simulations} for more details.
     }
\label{fig:flatcase}
\end{figure*}

\subsubsection{Flat Case} \label{sec:flat}
In the flat case, increasing the weight on highlighting ($\eta$) can be shown to have some nice properties for both the platform and consumer welfare, namely, it increases engagement, while also reducing both polarization and misinformation. Consider what we referred to as the {\em flat case} for the active types in the previous section. That is, 
suppose a fixed share ($p_A$) of individuals who read a given article are also willing to highlight it, provided the news item's signal is sufficiently close to the individual's signal ($y_m \in H(x_n)$). 
Then as $\eta$ increases, articles that get highlighted increase in total popularity and hence go up higher in the ranking, meaning that they are in turn also more likely to get clicked on. Since both the news items' and the individuals' signals are normally distributed, there is a relatively higher mass of individuals with signals around the truth ($\theta$) and so, such individuals are more likely to read and highlight articles closer to the truth. This pushes them further up in the ranking. 
Hence, higher values of $\eta$ will tend to concentrate clicking around the truth.
This decreases polarization and misinformation, and, because there are relatively more individuals with signals around the truth, due to their normal distribution, it also increases engagement. 

Thus increasing $\eta$ in the flat case directly increases what Facebook calls {\em meaningful social interactions}, and at the same time concentrates clicking around the truth, thereby decreasing misinformation and polarization.  
This is illustrated by Figure~\ref{fig:flatcase}, where the panel on the top right shows the (simulated) clicking distribution for a larger $\eta$ more concentrated around the truth, than for the corresponding graph on the top left, with smaller  $\eta$; the bottom graphs show the highlighting distribution, which is a key part of engagement and which is also more concentrated around the truth, moreover, it can be checked that the overall number of highlights is higher in the case depicted in the graph on the bottom right with larger $\eta$, than in the one on the bottom left with smaller $\eta$.

\subsubsection{Non-Flat Case: Crowding Out the Truth} \label{sec:nonflat}
In the non-flat case, increasing the weight on highlighting ($\eta$) can have desirable properties for the platform, namely, higher engagement, but not necessarily for users since it can result in higher misinformation and higher polarization.
To see this, consider now what we referred to as the {\em non-flat case} for the active types in the previous section. That is, suppose the share of individuals ($p_A$) who read an article and decide to highlight it is non-flat and takes the form as described in Eq.~(\ref{eq:BMA}).
Individuals with more extreme signals are more likely to highlight an article, provided it is sufficiently close to their own signal ($y_m \in H(x_n)$). But because the individuals' own signals are normally distributed, this makes the highlighting distribution bimodal with the two modes far away from each other and from the truth (as shown by the red lines in Figure~\ref{fig_highlightdistr}). And since increasing $\eta$ makes the clicking distribution inherit the basic shape of the highlighting distribution, this implies that, in the non-flat case, the clicking distribution goes from being concentrated around the truth (for smaller $\eta$) to being increasingly bimodal with the two modes further away from each other (for larger $\eta$). Importantly, increasing $\eta$, leads to higher engagement since more individuals willing to highlight items will be clicking on items they are actually interested in highlighting. But, at the same time, it also leads to more polarization and misinformation, as individuals are less likely to click on items near the truth ($y$'s $\approx \theta$) and  more likely to click on items further away from the truth  ($y$'s $\approx -x^*, x^*$; see Fig.~\ref{fig_highlightdistr}). This is the phenomenon we refer to as {\em crowding out the truth}.

Thus increasing $\eta$ in the non-flat case,  increases {\em meaningful social interactions}, but also moves clicking away from the truth leading to higher misinformation and polarization. 
This is illustrated by Figure~\ref{fig:non-flatcase}, where the panel on the top right shows the (simulated) clicking distribution for a larger $\eta$ strongly bimodal and farther away from the truth, as compared to the corresponding graph on the top left, with a smaller $\eta$; again, the bottom graphs show the highlighting distribution, it can be checked that the overall level of highlights is higher for the graph on the bottom right with larger $\eta$ than the one on the bottom left with smaller $\eta$.

\begin{figure*}
  \centering
\includegraphics[width=.49\linewidth]{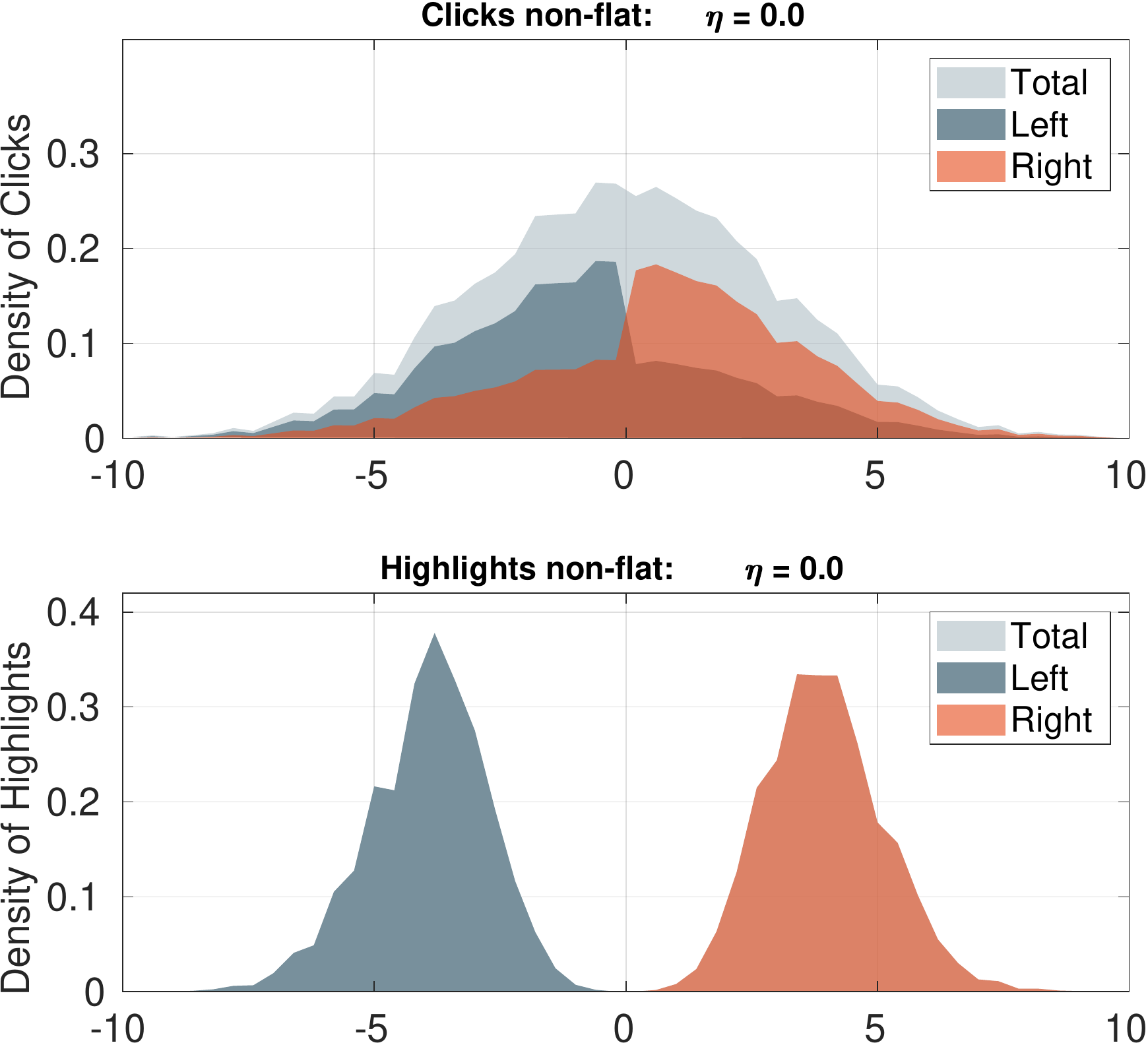}
\includegraphics[width=.49\linewidth]{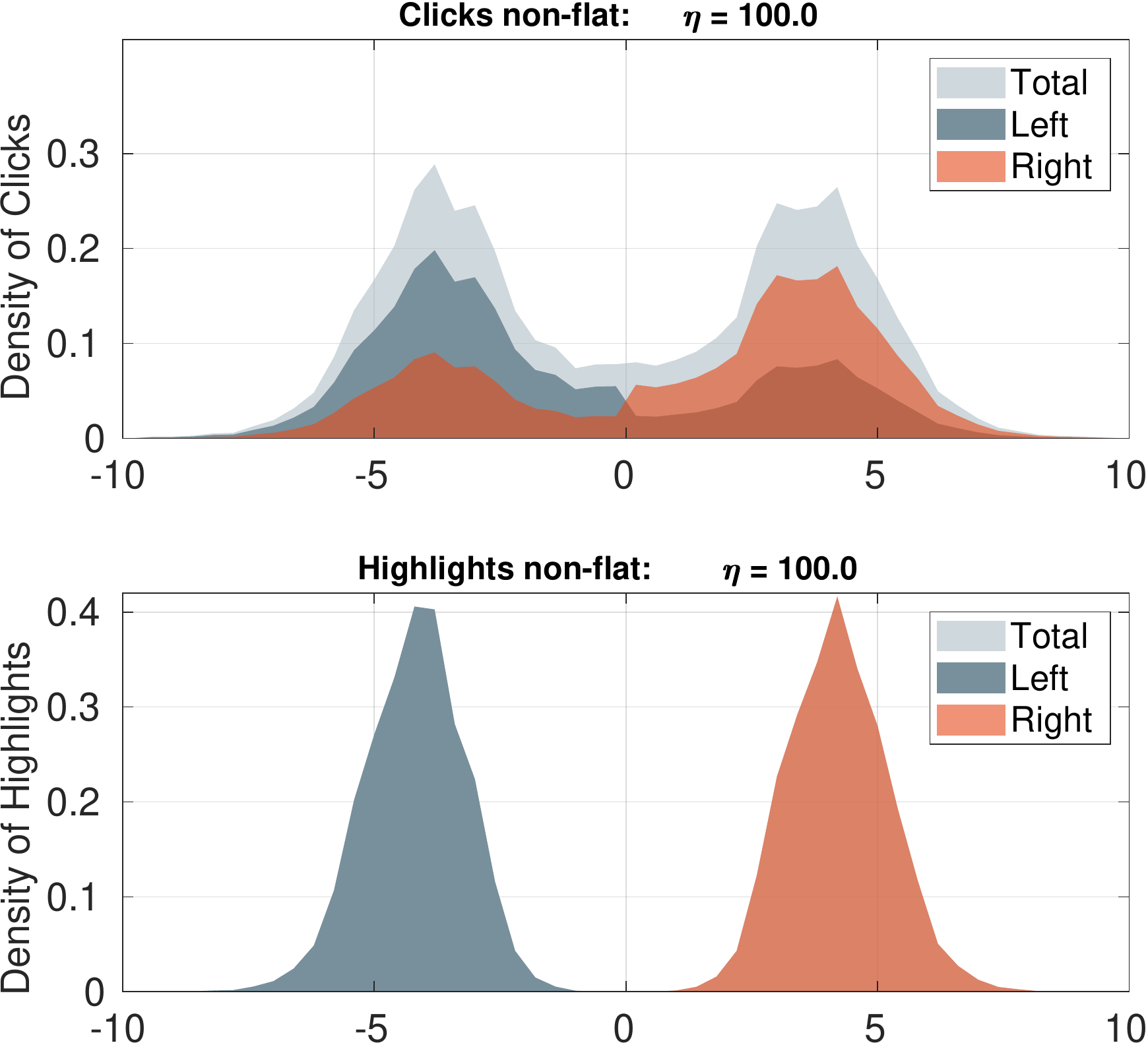}
     \caption{Users’ clicking (top) and highlighting (bottom) behavior for $\eta=0$~(left) and for $\eta=100$~(right) in the \emph{non-constant (non-flat)} highlighting case.
     For increasing values of $\eta$, both polarization and misinformation \emph{increase}. Polarization increases from an average value of $1.8$ (stev $0.5$) to $2.8$ (stdev $0.4$), and the misinformation increases from an average value of $2.4$ (stev $0.6$) to $3.6$ (stdev $0.4$). See Section~\ref{section:simulations} for more details.}
  \label{fig:non-flatcase}
\end{figure*}

In the subsections that follow, we look at the above phenomena more formally and also connect them to welfare evaluations for the platform and consumers.

\subsection{Analytical results}

We here look in more detail at the model and what it implies in terms of the key indices presented in the previous section. In order to compute those it is important to be able to characterize at least to some degree of approximation the actual clicking and highlighting behavior of the individuals, while keeping track of the dynamic feedback between clicking, highlighting and ranking. We do this by characterizing the limit clicking and highlighting distributions, since these ultimately determine what to expect in terms of opinions and engagement.

A central feature of the ranking algorithm is its dependence on the popularity of the different items. In order to fix ideas and more clearly explain the basic mechanisms at work, we will make an assumption that directly relates the expected rank of an item to its expected popularity, absent ranking. This will then allow us to analytically characterize in a particularly transparent way the limit clicking and highlighting distributions. Section \ref{section:simulations} provides numerical simulation results when relaxing such assumption. In particular, the key comparative statics results that we derive in this section (in the case of linear approximation of the expected ranking) are consistent with the ones shown by the numerical simulations (where no such approximation is imposed). 

\subsubsection{Limit Clicking and Highlighting Behavior}

Define the {\em expected popularity of an item with signal $y_m=y$, absent ranking}, as the sum of the expected clicking and highlighting propensities, absent ranking, but weighted by $\eta$:
\begin{equation} \label{eq:pi}
\pi(y) = 
\frac{1+ \eta \cdot \mu_H(y)}{M \cdot (1+ \eta \cdot \bar{\mu}_H) + \eta \cdot (\mu_H(y) - \bar{\mu}_H )}, 
\end{equation}
where 
\[ \mu_H(y) = \int_{ x \in H^{-1}(y)}  p_A(x) f(x) dx   \hspace{.1in} \mbox{  and  } \hspace{.1in}  
H^{-1}(y)= \{ x \in \R \, | \, y \in H(x) \},  \]
where $f$ is the density of the individuals' signals, and so that $\bar{\mu}_H = \int \mu_H(y) g(y) dy$, where $g$ is the density of the news items' signals.\footnote{Eq.~(\ref{eq:pi}) is derived from:\[ \pi(y_m) = \frac{\int \sum_{k \in \{ C, E, I \}} p_k  \cdot \varphi_{n,m} \cdot \left( 1 + \eta \cdot p_A(x_n) \cdot 1_{\{ x_n \in H(y_m) \}} \right) f(x) dx}{\sum_{m' \in M} \int \sum_{k \in \{ C, E, I \}} p_k  \cdot \varphi_{n,m'} \cdot \left( 1 + \eta \cdot p_A(x_n) \cdot 1_{\{ x_n \in H(y_{m'}) \}} \right) f(x) dx} , \]
taking averages over $T$ repetitions for $T\rightarrow \infty$.}

An assumption that we maintain in this and the following subsections is that the expected rank of a given news item with signal $y \in \R$ is approximated by a linear decreasing function of the expected popularity of that item, absent ranking:
\begin{equation} \label{eq:exprank}
    r(y) \approx \zeta_0 - \zeta_1 \cdot \pi(y),
\end{equation}
where $\zeta_0, \zeta_1 >0$ are constants.\footnote{Obtaining the expected value for the rank $r(y_m)$ of a given item $y_m$ drawn among $M > 2$ items is an open problem that requires computing the distribution over the limit rankings of a relatively complicated process. \cite{analytis_2022} study a related problem (without highlighting) and are able to derive such a distribution for the case of two items $M=2$. Indeed, it can be checked that in that case, for our model with no highlighting, there is an exact linear relationship between the expected rank and the clicking propensity as assumed in Eq.~(\ref{eq:exprank}).}  Figure~\ref{fig:sim_linear} shows that the assumption is quite accurate for simulations with various values of $\eta$ in the flat and non-flat cases. Moreover, as pointed out before, the simulation results present in Section~\ref{section:simulations}---obtained without imposing such approximation---further suggest that the assumption is relatively innocuous within our overall framework in terms of deriving our main qualitative results.

\begin{figure*}[t]
  \centering
  \includegraphics[width=.3\linewidth]{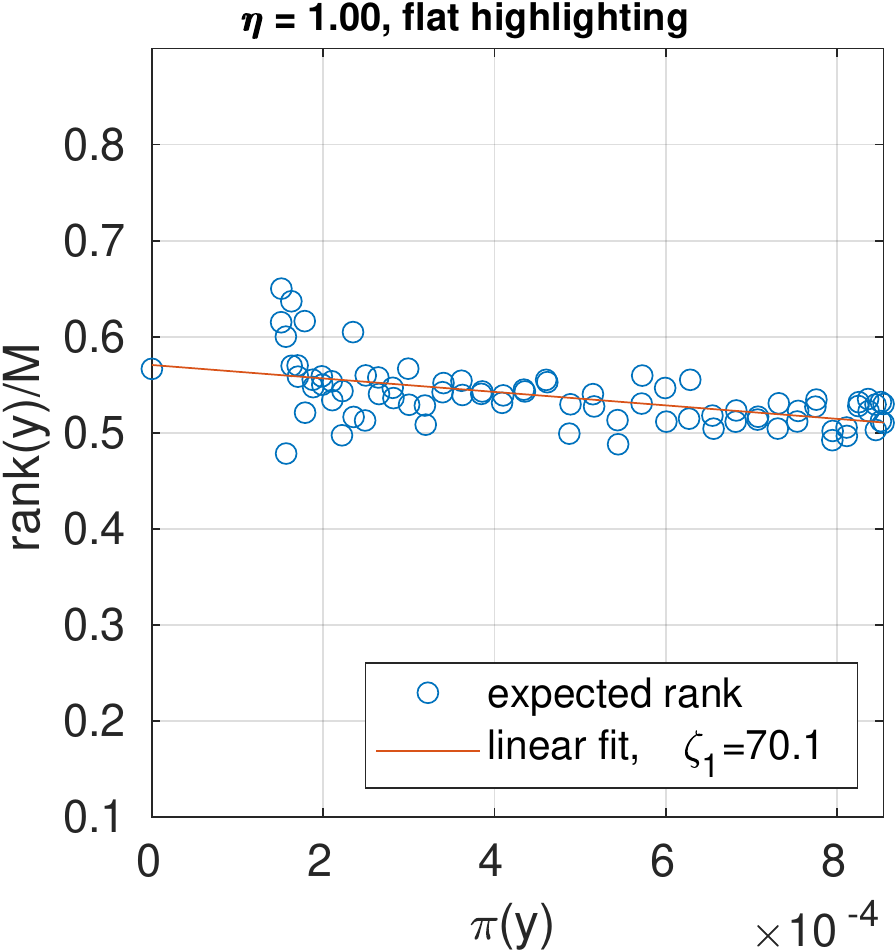}
  \includegraphics[width=.3\linewidth]{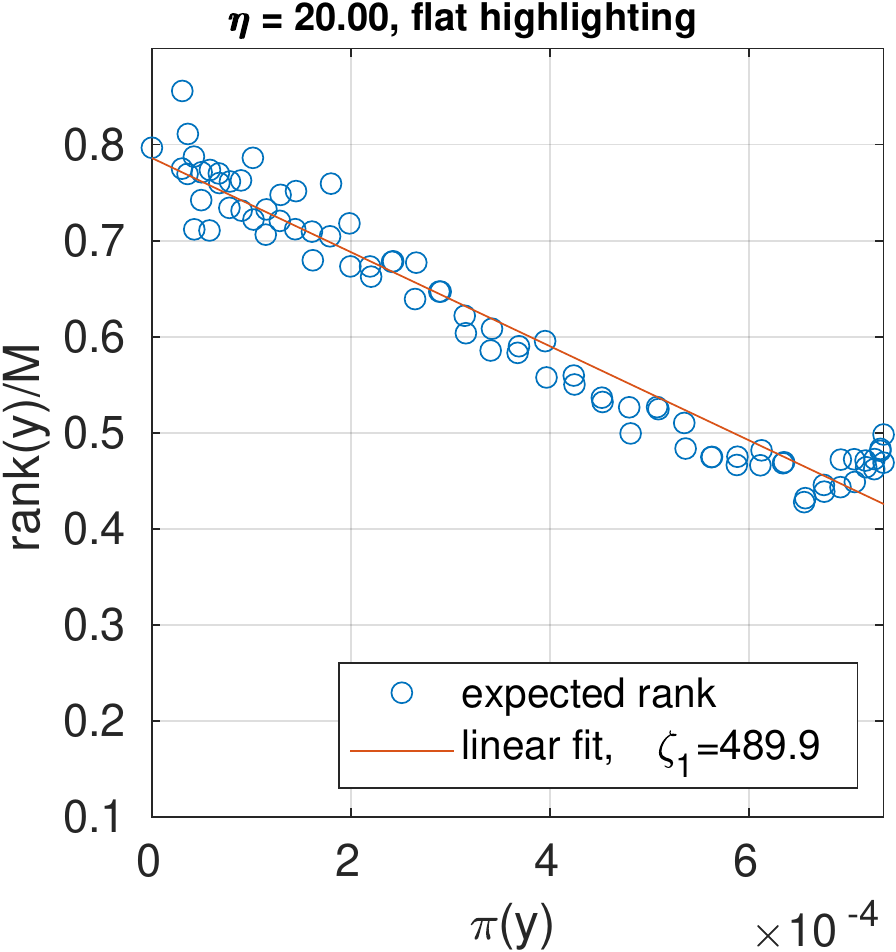}
  \includegraphics[width=.3\linewidth]{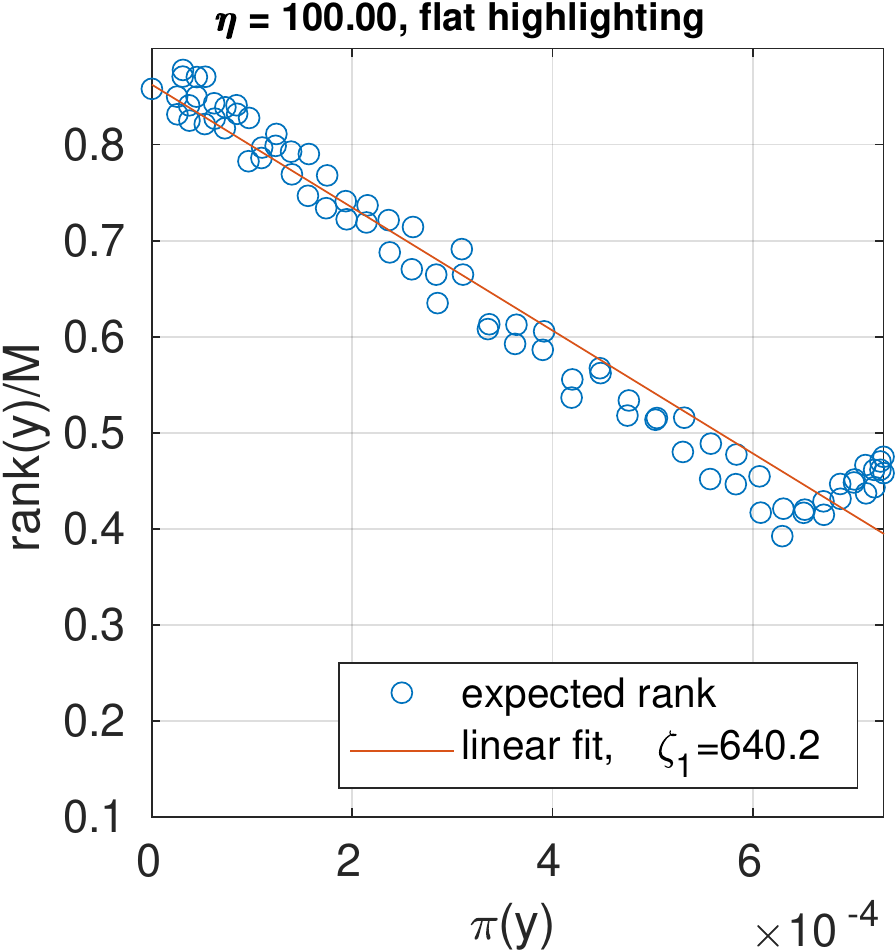}
  \includegraphics[width=.3\linewidth]{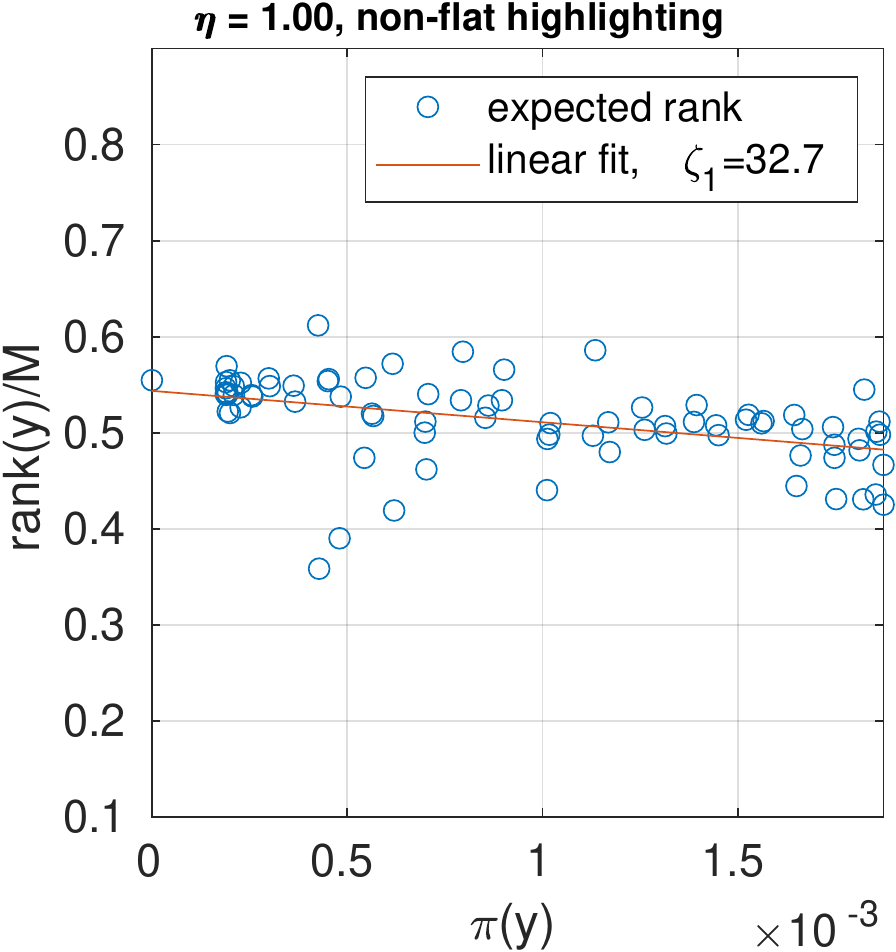}
  \includegraphics[width=.3\linewidth]{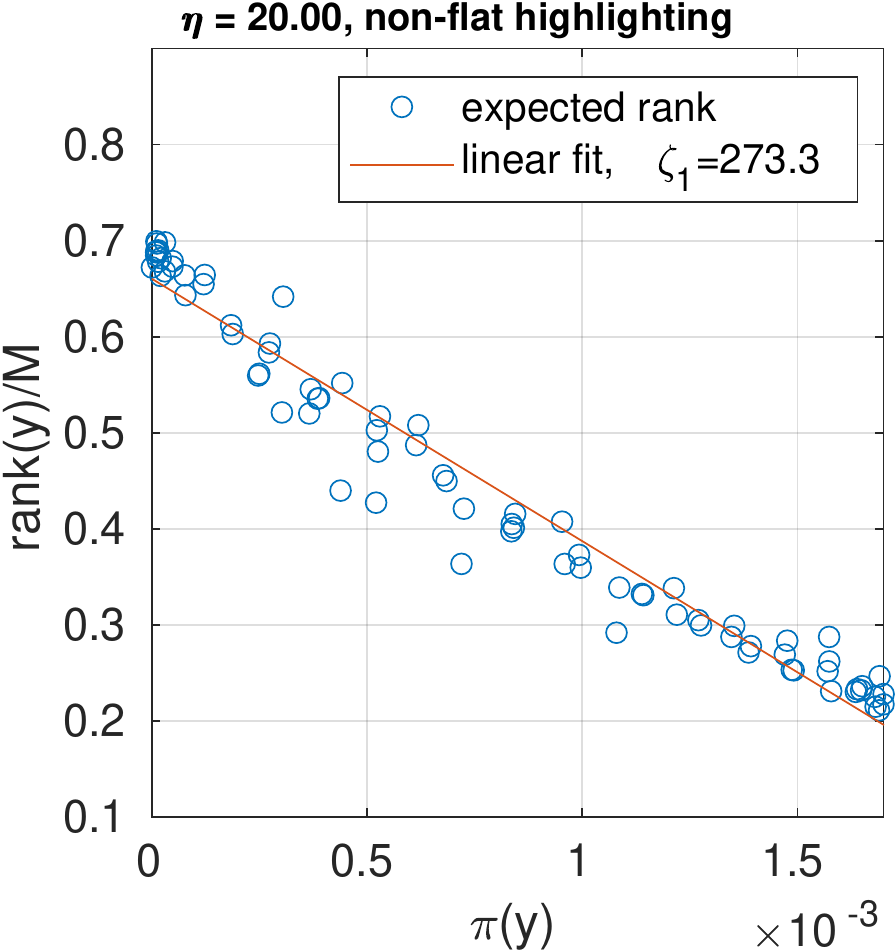}
  \includegraphics[width=.3\linewidth]{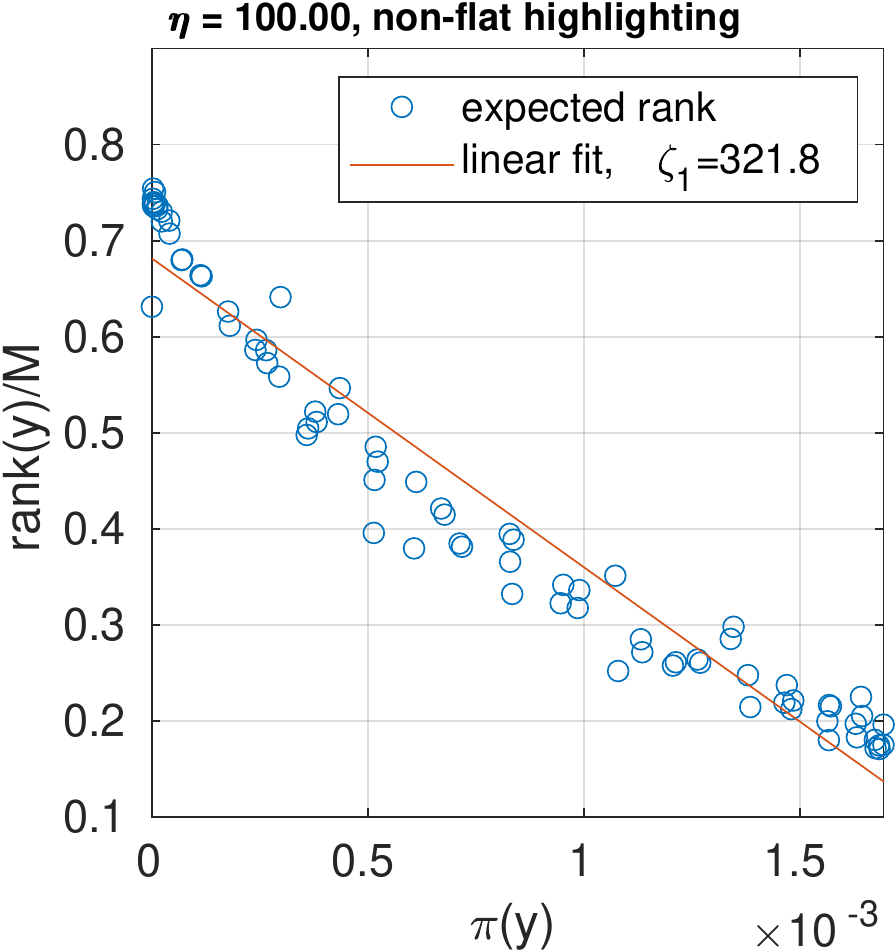}    
    \caption{Linear dependence between the expected rank (blue circles, obtained from simulations) and the expected popularity $\pi(y)$ as in Eq.~\eqref{eq:pi}.
    Red line denotes the best linear fit. 
    To compute each blue dot, we binned the item's signals into $81$ bins (bin-size = $0.2$) and compute the mean popularity and the corresponding mean rank
    from $T=10^3$ experiments, each of them with different $M=20$ item's signals and different $N=5\cdot 10^3$ individual's signals.
    }
    \label{fig:sim_linear}
\end{figure*}

With this assumption it is possible to characterize both the limit clicking and the limit highlighting distributions. Such limit distributions represent expectations of $T$ repetitions (for $T \rightarrow \infty$) of the process described in Section \ref{section:model}. That is, in order to derive the comparative statics properties of our model, we study the limit clicking and highlighting distributions which capture engagement, misinformation and polarization in terms of averages of outcomes based on $T$ possible repeated random draws of the individuals and news items' signals.

\begin{lemma}[Limit Clicking and Highlighting Distributions] \label{lemma:LCD}
Assume Eq.~(\ref{eq:exprank}), then the limit clicking distribution can be approximated as:
\begin{equation}
 LCD(y) \approx  \Lambda_{\beta} ( \pi(y) ) \cdot g(y),    
\end{equation}
 where, $\Lambda_{\beta}$ is a linear function with $\Lambda_{\beta}'>0$ for $\beta>1$, $\pi$ is defined in Eq.~(\ref{eq:pi}), and $g$ is the density of the news items' signals.
Accordingly, the limit highlighting distribution can be approximated as:
\begin{equation}
 LHD(y) \approx \mu_H (y) \cdot LCD(y).
\end{equation}
\end{lemma}
The lemma shows a basic feature of our ranking-based dynamics, namely, that the expected traffic on a given item is driven by its expected popularity, absent ranking. The presence of attention bias enters $\Lambda_{\beta}$ by making it strictly increasing function of $\pi(y)$ for any $\beta>1$. Thus a higher $\pi(y)$ increases the rank of an item with signal $y$, thereby increasing its traffic, the more so, the greater $\beta$. The intuition for the proof follows essentially from the linearity assumption in Eq.~(\ref{eq:exprank}) combined with the functional form of the clicking probabilities assumed in Eq.~(\ref{eq:clickprob}).

\subsubsection{Popularity Based Ranking without Personalization}

In this subsection we assume there is no personalization (so $\lambda = 1$) and focus instead on the comparative statics relative to the popularity parameter for the highlights ($\eta$). As is clear from Eq.~(\ref{eq:clickpop}), the popularity variable ($\kappa_{n,m}$) depends on both clicks and highlights, where highlights are weighted by $\eta$. This carries over to the expected popularity variable ($\pi)$ (see Eq.~(\ref{eq:pi})). Therefore, as $\eta$ increases, the expected popularity of an item and hence its expected traffic is increasingly driven by the highlighting propensity. This observation is a central message of the paper and has important consequences for how the parameter $\eta$ affects engagement, misinformation and polarization, as the following result shows:

\begin{prop} \label{prop:main1}
Assume Eq.~(\ref{eq:exprank}).
If individuals’ highlighting behavior is non-flat ($p_A$ as in Eq.~(\ref{eq:BMA})), then, increasing the weight on highlighting (higher $\eta$) increases user engagement, misinformation and polarization. 
If, instead, the highlighting behavior is flat ($p_A$ constant), then the above results need not hold; a higher $\eta$ increases user engagement and decreases misinformation and polarization. 
\end{prop}

Thus, in the case of non-flat highlighting behavior, increasing $\eta$, increases engagement, but also has the adverse effects for individuals by increasing misinformation and polarization. This is not the case when highlighting behavior is flat, where a higher $\eta$ increases engagement, while decreasing misinformation and polarization, albeit slightly.\footnote{The intuition for the result is sketched in Subsections~\ref{sec:flat} and~\ref{sec:nonflat} above, for the flat and non-flat cases, respectively. Assuming Eq.~(\ref{eq:exprank}) allows for a direct numerical computation of the effects using the analytically derived expressions for the limit clicking and highlighting distribution.}

It is possible to interpret the results of Proposition~\ref{prop:main1} in light of the evidence provided by \cite{bakshy2015exposure}. As discussed above, \cite{bakshy2015exposure} point out that in the case of ``hard" news (e.g, national, political), the propensity to highlight content is indeed higher for individuals with a more extreme prior, whereas the same does not apply to ``soft" news (e.g., entertainment). Proposition~\ref{prop:main1} suggests that social media platforms have an incentive to choose a high level of $\eta$ as this results in a high level of engagement across all types of news contents. Yet, while this is not so much a concern for users' welfare in the case of ``soft" news, we show it might have detrimental effects on misinformation and polarization when it comes to political news contents.

\subsubsection{Popularity Based Personalized Ranking}

We now allow the ranking to be personalized, based on the sign of the individuals' signals as in Section~\ref{model:personalization}. 
It is not difficult to see that Proposition~\ref{prop:main1} continues to hold for any degree of personalization ($\lambda \in [0,1]$). The next proposition addresses the question of what is effect of the personalization parameter ($\lambda$) on engagement, polarization and misinformation. 

\begin{prop} \label{prop:personalization}
Assume Eq.~(\ref{eq:exprank}) and fix $\eta \ge 0$ arbitrarily. Then increasing personalization (lower $\lambda$) increases user engagement and polarization, both when individuals' highlighting behavior is flat and when it is non-flat ($p_A$ as in Eq.~(\ref{eq:BMA})). 
\end{prop}

The fact that more personalization increases polarization is straightforward. Decreasing $\lambda$ makes the rankings of the two groups increasingly less correlated, which in turn makes users in each group more likely to click on items carrying a signal of the same sign as their own. This directly increases the polarization measure $POL$. To see the effect on engagement, note first that users that are active types only share items that are close enough to their own signal ($y_m \in H(x_n)$). As $\lambda$ decreases and the rankings become less correlated, users are more likely to see items that have signals closer to their own more prominently ranked, and are in turn also more likely to click on them. But since items that are more prominently ranked are more likely to be in the set $H(x_n)$, they are also more likely to be highlighted. Overall, whether highlighting behavior is flat or non-flat, a lower $\lambda$ (more personalization) contributes to an increase in $ENG$. 

One effect that the personalization parameter $\lambda$ does not have in our model, differently from the highlighting parameter $\eta$, is that it does not significantly impact misinformation. This is due to the fact that it mainly contributes towards interchanging clicks made from one group on items with signals of the opposite sign with clicks made by individuals from the other group, who have signals of the same sign. While this contributes to increasing polarization it does not really affect misinformation.

\subsubsection{Towards a Socially Efficient Ranking}

Consider the welfare index ($W_{\psi}$) defined in Section \ref{sec:indexes}, Eq~(\ref{eq:welfare}). From the analysis of the previous sections, we can show:
\begin{prop} \label{prop:welfare}
Assume Eq.~(\ref{eq:exprank}).
If individuals’ highlighting behavior is non-flat ($p_A$ as in Eq.~(\ref{eq:BMA})), then, for small values of $\psi$, ($\psi \approx 0$), social welfare  ($W_{\psi}$)  is maximized at $(\eta, \lambda) \approx (0, 1)$, while for large values of $\psi$, ($\psi \approx 1$), social welfare is maximized at $(\eta, \lambda) \approx (\infty, 0)$. 

If instead individuals’ highlighting behavior is flat ($p_A$ constant), then, for small values of $\psi$, ($\psi \approx 0$), social welfare is maximized at $(\eta, \lambda) \approx (\infty, 1)$, while for large values of $\psi$, ($\psi \approx 1$), social welfare is maximized at $(\eta, \lambda) \approx (\infty, 0)$. 
\end{prop}

The above proposition presents the key result of the paper. In the empirically relevant case of non-flat propensity to highlight, there is a clear dichotomy between the desirable weight assigned by the ranking algorithm to the highlights, from the perspective of the platform and from the one of users (and, more generally, of public policy). This resonates with the reports leaked by Facebook’s whistle-blowers, which suggested the conflicting welfare effects created by platform’s 2018 “Meaningful social interactions” update, which boosted the weight given to content sharing in the ranking algorithms. Indeed, while this change increased the overall users’ engagement on Facebook, it seems to have also led to an increase in misinformation and polarization, as predicted by Proposition \ref{prop:welfare}. Section \ref{sec:empirical} presents direct empirical evidence in this regard.

\section{Engagement, Misinformation, Polarization and Social Welfare: Numerical Simulations} \label{section:simulations}

In this section, we provide simulation results for the more general case where no restriction on the linearity of the expected ranking is imposed.  
We run $T=4,000$ independent simulations of the basic model with both non-flat and flat highlighting propensities. Each run corresponds to $M=20$ different news items signals $y_m\sim N(\theta=0,\sigma_y^2=9)$ and to $N=10^5$ individuals signals $x_n\sim N(\theta=0,\sigma_x^2=9)$.
When reporting the key evaluation indices, we only consider the last $2,000$ clicks. This avoids dependence on the initialization of the ranking. As in Section \ref{section:results} we also set $\widehat{\theta}=\theta$.
The proportions of confirmatory, exploratory, and indifferent clicking types are set to $p_C=0.7$, $p_E=0.15$, and $p_I=0.15$, respectively, and their corresponding propensities to $\gamma_C=0.8$, $\gamma_E=0.4$, and $\gamma_I=0.5$, respectively.
The value of $\beta$ determining the attention bias is set to~$\beta=1.25$ and the value of $\alpha$ for the non-flat highlighting probability $p_A$ is $4$. Results are robust to choosing different values of the parameters.

\begin{figure*}[!t]
  \centering
  \includegraphics[width=.45\linewidth]{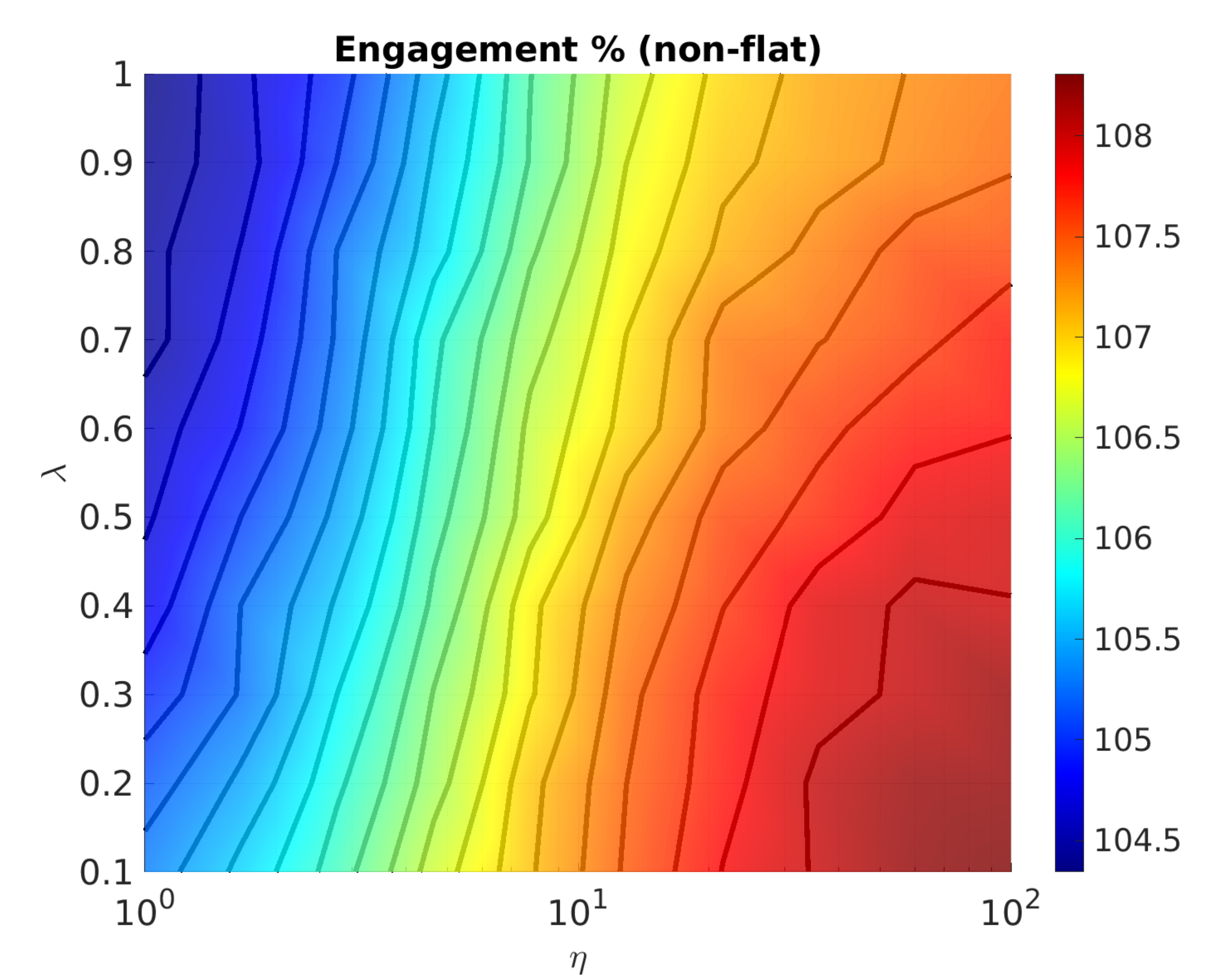}
  \includegraphics[width=.45\linewidth]{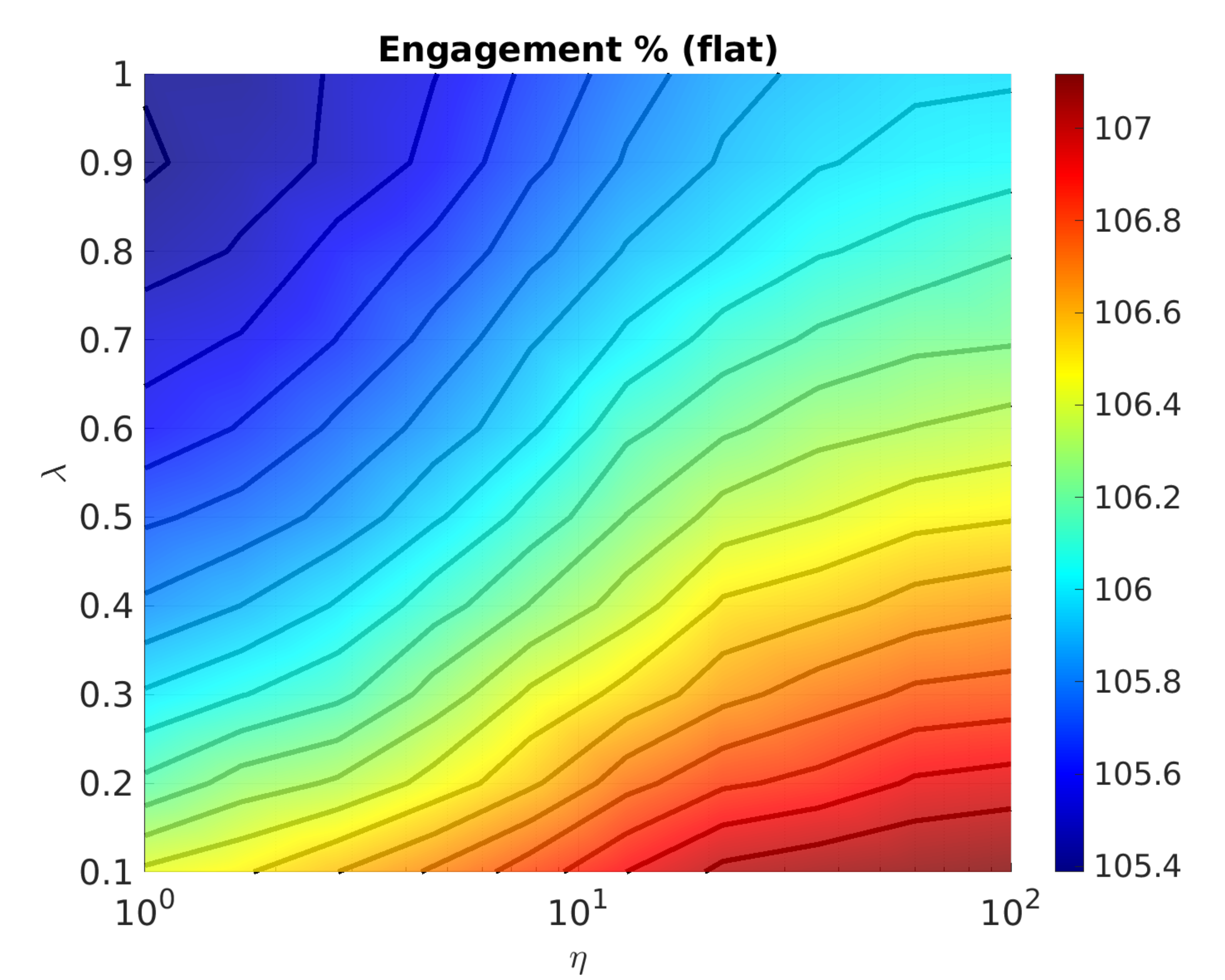}
    \caption{User engagement as a function of personalization $\lambda$ and highlighting weight $\eta$ for non-flat (left) and flat (right) individuals' highlighting behavior.}\label{fig:sim_eng}
\end{figure*}
\begin{figure*}[!t]
  \centering
  \includegraphics[width=.45\linewidth]{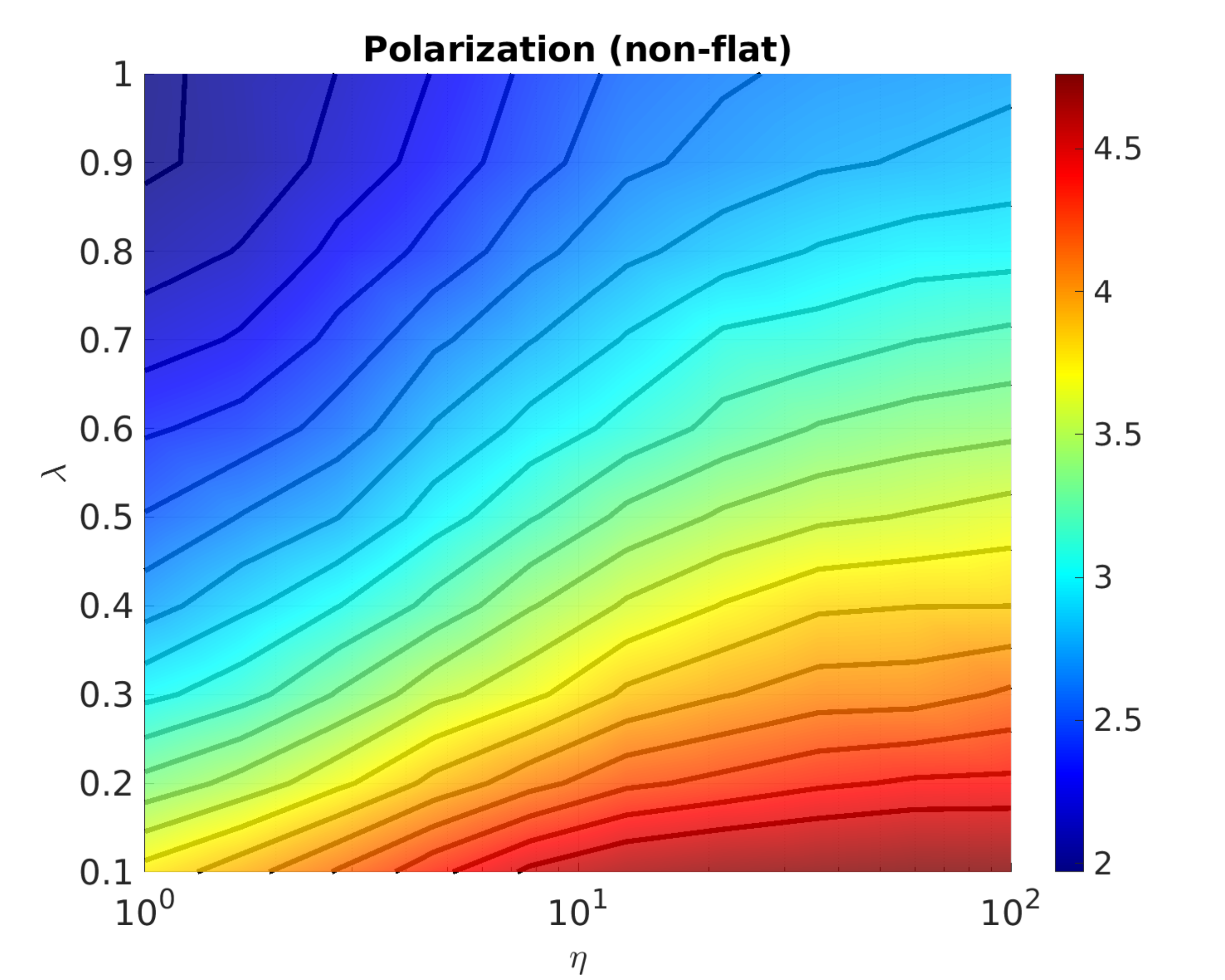}
  \includegraphics[width=.45\linewidth]{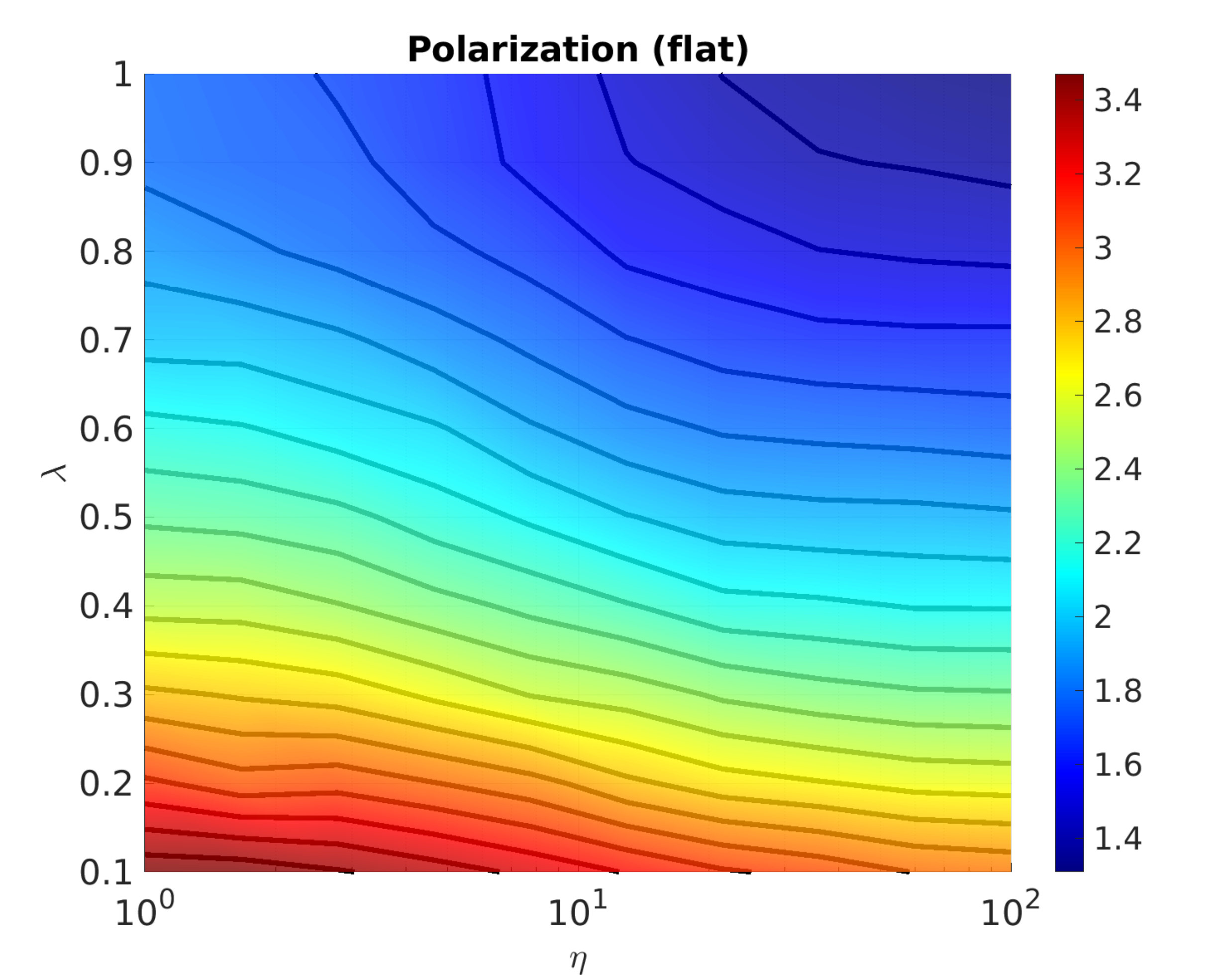}
    \caption{User polarization as a function of personalization $\lambda$ and highlighting weight $\eta$ for non-flat (left) and flat (right) individuals' highlighting behavior.}\label{fig:sim_pol}
\end{figure*}
\begin{figure*}[!t]
  \centering
  \includegraphics[width=.45\linewidth]{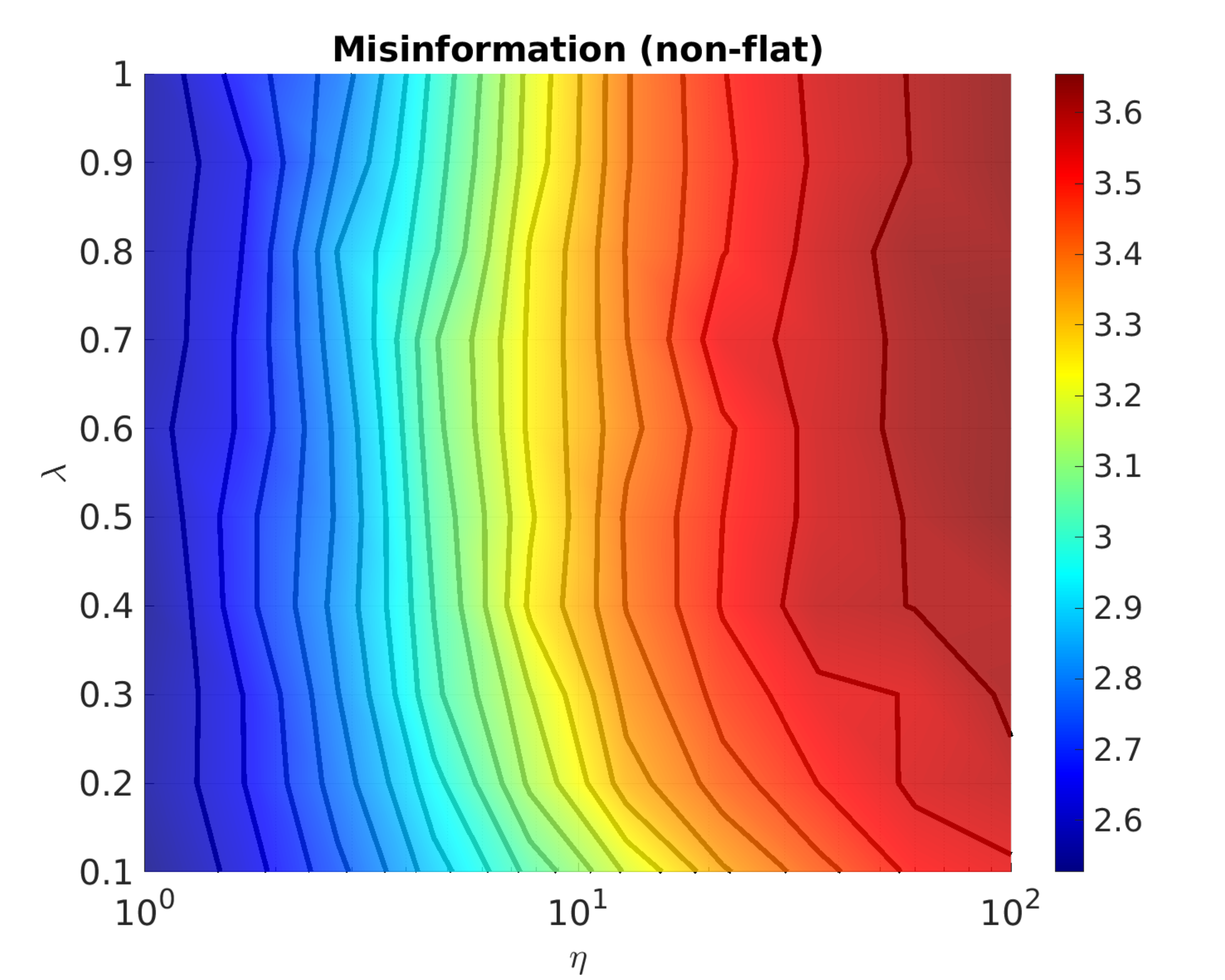}
  \includegraphics[width=.45\linewidth]{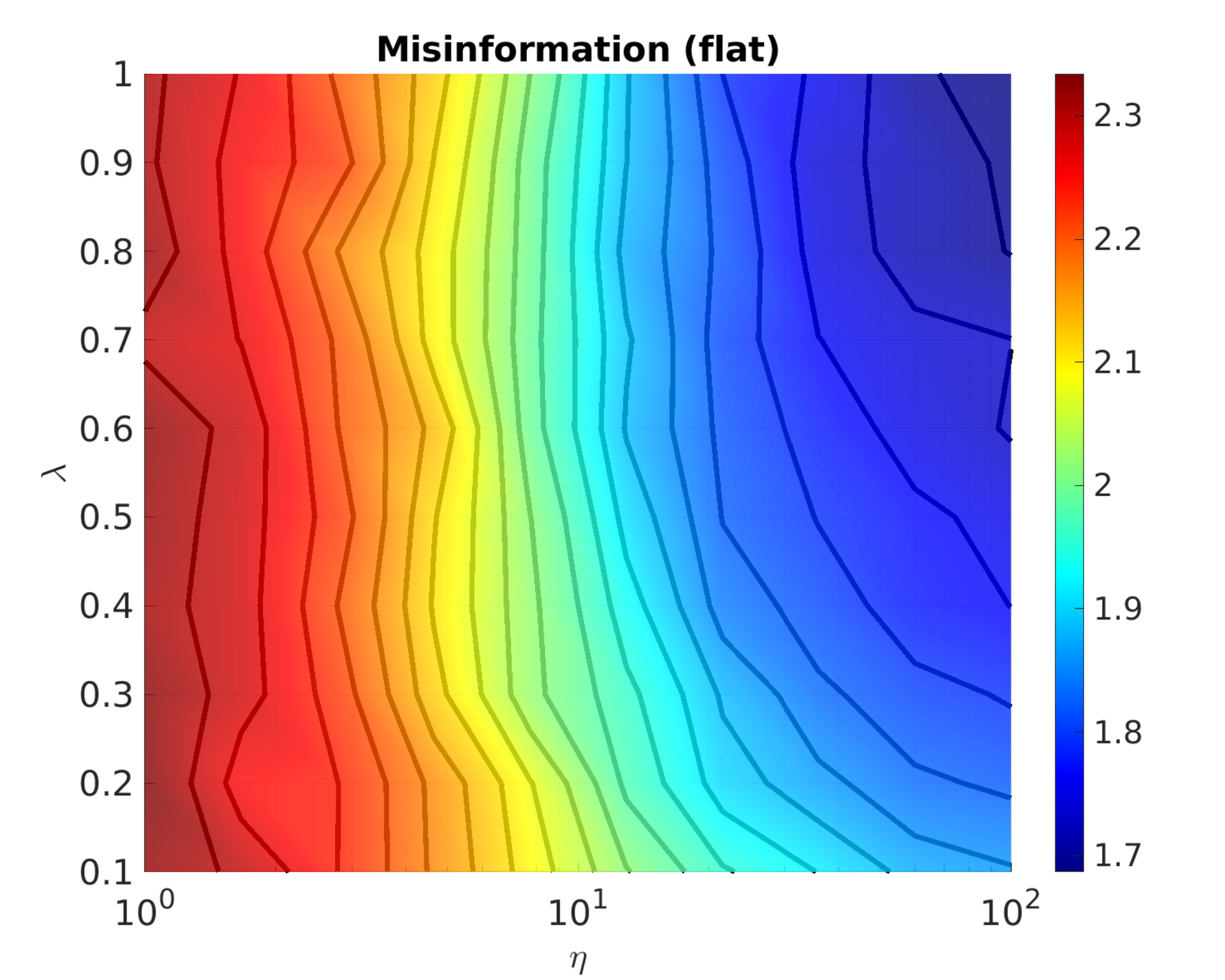}
    \caption{Misinformation as a function of personalization $\lambda$ and highlighting weight $\eta$ for non-flat (left) and flat (right) individuals' highlighting behavior.}\label{fig:sim_mis}
\end{figure*}

The following figures summarize the simulated effects of $\eta$ and $\lambda$ on the key evaluation indices for the non-flat (left) and the flat (right) cases, where the reported measure is an average among all $T$ independent simulations. Figure~\ref{fig:sim_eng} shows total user engagement ($ENG$). We observe that increasing $\eta$ and increasing personalization (decreasing $\lambda$) results in an increase of engagement in both non-flat and flat cases. The dependence on $\lambda$ is more pronounced in the flat case.

Figure~\ref{fig:sim_pol} shows results of user polarization ($POL$). 
In agreement with the analytical results, we observe that the effect of increasing $\eta$ is different depending on the highlighting propensities. In the non-flat case, increasing $\eta$ results in an increase of polarization, whereas in the flat case, it has the opposite effect. However, increasing personalization (decreasing $\lambda$) results in an increase in polarization, for both types of highlighting propensities.

Figure~\ref{fig:sim_mis} shows results of misinformation ($MIS$).
Again, increasing $\eta$ has opposite effects in non-flat and flat scenarios, resulting in an increase of misinformation in the non-flat case, and a decrease the flat case. In contrast to the polarization results, we only observe a weak dependence of misinformation on the degree of personalization $\lambda$, which is only noticeable for higher values of $\eta$.

\begin{figure*}[t]
  \centering
  \includegraphics[width=.45\linewidth]{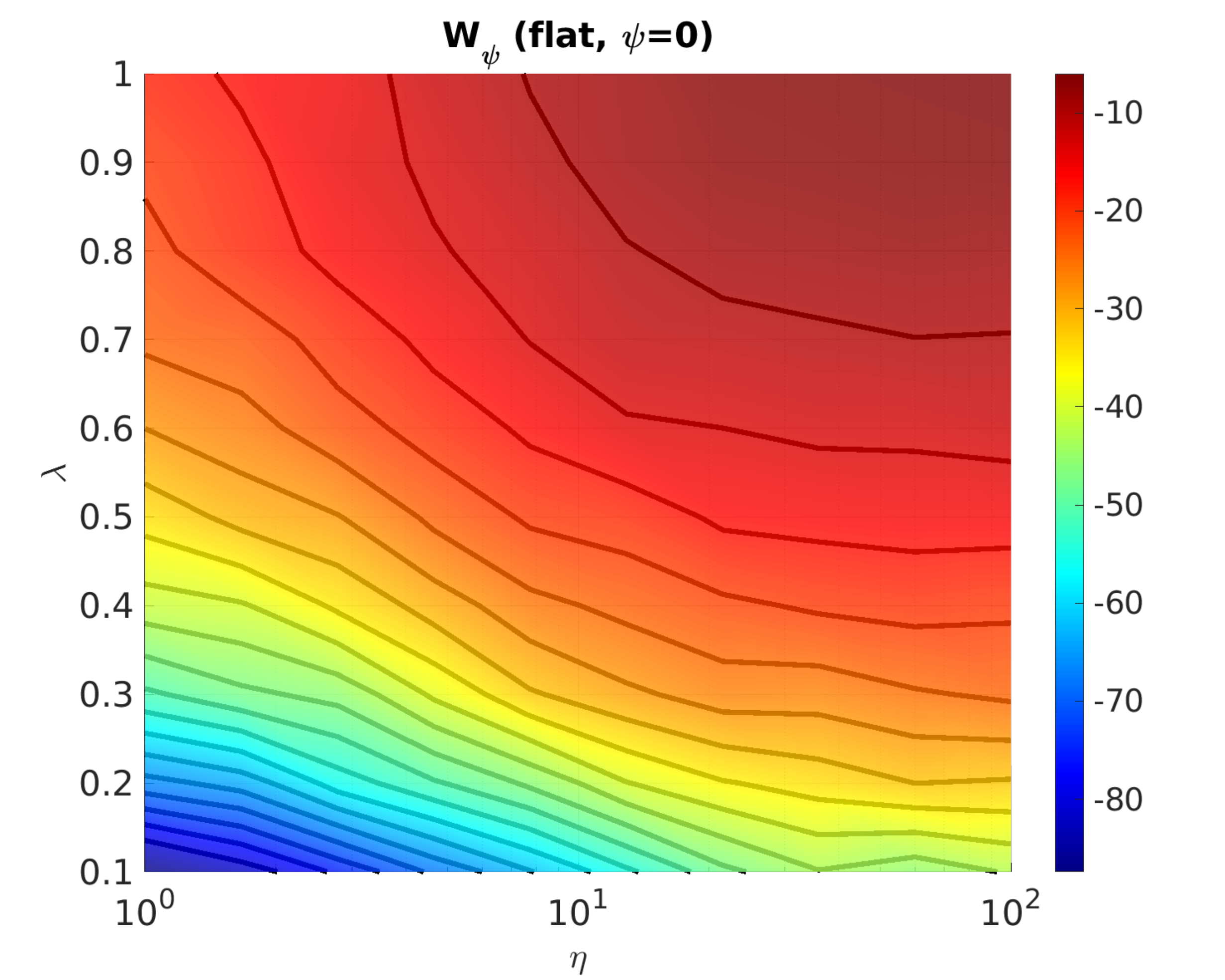}
  \includegraphics[width=.45\linewidth]{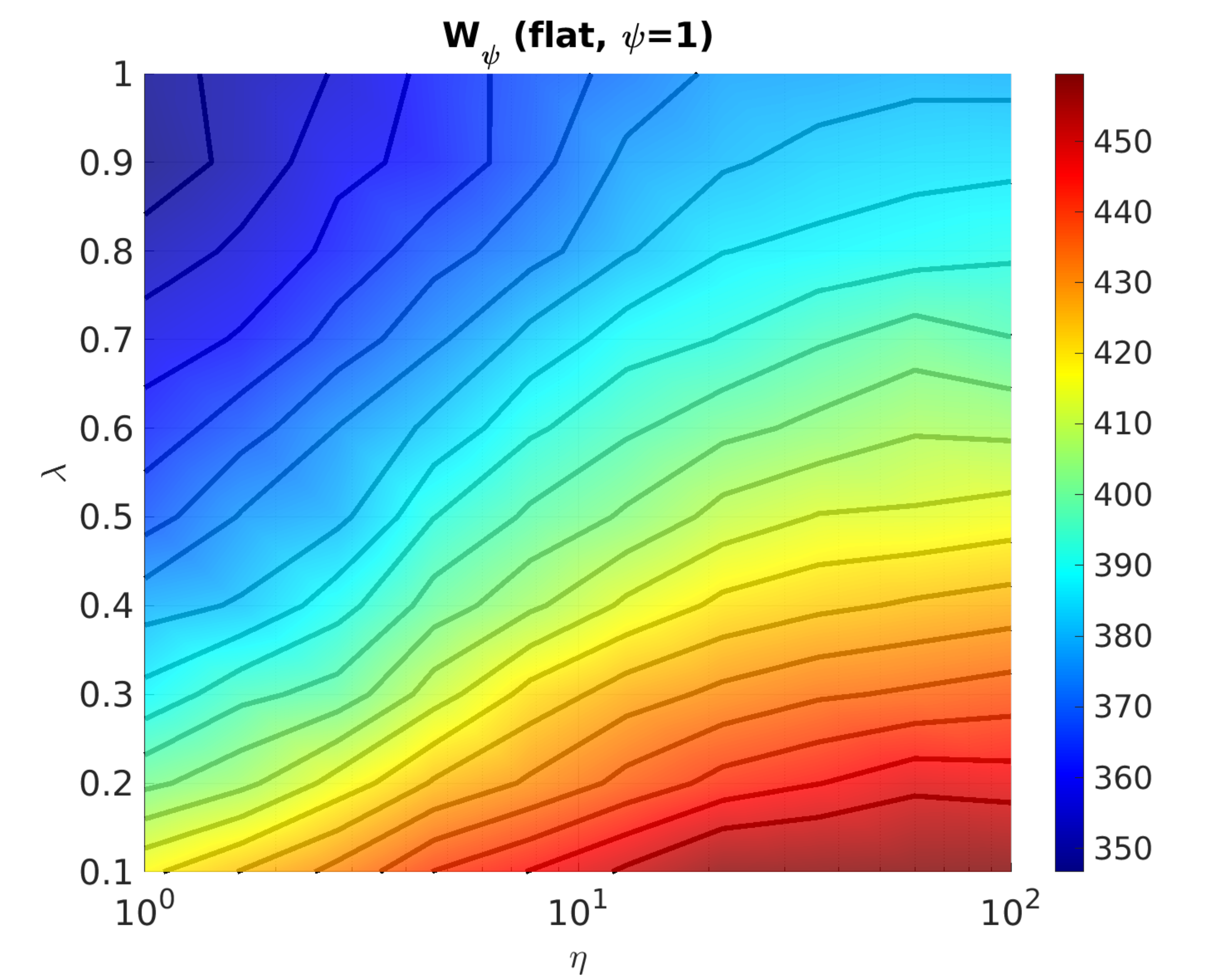}
 \includegraphics[width=.45\linewidth]{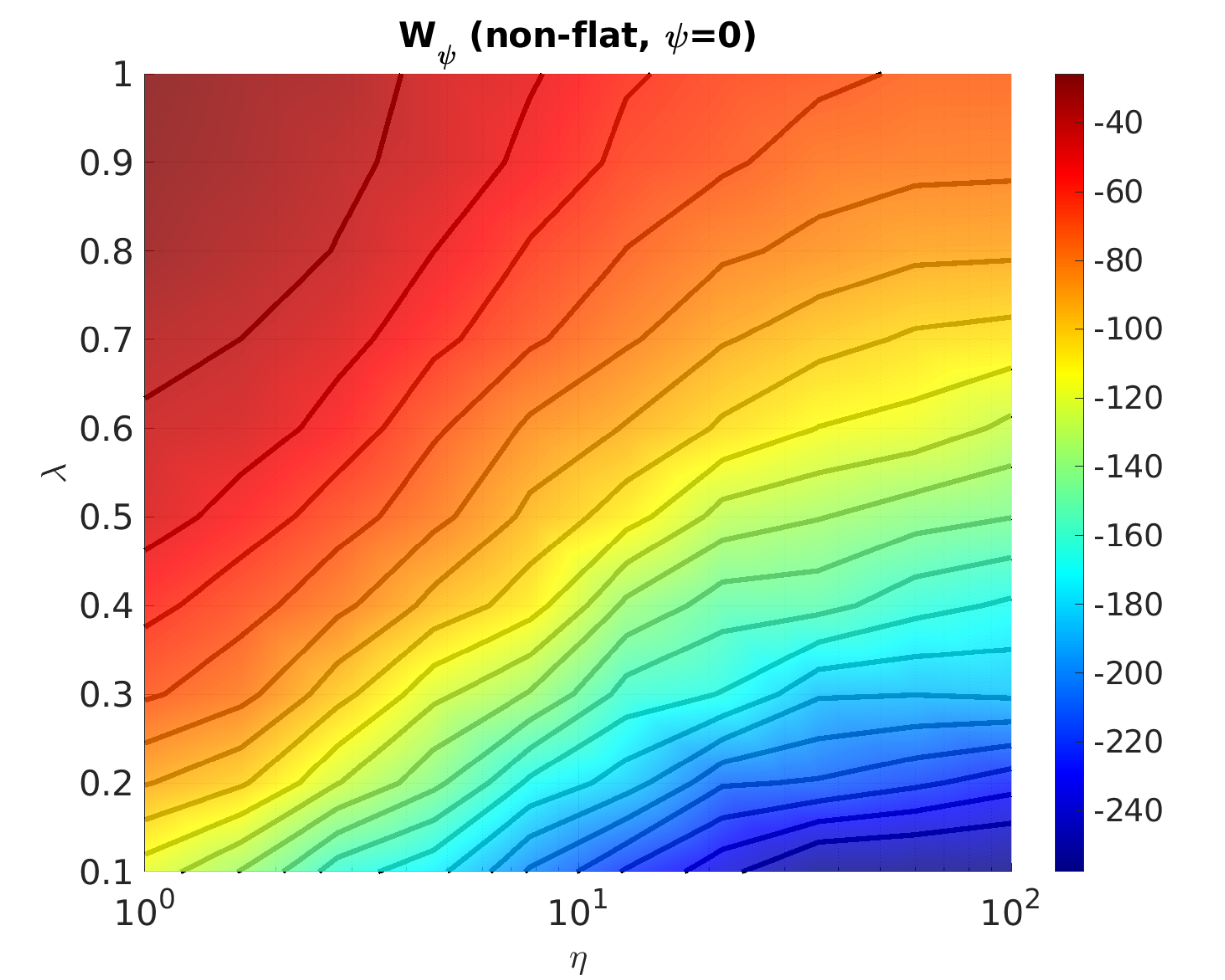}
  \includegraphics[width=.45\linewidth]{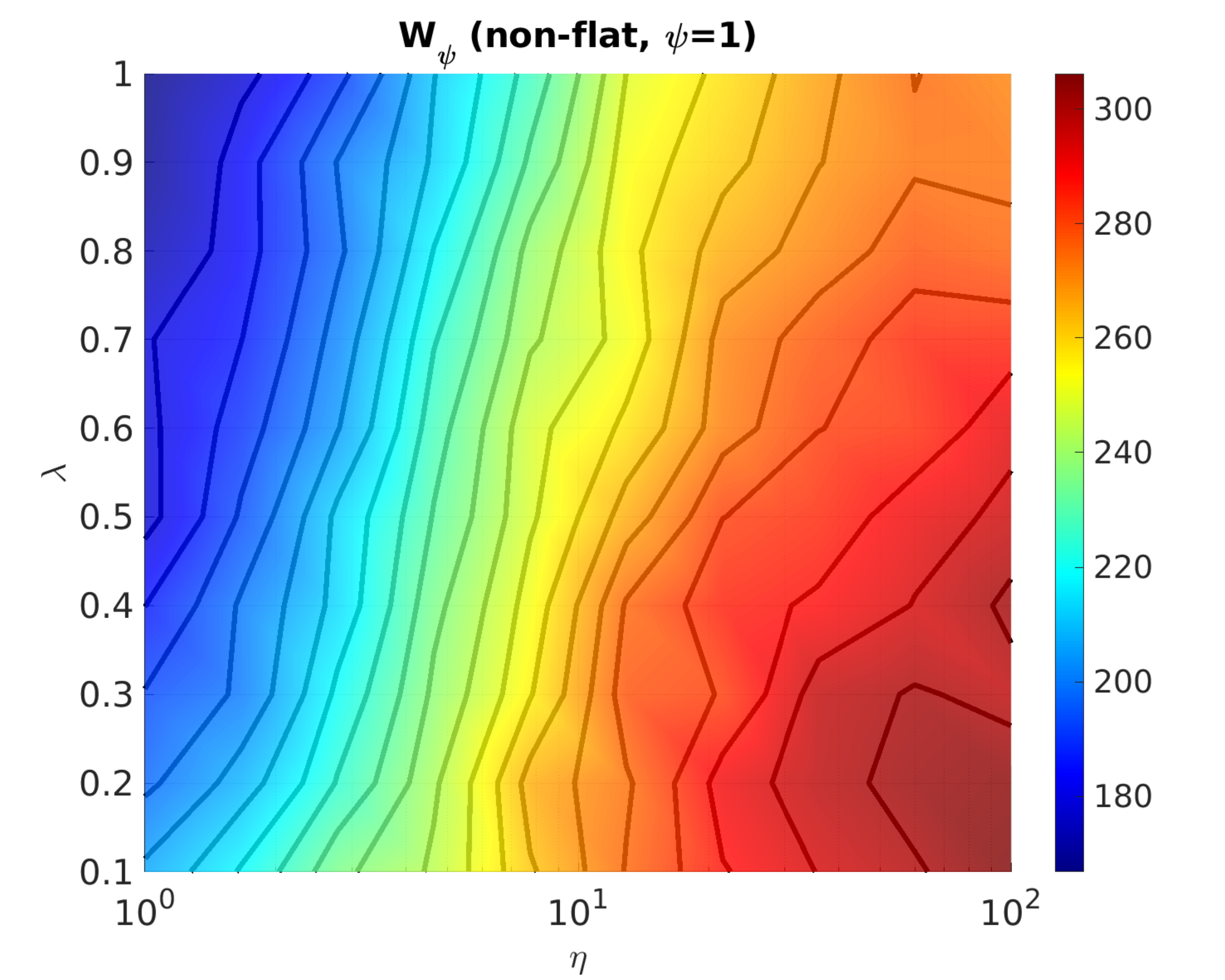}
    \caption{Impact of highlighting and personalization on social welfare, Eq.~\eqref{eq:welfare},
    in the flat case (top) and the non-flat case (bottom), for users's welfare $\psi=0$
    (left) and platform's welfare $\psi=1$ (right).}\label{fig:sim_welfare}
\end{figure*}

Finally, we show results of the welfare index $W_\psi(\eta,\lambda)$.
Figure~\ref{fig:sim_welfare} shows $W_\psi(\eta,\lambda)$ for different values of the weight $\psi$ which controls for the relative importance of the platform’s welfare (high $ENG$) relative
to the users’ welfare (low $MIS$ and $POL$). We observe that the values of $\eta$ and $\lambda$ for which the welfare index is maximized coincide with those stated in 
Proposition~\ref{prop:welfare} for both flat and non-flat cases.

\section{Meaningful Social Interactions and Political Polarization: Empirical Evidence}\label{sec:empirical}

The theoretical predictions of our model discussed in Section \ref{section:results} (Proposition~\ref{prop:main1}) suggest that an increase in the weight given by the ranking algorithm to the ``highlights'' (an increase in $\eta$) will result in individuals being more exposed to extremist contents and, in turn, in a higher level of political polarization.  To connect this prediction to observational data, we exploit Facebook's ``Meaningful Social Interaction'' (MSI) update implemented in January 2018, which boosted the weight given to comments and shares in Facebook's ranking algorithm.\footnote{See \url{https://www.facebook.com/business/news/news-feed-fyi-bringing-people-closer-together} and \url{www.edition.cnn.com/2021/10/27/tech/facebook-papers-meaningful-social-interaction-news-feed-math}.} In particular, our theoretical framework suggests that---if the propensity to highlight contents is higher for people with more extreme priors (non-flat case)---we should observe an increase in extremism and political polarization following such a change in the algorithm. In what follows, we provide empirical evidence in support of such predictions by leveraging a survey dataset from Italy containing rich information on political preferences around the time of the change in Facebook's algorithm.

\subsection{Data\label{s:data}}

The dataset comes from the \textit{Polimetro} (i.e., Political meter) surveys run by the leading Italian public opinion polling company \textit{Ipsos}. The  \textit{Polimetro} contains weekly/monthly interviews on a representative sample of the Italian voting population (i.e., aged 18 or above).  

In particular, for the purpose of our analysis, the survey asks questions on the main sources of information used by an individual to form a political opinion (i.e., newspapers, radio news, tv news, friends, internet, etc). It is important to notice that around the time of its MSI update, Facebook was by far the first social network in Italy with a 60\% penetration rate and 34 million active users per month compared with 33\% and 23\% penetration rate for Instagram and Twitter, respectively (Hootsuite, 2018). 
Hence, while the Ipsos survey does not directly ask questions about Facebook use, it is possible to proxy the exposure to Facebook contents with the use of internet to form a political opinion. 

Furthermore, besides providing information on the socio-demographic characteristics of the respondents, the \textit{Polimetro} asks questions regarding the ideological position of the respondent on the left-right scale and on the probability of voting for each party. Accordingly, we make use of these questions to construct two main outcome variables. The first one is a dummy variable taking value zero if a respondent self-identifies with a moderate political position (center, center-left or center-right) and one if she instead identifies with a more extremist position (left, right, extreme-left, extreme-right). This variable is thus meant to capture a simple measure of political extremism. The second one is a measure of affective polarization. In particular, affective polarization ``captures the extent to which citizens feel sympathy towards partisan in-groups and antagonism towards partisan out-groups'' (Wagner 2021, page 1). Since Italy is a multi-party political system, we follow Alvarez and Nagler (2004) and Wagner (2021) and define a measure of Weighted Affective Polarization (WAP) for individual $i$ as:
\begin{equation}
WAP=\sqrt{\sum_{p=1}^P v_p*\mid symp_{ip}-\overline{symp_{i}}\mid} \, ,
\end{equation}
where $v_p$ is the vote share of party $p$ (measured as a proportion ranging from 0 to 1), $symp_{ip}$ is measured with the probability attached by individual $i$ to voting for party $p$ (ranging from 0 to 10), and $\overline{symp_{i}}$ is individual $i$'s weighted average party sympathy score. That is:

\begin{equation}
\overline{symp_{i}}=\sum_{p=1}^P v_p* symp_{ip} \, .
\end{equation}

\subsection{Empirical strategy}\label{s:identification}

We implement a Differences-in-Differences empirical model to assess whether the change in the Facebook algorithm implemented in January 2018 via the introduction of the ``Meaningful Social Interaction'' weights had a causal impact on self-declared ideological extremism and affective polarization. Specifically, we look at such outcomes in the group of people leaving in a given municipality who use internet to form a political opinion and who were interviewed after the Meaningful Social Interaction (MSI) algorithm was introduced (i.e., January-June 2018) and then compare it with the ones of the group of people also using internet to form an opinion who were interviewed before such a change (i.e., June-December 2017) and at the same time with the group of people interviewed after such change in the algorithm that were not using internet as one of the main sources to form an opinion. Accordingly, we estimate the following econometric specification:
\begin{eqnarray} \label{eq:baseline}
{\tt
Y_{i,m,t}} &=& \alpha + \beta_1 \text{\tt (Opinion via internet}_{i,m,t} \times \text{\tt Post MSI)}  \nonumber \\ 
&& + \, \, \beta_2 \text{ \tt Opinion via internet}_{i,m,t} + \beta_3 \text{ \tt Post MSI}+ \alpha_{m} + {\tt X_{i,t}} + \varepsilon_{i,m,t}
\end{eqnarray}
where $Y_{i,m,t}$ represents the outcome of interest relative to individual $i$, leaving in municipality $m$ interviewed in the survey wave $t$ (i.e., probability of declaring a non-moderate political ideology or weighted affective polarization). $\alpha_m$ captures municipality fixed effects. $\beta_1$ is the parameter of interest. In more demanding specifications, we also include either time fixed effects (i.e., survey-wave fixed effects) or province-by-time fixed (which account for any unobservable shock at the province-time level).  $X_{i,t}$ represents a vector of socio-demographic control variables including the respondent's age (and age squared), gender, number of resident family members, level of education, type of occupation and religiosity. Observations are weighted according to the sampling weights provided by Ipsos and thus the results are representative of the Italian voting age population. 

\subsection{Results}\label{s:results}

Table \ref{tab:baseline} shows our baseline results on the effect of the introduction of Facebook's MSI update on the probability that an individual using internet to form an opinion holds a non-moderate political position. Column 1 provides estimates when including municipality fixed effects only (besides individual level controls). Column 2 includes also date-of-interview fixed effects accounting for possible overall time-varying patterns in ideological positions. Column 3 includes fixed effects both at the municipality and at the province-by-date-of-interview level, thus accounting for any province-time variation in political preferences. Columns 1-3 present estimates when clustering standard error at the regional levels (which in Italy correspond to electoral districts for the upper chamber). Column 4 provides evidence that results are robust when clustering standard error at a finer geographical level (provinces). The most demanding specifications (columns 3 and 4) suggest that in the period after the MSI implementation, individuals using internet to form a political opinion had a higher probability of holding a non-moderate ideology. The effect is sizeable accounting for around one standard deviation increase in such probability.

\begin{table}[h!]
\begin{center} 
{\fontsize{9}{9}\selectfont
\centering \caption{MSI and non-moderate ideological position  \label{tab:baseline}}
\begin{tabular}{lcccc} \hline\hline
 & (1) & (2)& (3)& (4)  \\
 & Non-moderate & Non-moderate  &Non-moderate &Non-moderate  \\  
 &Ideology&Ideology&Ideology&Ideology\\\hline
 &   & &   &  \\
 \text{Opinion via internet} $\times$ Post MSI & 0.062*** & 0.058*** & 0.051*** & 0.051*** \\
 & (0.016) & (0.015) & (0.014) & (0.018) \\
\text{Opinion via internet} & -0.012 & -0.006 & -0.012 & -0.012 \\
 & (0.020) & (0.020) & (0.024) & (0.022) \\
Post MSI & -0.017* &  &  &  \\
 & (0.009) &  &  &  \\
 &  &  &  &  \\
Observations & 25,690 & 25,690 & 25,690 & 25,690 \\
Mean outcome & 0.36 & 0.36 & 0.36 & 0.36 \\
SD outcome & 0.48 & 0.48 & 0.48 & 0.48 \\
&   & &   &  \\
Municipality FE & YES & YES & YES & YES \\
Date of interview FE & NO & YES & NO & NO \\
Province-Date of interview FE & NO & NO & YES & YES \\
&   & &   &  \\
 Cluster SE & Region & Region & Region & Province \\ \hline\hline
\end{tabular}
}
\begin{minipage}{15cm}
\vspace{.1cm}
\scriptsize{\textbf{Note:} Time horizon: June 2017-June 2018.  All estimates include the following control variables: age, age squared, gender, number of resident family members, level of education, type of occupation and religiosity of the respondent. Observations are weighted according to the sampling weights provided by Ipsos and thus the results are representative of the Italian voting age population. Robust Standard Errors in parenthesis. *** p$<$0.01, ** p$<$0.05, * p$<$0.1}
\end{minipage}
 \end{center}
\end{table}

We now turn to the analysis on affective polarization. Table \ref{tab:baseline_affective_pol} presents our results.\footnote{The lower number of observations relative to Table \ref{tab:baseline} is due to the fact that the questions used as proxies of sympathy score for the different parties are asked less frequently (i.e., in fewer surveys) with respect to the one on the self-decleared ideological position.} The results present in Columns 1-4 show a positive, statistically significant and robust effect. That is, in the period after the MSI algorithmic update, individuals using internet to form a political opinion had a higher level of affective polarization. Also in this case,  the effect is sizeable accounting for around 1.2 of a standard deviation increase in affective polarization.
All in all, Tables \ref{tab:baseline} and \ref{tab:baseline_affective_pol} provide evidence in support of the key theoretical predictions of our model.\footnote{Appendix Tables \ref{tab:rob_extremist} and \ref{tab:rob_affective_pol} show that the results are robust to excluding observations in the pre-electoral period (January-March 2018).}

\begin{table}[h!]
\begin{center} 
{\fontsize{9}{9}\selectfont
\centering \caption{MSI and Affective Polarization  \label{tab:baseline_affective_pol}}
\begin{tabular}{lcccc} \hline\hline
 & (1) & (2)& (3)& (4)  \\
 & Affective & Affective &Affective &Affective  \\  
 &Polarization&Polarization&Polarization&Polarization\\\hline
 &   & &   &  \\
 \text{Opinion via internet} $\times$ Post MSI  & 0.054** & 0.055** & 0.073*** & 0.073*** \\
 & (0.024) & (0.024) & (0.019) & (0.025) \\
\text{Opinion via internet}& -0.012 & -0.011 & -0.006 & -0.006 \\
 & (0.023) & (0.022) & (0.023) & (0.025) \\
Post MSI & 0.118*** &  &  &  \\
 & (0.020) &  &  &  \\
 &  &  &  &  \\
Observations & 14,499 & 14,499 & 14,499 & 14,499 \\
Mean outcome & 1.29 & 1.29 & 1.29 & 1.29 \\
SD outcome & 0.61 & 0.61 & 0.61 & 0.61 \\

&   & &   &  \\
Municipality FE & YES & YES & YES & YES \\
Date of interview FE & NO & YES & NO & NO \\
Province-Date of interview FE & NO & NO & YES & YES \\
&   & &   &  \\
 Cluster SE & Region & Region & Region & Province \\ \hline\hline
\end{tabular}
}
\begin{minipage}{13.5cm}
\vspace{.1cm}
\scriptsize{\textbf{Note:} Time horizon: June 2017-June 2018. All estimates include the following control variables: age, age squared, gender, number of resident family members, level of education, type of occupation and religiosity of the respondent. Observations are weighted according to the sampling weights provided by Ipsos and thus the results are representative of the Italian voting age population. Robust Standard Errors in parenthesis. *** p$<$0.01, ** p$<$0.05, * p$<$0.1}
\end{minipage}
 \end{center}
\end{table}

\section{Conclusion}
Social media platforms such as Facebook, Twitter or Instagram are \textquotedblleft algorithmic gatekeepers\textquotedblright\  \citep{napoli2015social,tufekci2015algorithmic}: their ranking algorithm determines the order in which items are to be displayed to a given user. This paper provides a simple theoretical framework showing how polarization and misinformation may emerge naturally from very basic aspects of popularity and personalization of the algorithm combined with basic well-documented behavioral traits of the users, and the dynamic feedback between algorithm and users. 

In particular, we point out the existence of a trade-off between the platform's welfare and the users' welfare when fine-tuning its ranking algorithm. Changes in parameters of the ranking algorithm (popularity and personalization) that increase platform engagement may have detrimental effects in terms of misinformation (\textit{crowding-out the truth}) and/or polarization. Our results are consistent with the evidence provided by the empirical literature assessing the impact of personalization on political polarization (e.g., \citealt{levy2021,huszar_2022}). Most importantly, by exploiting the 2018 Facebook MSI algorithmic ranking update and leveraging a rich survey dataset from Italy, we also provide direct empirical evidence corroborating the detrimental impact on political polarization created by a boost in the weight given by the ranking algorithm to \textquotedblleft highlighted" contents, as predicted by our model.

Our paper provides academic guidance to the public debate on the potential undesirable consequences of algorithmic gatekeepers on social welfare.
We conclude by acknowledging that the model does not embed other important features of social media such as endogenous networks or fact-checking. Complementary research \citep{acemoglu_2022} points out that these additional features may lead to further reinforcing the trade-off between platform engagement and social welfare. All in all, the insights from this line of research provide a \textquotedblleft theory of harm" indirectly endorsing the recent attempt by the European Union to regulate digital platforms.\footnote{See \url{https://digital-strategy.ec.europa.eu/en/policies/digital-services-act-package}.}

Future research combining endogenous dynamic algorithmic ranking and endogenous belief and network formation may provide additional insights to guide public regulators and social media platforms in their efforts to reduce the negative impact of ranking dynamics on social media users and on our society at large.

\newpage

\bibliographystyle{aer}
\bibliography{bibliography_updated.bib}

@misc{amnesty_2022,
	Author = {{Amnesty International}},
	Howpublished = {\url{www.amnesty.org/en/documents/ASA16/5933/2022/en/}},
	Note = {[Online; accessed 01-October-2022]},
	Title = {{The Social Atrocity. Meta and the Right to Remedy for the Rohingya}},
	Year = {2022}}

@article{ortoleva_2015,
  title={Overconfidence in political behavior},
  author={Ortoleva, Pietro and Snowberg, Erik},
  journal={American Economic Review},
  volume={105},
  number={2},
  pages={504--35},
  year={2015}
}

@InProceedings{tabibian_2020,
  title = {On the design of consequential ranking algorithms},
  author =       {Tabibian, Behzad and G\'{o}mez, Vicen\c{c} and De, Abir and Sch\"{o}lkopf, Bernhard and Gomez Rodriguez, Manuel},
  booktitle = {Proceedings of the 36th Conference on Uncertainty in Artificial Intelligence (UAI)},
  pages = {171--180},
  year = {2020},
  volume = {124},
  series = {Proceedings of Machine Learning Research},
  month = {03--06 Aug},
  publisher =    {PMLR}
}

@article{tufekci_2018,
  title={How social media took us from Tahrir Square to Donald Trump},
  author={Tufekci, Zeynep},
  journal={MIT Technology Review},
  volume={14},
  pages={18},
  year={2018}
}

@misc{van_gils_2020,
  title={Big data and democracy},
  author={van Gils, Freek and M{\"u}ller, Wieland and Prufer, Jens},
  year={2020},
  howpublished={TILEC Discussion Paper No. DP 2020-003}
}

@article{levy2019echo,
  title={Echo chambers and their effects on economic and political outcomes},
  author={Levy, Gilat and Razin, Ronny},
  journal={Annual Review of Economics},
  volume={11},
  pages={303--328},
  year={2019},
  publisher={Annual Reviews}
}

@misc{cnn_2021,
	Author = {CNN},
	Howpublished = {\url{www.edition.cnn.com/2021/10/27/tech/facebook-papers-meaningful-social-interaction-news-feed-math}},
	Note = {[Online; accessed 01-July-2022]},
	Title = {{Likes, anger emojis and RSVPs: the math behind Facebook's News Feed — and how it backfired}},
	Year = {2021}}

@article{lauer_2021,
  title={Facebook’s ethical failures are not accidental; they are part of the business model},
  author={Lauer, David},
  journal={AI and Ethics},
  volume={1},
  number={4},
  pages={395--403},
  year={2021},
  publisher={Springer}
}

@article{grinberg_2019,
  title={Fake news on Twitter during the 2016 US presidential election},
  author={Grinberg, Nir and Joseph, Kenneth and Friedland, Lisa and Swire-Thompson, Briony and Lazer, David},
  journal={Science},
  volume={363},
  number={6425},
  pages={374--378},
  year={2019},
  publisher={American Association for the Advancement of Science}
}

@article{hopp_2020,
  title={Why do people share ideologically extreme, false, and misleading content on social media? A self-report and trace data--based analysis of countermedia content dissemination on Facebook and Twitter},
  author={Hopp, Toby and Ferrucci, Patrick and Vargo, Chris J},
  journal={Human Communication Research},
  volume={46},
  number={4},
  pages={357--384},
  year={2020},
  publisher={Oxford University Press}
}

@article{tucker_2018,
  title={Social media, political polarization, and political disinformation: A review of the scientific literature},
  author={Tucker, Joshua A and Guess, Andrew and Barber{\'a}, Pablo and Vaccari, Cristian and Siegel, Alexandra and Sanovich, Sergey and Stukal, Denis and Nyhan, Brendan},
  journal={Political polarization, and political disinformation: a review of the scientific literature (March 19, 2018)},
  year={2018}
}

@techreport{bursztyn_2019,
  title={Social media and xenophobia: evidence from Russia},
  author={Bursztyn, Leonardo and Egorov, Georgy and Enikolopov, Ruben and Petrova, Maria},
  year={2019},
  institution={National Bureau of Economic Research}
}

@techreport{pew_2019,
  title={National Politics on Twitter:
Small Share of U.S. Adults Produce Majority of Tweet},
  author={Pew},
  year={2019},
  institution={Pew Research Center}
}

@article{mosleh_2020,
  title={Self-reported willingness to share political news articles in online surveys correlates with actual sharing on Twitter},
  author={Mosleh, Mohsen and Pennycook, Gordon and Rand, David G},
  journal={Plos one},
  volume={15},
  number={2},
  pages={e0228882},
  year={2020},
  publisher={Public Library of Science San Francisco, CA USA}
}

@article{azzimonti_2022,
  title={Social media networks, fake news, and polarization},
  author={Azzimonti, Marina and Fernandes, Marcos},
  journal={European Journal of Political Economy},
  pages={102256},
  year={2022},
  publisher={Elsevier}
}

@article{muller_2020,
  title={From hashtag to hate crime: Twitter and anti-minority sentiment},
  author={M{\"u}ller, Karsten and Schwarz, Carlo},
  journal={Available at SSRN 3149103},
  year={2020}
}

@article{muller_2021,
  title={Fanning the flames of hate: Social media and hate crime},
  author={M{\"u}ller, Karsten and Schwarz, Carlo},
  journal={Journal of the European Economic Association},
  volume={19},
  number={4},
  pages={2131--2167},
  year={2021},
  publisher={Oxford Academic}
}

@article{levy2021,
  title={Social media, news consumption, and polarization: Evidence from a field experiment},
  author={Levy, Ro'ee},
  journal={American economic review},
  volume={111},
  number={3},
  pages={831--70},
  year={2021}
}

@article{allcott_2020,
  title={The welfare effects of social media},
  author={Allcott, Hunt and Braghieri, Luca and Eichmeyer, Sarah and Gentzkow, Matthew},
  journal={American Economic Review},
  volume={110},
  number={3},
  pages={629--76},
  year={2020}
}

@misc{ditella_2021,
  title={Does Social Media Cause Polarization? Evidence from Access to Twitter Echo Chambers During the 2019 Argentine Presidential Debate},
  author={Di Tella, Rafael and G{\'a}lvez, Ramiro and Schargrodsky, Ernesto},
  howpublished={NBER Working Paper (w29458).},
  year={2021}
}

@TechReport{acemoglu_2022,
  author={Acemoglu, Daron and Ozdaglar, Asuman and Siderius, James},
  title={{A Model of Online Misinformation}},
  year=2022,
  month=Jan,
  institution={C.E.P.R. Discussion Papers},
  type={CEPR Discussion Papers},
  url={https://ideas.repec.org/p/cpr/ceprdp/16932.html},
  number={16932},
  doi={},
}

@TechReport{dujgar2022,
  author={Dujeancourt, Erwan and Marcel Garz},
  title={{The Effects of Algorithmic Content Selection on User Engagement with News on Twitter}},
  year=2022,
  month=Mar,
  institution={J\"{o}nk\"{o}ping International Business School},
  type={},
  url={https://www.marcelgarz.com/wp-content/uploads/Twitter-algorithm-Mar-2022.pdf},
  number={},
  doi={},
}

@TechReport{analytis_2022,
  author={Pantelis P. Analytis and Francesco Cerigioni and Alexandros Gelastopoulos and Hrvoje Stojic},
  title={{Sequential choice and selfreinforcing rankings}},
  year=2022,
  month=Feb,
  institution={Department of Economics and Business, Universitat Pompeu Fabra},
  type={Economics Working Papers},
  url={https://ideas.repec.org/p/upf/upfgen/1819.html},
  number={1819},
  doi={},
}

@article{mullainathan_2005,
  title={The market for news},
  author={Mullainathan, Sendhil and Shleifer, Andrei},
  journal={American economic review},
  volume={95},
  number={4},
  pages={1031--1053},
  year={2005}
}

@article{bernhardt_2008,
  title={Political polarization and the electoral effects of media bias},
  author={Bernhardt, Dan and Krasa, Stefan and Polborn, Mattias},
  journal={Journal of Public Economics},
  volume={92},
  number={5-6},
  pages={1092--1104},
  year={2008},
  publisher={Elsevier}
}

@article{gentzkow_2010,
  title={What drives media slant? Evidence from US daily newspapers},
  author={Gentzkow, Matthew and Shapiro, Jesse M},
  journal={Econometrica},
  volume={78},
  number={1},
  pages={35--71},
  year={2010},
  publisher={Wiley Online Library}
}

@incollection{gentzkow_2015,
title = {Chapter 14 - Media Bias in the Marketplace: Theory},
editor = {Simon P. Anderson and Joel Waldfogel and David Strömberg},
series = {Handbook of Media Economics},
publisher = {North-Holland},
volume = {1},
pages = {623-645},
year = {2015},
booktitle = {Handbook of Media Economics},
issn = {2213-6630},
doi = {https://doi.org/10.1016/B978-0-444-63685-0.00014-0},
url = {https://www.sciencedirect.com/science/article/pii/B9780444636850000140},
author = {Matthew Gentzkow and Jesse M. Shapiro and Daniel F. Stone}
}

@article{sobbrio_2014,
  title={Citizen-editors' endogenous information acquisition and news accuracy},
  author={Sobbrio, Francesco},
  journal={Journal of Public Economics},
  volume={113},
  pages={43--53},
  year={2014},
  publisher={Elsevier}
}

@inproceedings{germano_2019,
  title={The few-get-richer: a surprising consequence of popularity-based rankings},
  author={Germano, Fabrizio and G{\'o}mez, Vicen{\c{c}} and Le Mens, Ga{\"e}l},
  booktitle={The World Wide Web Conference},
  pages={2764--2770},
  year={2019}
}

@misc{rudiger_2013,
  author={Rudiger, Jesper},
  title={Cross-checking the media},
  howpublished = "EUI MWP, 2013/17",
  year         = "2013",
}

@article{athey_2018,
  title={The impact of consumer multi-homing on advertising markets and media competition},
  author={Athey, Susan and Calvano, Emilio and Gans, Joshua S},
  journal={Management science},
  volume={64},
  number={4},
  pages={1574--1590},
  year={2018},
  publisher={INFORMS}
}

@article{germano_2020,
title = {Opinion dynamics via search engines (and other algorithmic gatekeepers)},
journal = {Journal of Public Economics},
volume = {187},
pages = {104188},
year = {2020},
issn = {0047-2727},
author = {Fabrizio Germano and Francesco Sobbrio},
}

@article{krasa_2009,
  title={Is mandatory voting better than voluntary voting?},
  author={Krasa, Stefan and Polborn, Mattias K},
  journal={Games and Economic Behavior},
  volume={66},
  number={1},
  pages={275--291},
  year={2009},
  publisher={Elsevier}
}

@article{krishna_2011,
  title={Overcoming ideological bias in elections},
  author={Krishna, Vijay and Morgan, John},
  journal={Journal of Political Economy},
  volume={119},
  number={2},
  pages={183--211},
  year={2011},
  publisher={University of Chicago Press Chicago, IL}
}

@article{galasso_2011,
  title={Competing on good politicians},
  author={Galasso, Vincenzo and Nannicini, Tommaso},
  journal={American political science review},
  volume={105},
  number={1},
  pages={79--99},
  year={2011},
  publisher={Cambridge University Press}
}

@article{garz_2020,
  title={Partisan selective engagement: Evidence from Facebook},
  author={Garz, Marcel and S{\"o}rensen, Jil and Stone, Daniel F},
  journal={Journal of Economic Behavior \& Organization},
  volume={177},
  pages={91--108},
  year={2020},
  publisher={Elsevier}
}

@article{huszar_2022,
	author = {Husz{\'a}r, Ferenc and Ktena, Sofia Ira and O{\textquoteright}Brien, Conor and Belli, Luca and Schlaikjer, Andrew and Hardt, Moritz},
	title = {Algorithmic amplification of politics on Twitter},
	volume = {119},
	number = {1},
	year = {2022},
	publisher = {National Academy of Sciences},
	issn = {0027-8424},
	URL = {https://www.pnas.org/content/119/1/e2025334119},
	journal = {Proceedings of the National Academy of Sciences}
}

@article{polarization,
author = {Iyengar, Shanto and Lelkes, Yphtach and Levendusky, Matthew and Malhotra, Neil and Westwood, Sean J.},
title = {The Origins and Consequences of Affective Polarization in the United States},
journal = {Annual Review of Political Science},
volume = {22},
number = {1},
pages = {null},
year = {2019},
}

@article{vosoughi2018spread,
  title={The spread of true and false news online},
  author={Vosoughi, Soroush and Roy, Deb and Aral, Sinan},
  journal={Science},
  volume={359},
  number={6380},
  pages={1146--1151},
  year={2018},
  publisher={American Association for the Advancement of Science}
}

@article{bakshy2015exposure,
  title={Exposure to ideologically diverse news and opinion on Facebook},
  author={Bakshy, Eytan and Messing, Solomon and Adamic, Lada A},
  journal={Science},
  volume={348},
  number={6239},
  pages={1130--1132},
  year={2015},
  publisher={American Association for the Advancement of Science}
}

@article{hao_2017,
author = "Hao Liao and Manuel Sebastian Mariani and Mat\'us Medo and Yi-Cheng Zhang and Ming-Yang Zhou",
title = "Ranking in evolving complex networks",
journal = "Physics Reports",
volume = "689",
pages = "1--54",
year = "2017",
}

@article{choo_et_al,
	Author = {Cho, Junghoo and Roy, Sourashis and Adams, Robert E},
	Journal = {SIGMOD},
	Title = {{Page Quality: In Search of an Unbiased Web Ranking}},
	Volume = {14},
	Year = {2005}}

@article{napoli2015social,
	Author = {Napoli, Philip M},
	Journal = {Telecommunications Policy},
	Number = {9},
	Pages = {751--760},
	Publisher = {Elsevier},
	Title = {Social media and the public interest: Governance of news platforms in the realm of individual and algorithmic gatekeepers},
	Volume = {39},
	Year = {2015}}

@article{white2015belief,
	Author = {White, Ryen W and Horvitz, Eric},
	Journal = {ACM Transactions on Information Systems (TOIS)},
	Number = {4},
	Pages = {18},
	Publisher = {ACM},
	Title = {Belief Dynamics and Biases in Web Search},
	Volume = {33},
	Year = {2015}}

@article{epstein2015search,
	Author = {Epstein, Robert and Robertson, Ronald E},
	Journal = {Proceedings of the National Academy of Sciences},
	Number = {33},
	Pages = {E4512----E4521},
	Publisher = {National Acad Sciences},
	Title = {{The Search Engine Manipulation Effect \uppercase{(SEME)} and its Possible Impact on the Outcomes of Elections}},
	Volume = {112},
	Year = {2015}}

@article{flaxman_et_al,
  title={Filter bubbles, echo chambers, and online news consumption},
  author={Flaxman, Seth and Goel, Sharad and Rao, Justin M},
  journal={Public opinion quarterly},
  volume={80},
  number={S1},
  pages={298--320},
  year={2016},
  publisher={Oxford University Press}
}

@article{glick_et_al,
	Author = {Glick, Mark and Richards, Greg and Sapozhnikov, Margarita and Seabright, Paul},
	Journal = {Review of Industrial Organization},
	Month = {September},
	Number = {2},
	Pages = {99-119},
	Title = {{How Does Ranking Affect User Choice in Online Search?}},
	Volume = {45},
	Year = {2014}}

@article{goldman2006search,
	Author = {Goldman, Eric},
	Journal = {Yale Journal of Law \& Technology},
	Pages = {6--8},
	Title = {{Search Engine Bias and the Demise of Search Engine Utopianism}},
	Year = {2006}}

@article{granka,
	Author = {Granka, Larua A},
	Journal = {The Information Society},
	Pages = {364--374},
	Title = {{The Politics of Search: A Decade Retrospective}},
	Volume = {26},
	Year = {2010}}

@article{grimmelmann,
	Author = {Grimmelmann, James},
	Journal = {New York Law School Law Review},
	Pages = {939--950},
	Title = {{The Google Dilemma}},
	Volume = {53},
	Year = {2009}}

@article{hargittai2004changing,
	Author = {Hargittai, Eszter},
	Journal = {Community practice in the network society: local action/global interaction},
	Publisher = {Psychology Press},
	Title = {{The Changing Online Landscape}},
	Year = {2004}}

@book{hindman,
	Author = {Hindman, Matthew},
	Publisher = {Princeton University Press},
	Title = {{The Myth of Digital Democracy}},
	Year = {2009}}

@article{lazer2015rise,
	Author = {Lazer, David},
	Journal = {Science},
	Number = {6239},
	Pages = {1090--1091},
	Title = {{The Rise of the Social Algorithm}},
	Volume = {348},
	Year = {2015}}

@article{menczer2006googlearchy,
	Author = {Menczer, Filippo and Fortunato, Santo and Flammini, Alessandro and Vespignani, Alessandro},
	Journal = {IEEE Spectrum Online},
	Title = {{Googlearchy or Googleocracy}},
	Year = {2006}}

@article{NovareseWilson2013,
	Author = {Novarese, Marco and Wilson, Chris},
	Journal = {Working Paper},
	Title = {{Being in the Right Place: A Natural Field Experiment on List Position and Consumer Choice}},
	Year = {2013}}

@article{pan_et_al,
	Author = {Pan, Bing and Hembrooke, Helene and Joachims, Thorsten and Lorigo, Lori and Gay, Geri and Granka, Laura},
	Journal = {Journal of Computer-Mediated Communication},
	Pages = {801--823},
	Title = {{In Google We Trust: Users' Decisions on Rank, Position, and Relevance}},
	Volume = {12},
	Year = {2007}}

@book{pariser,
	Author = {Pariser, Eli},
	Publisher = {Penguin Books},
	Title = {{The Filter Bubble: How the New Personalized Web Is Changing What We Read and How We Think}},
	Year = {2011}}

@book{putnam2001,
	Author = {Putnam, Robert D},
	Publisher = {New York: Simon and Schuster},
	Title = {{Bowling Alone: {The} Collapse and Revival of American Community}},
	Year = {2001}}

@book{sunstein2009,
	Author = {Sunstein, Cass R},
	Publisher = {Princeton University Press},
	Title = {{Republic.com 2.0}},
	Year = {2009}}

@article{tufekci2015algorithmic,
	Author = {Tufekci, Zeynep},
	Journal = {J. on Telecomm. \& High Tech. L.},
	Pages = {203},
	Publisher = {HeinOnline},
	Title = {{Algorithmic Harms beyond Facebook and Google: Emergent Challenges of Computational Agency}},
	Volume = {13},
	Year = {2015}}

@article{yom_et_al,
	Author = {Yom-Tov, Elad and Dumais, Susan and Guo, Qi},
	Journal = {Social Science Computer Review},
	Pages = {1--10},
	Title = {{Promoting Civil Discourse Through Search Engine Diversity}},
	Year = {2013}}

\pagebreak{}

\clearpage

\appendix
\newpage
\clearpage
\renewcommand*\thesection{Appendix}
\setcounter{table}{0}
\setcounter{figure}{0}
\renewcommand{\thetable}{A.\arabic{table}} 
\renewcommand{\thefigure}{A.\arabic{figure}}

\subsection{Proofs}

{\bf Proof of Lemma \ref{lemma:LCD}.}
From Eqs.~(\ref{eq:clickprob}) and~(\ref{eq:clickpop}), we have that clicks on item $m$ get updated according to: 
\begin{eqnarray*}
\widehat{\kappa}_{n,m} - \widehat{\kappa}_{n-1, m}  
=  \frac{\beta^{M- r_{n,m}} \varphi_{n,m}}{\sum_{m' \in M} \beta^{M- r_{n,m'}\varphi_{n,m'}}} .
\end{eqnarray*}
Taking expectations of an average run out of $T$ runs, we have $\varphi_{n,m} \approx 1/(2[m])$ and can write:
\begin{eqnarray*}
\EXP{[\widehat{\kappa}_{n,m} - \widehat{\kappa}_{n-1, m}]}
&=& 
\left\{
\begin{array}
[cl]{cl}
\frac{\beta^{M- r_{n,m}}/(2M_{+})}{\sum_{m' \in M_{+}} \beta^{M- r_{n,m}}/(2M_{+}) + \sum_{m' \in M_{-}} \beta^{M- r_{n,m}}/(2M_{-})} &  \text{ if }  m \in M_{+} 
\\  \\
\frac{\beta^{M- r_{n,m}}/(2M_{-})}{\sum_{m' \in M_{-}} \beta^{M- r_{n,m}}/(2M_{+}) + \sum_{m' \in M_{+}} \beta^{M- r_{n,m}}/(2M_{-})} &  \text{ if }  m \in M_{-} 
\end{array}
\right. \\
&\stackrel{(1)}{\approx}&   \frac{\beta^{M- r_{n,m}}}{\sum_{m' \in M_{+}} \beta^{M- r_{n,m'}} + \sum_{m' \in M_{-}} \beta^{M- r_{n,m'}}}  \hspace{.18in} \text{ for all } m \in M \\
&=&  \frac{\beta^{M- r_{n,m}}}{\sum_{m' \in M} \beta^{M-m'}} 
=  \frac{\beta^{M} \beta^{\frac{M- r_{n,m}}{M}}}{\beta^{M} \sum_{m' \in M} \beta^\frac{M-m'}{M}} 
=  \frac{\beta^{\frac{M- r_{n,m}}{M}}}{\sum_{m' \in M} \beta^\frac{M-m'}{M}} ,
\end{eqnarray*}
where (1) follows from $\Pr (\mbox{sgn}(x_n) = \mbox{sgn}(y_m)) \approx \Pr (\mbox{sgn}(x_n) \ne \mbox{sgn}(y_m)) \approx 1/2$ 
and from $[m] \approx M_{+} \approx M_{-} \approx M/2$, for $T$ sufficiently large (in the limit as $T \rightarrow \infty$).

From this we can write the expected probability of an item with signal $y_m=y$ being clicked in an average run out of $T$ runs (again as $T\rightarrow \infty$) as:
\begin{eqnarray*}
\EXP{[\hat{\kappa}(y)]}
&=&  \frac{\beta^{\frac{M- r(y)}{M}}}{\sum_{m' \in M} \beta^\frac{M-m'}{M}} 
\stackrel{(1)}{=}  \frac{\beta^{\frac{M-\zeta_0 + \zeta_1 \cdot \pi(y)}{M}}}{\sum_{m' \in M} \beta^{\frac{M-m'}{M}}} 
\approx  \frac{1 + \log{\beta} \cdot \frac{M - \zeta_0 + \zeta_1 \cdot \pi(y)}{M}}{1 + \log{\beta} \cdot \sum_{m' \in M} \frac{M-m'}{M}} \, ,
\end{eqnarray*}
where (1) follows from applying Eq.~(\ref{eq:exprank}) to the expected rank $r(y)$.
Taking into account the distribution of the items' signals $g$, this readily implies as the limit clicking distribution:
\begin{equation}
LCD (y) = \Lambda_{\beta}(\pi(y)) \cdot g(y),
\end{equation}
where, for $z \ge 0$: 
\begin{equation} \label{eq:phi}
\Lambda_{\beta}(z) 
=  \frac{M + \log{\beta} \cdot (M - \zeta_0 + \zeta_1 \cdot z )}{M + \log{\beta} \cdot \sum_{m' \in M} (M-m')}
= \frac{M + \log{\beta} \cdot (M - \zeta_0 + \zeta_1 \cdot z ))}{M + \log{\beta} \cdot M(M-1)/2}, 
\end{equation}
so that $\Lambda_{\beta}'(z)$ is a constant and  $\Lambda_{\beta}' (z) \equiv \Lambda_{\beta}' =  \frac{ \zeta_1 \cdot \log{\beta} }{M + \log{\beta} \cdot M(M-1)/2} > 0$ since $\beta>1, \zeta_1 > 0$.

Similarly, we can write the limit highlighting distribution as:
\begin{equation}
LHD(y) = \mu_H(y) \cdot LCD(y).
\end{equation}
These distributions do not integrate to 1 but rather give a per capita probability of clicking and highlighting a given item with signal $y_m=y$.
\hfill $\Box$

\bigskip
\noindent
{\bf Proof of Proposition \ref{prop:main1}.} Set $\theta=\widehat{\theta}=0$ and fix $\beta>1$. To simplify notation we drop the subscript $\beta$ from the function $\Lambda_{\beta}$ and write just $\Lambda$.
Given Eq.~(\ref{eq:exprank}), we can apply Lemma~\ref{lemma:LCD} and write engagement ($ENG$), polarization ($POL$) and misinformation ($MIS$), respectively as:
\begin{eqnarray*}
ENG &=& \int \left( LCD(y) + LHD(y) \right) dy = \int \left(1 + \mu_H(y) \right) LCD(y) dy = \int \left(1 + \mu_H(y) \right) \Lambda(\pi(y)) f(y) dy \\
MIS &=& \int \left| y - 0 \right| LCD(y) dy =  \int \left| y \right|  \Lambda(\pi(y)) f(y) dy \\
POL &=&  \int \left| y LCD^R(y)  - y LCD^L (y) \right| dy  =  \int \left| y  \Lambda^R(\pi(y))  - y  \Lambda^L(\pi(y))  \right| f(y) dy,
\end{eqnarray*}
where for $g \in \{ L, R \}$, $LCD^g$ is the limit clicking distribution of individuals from group $g$ and (in the non-personalized case with $\lambda = 1$) can be written as:
\[ LCD^g(y)  =  \Lambda^g(\pi(y)) g(y) , \]
where $\Lambda^g (\pi(y))$ is now the expected probability an item with signal $y_m=y$ will be clicked on by an individual in group $g$. Note that while  clicking and highlighting by a given group is heavily dependent on the sign of the signal of the item, (that is, whether $m \in M_{+}$ or $m \in M_{-}$), both groups share the same ranking which depends on the total clicking and highlighting propensities of the two groups ($\pi(y)$).

From this we can compute the effect of a change in $\eta$ on he three variables. Suppose that highlighting behavior is non-flat. It suffices to compute:
\begin{eqnarray*}
\frac{\partial ENG}{\partial \eta} &=&  \frac{\partial }{\partial \eta} \int \left(1 + \mu_H(y) \right) \Lambda(\pi(y)) g(y) dy \\
&=&  \int  \frac{\partial (1 + \mu_H(y))}{\partial \eta} \Lambda(\pi(y)) g(y) dy +  \int \left(1 + \mu_H(y) \right)  \frac{\partial \Lambda(\pi(y))}{\partial \eta}  g(y) dy \\
&\stackrel{(1)}{=}& \int \left(1 + \mu_H(y) \right)\Lambda' \frac{\partial \pi(y)}{\partial \eta}  g(y) dy \\
&\stackrel{(2)}{=}& \int \left(1 + \mu_H(y) \right)\Lambda' \frac{(M-1)(\mu_H(y)-\bar{\mu}_H)}{\left(M(1+ \eta \bar{\mu}_H) + \eta ( \mu_H(y) - \bar{\mu}_H ) \right)^2}g(y)dy  \\
&\stackrel{(3)}{>}& 0,
\end{eqnarray*}
where (1) follows because $\frac{\partial \mu_H(y)}{\partial \eta}=0$ and $\frac{\partial g(y)}{\partial \eta} =0$, (2) follows from Eq.~(\ref{eq:pi}), and (3) follows since $\Lambda'(\pi(y)) = \Lambda' > 0$ and $\mu_H(y) \ge 0$ for all $y$ and, moreover, for $|y| > |y'|$ we have $\mu_H(y) > \mu_H(y')$ on a large enough mass of signals $y$ (strictly speaking until $-x^*, x^*$), and hence also for $\frac{1+ \mu_H(y)}{\left(M(1+\bar{\mu}_H) + \eta ( \mu_H(y) - \bar{\mu}_H ) \right)^2}$. The latter implies
$\int  \frac{(1 + \mu_H(y))(\mu_H(y)-\bar{\mu}_H)}{\left(M(1+\bar{\mu}_H) + \eta ( \mu_H(y) - \bar{\mu}_H ) \right)^2} g(y)dy > 0$. Since $\int (\mu_H(y)-\bar{\mu}_H) g(y)dy = 0$, increasing $\eta$ corresponds to shifting mass towards signals with higher absolute value, thereby strictly increasing engagement. Similarly:
\begin{eqnarray*}
\frac{\partial MIS}{\partial \eta} &=&  \frac{\partial }{\partial \eta} \int |y|  \Lambda(\pi(y)) g(y) dy 
=  \int |y| \frac{\partial \Lambda(\pi(y))}{\partial \eta} g(y) dy \\
&=&  \int |y|  \Lambda' \frac{(M-1)(\mu_H(y)-\bar{\mu}_H)}{\left(M(1+ \eta \bar{\mu}_H) + \eta ( \mu_H(y) - \bar{\mu}_H ) \right)^2} g(y)dy \\
&\stackrel{(1)}{>}& 0,
\end{eqnarray*}
where (1) follows again because $\Lambda'(\pi(y)) > 0$ and $\mu_H(y) \ge 0$, and for $|y| > |y'|$ we have $\mu_H(y) > \mu_H(y')$ on a large enough mass of signals $y$, ensuring that $\int \frac{|y|  (\mu_H(y)-\bar{\mu}_H)}{\left(M(1+ \eta \bar{\mu}_H) + \eta ( \mu_H(y) - \bar{\mu}_H ) \right)^2} g(y)dy > 0$.

Finally, to compute the effect on polarization, we need to keep track of clicking in the two groups. While there is a unique ranking (since $\lambda = 1$), individuals in the different groups nonetheless behave differently. 
\begin{eqnarray*}
\frac{\partial POL}{\partial \eta} &=& \frac{\partial }{\partial \eta}  \left| \int  y  \Lambda^R(\pi(y))  - y  \Lambda^L(\pi(y)) g(y) dy \right| \\
&=& \frac{\partial }{\partial \eta} \left| \int  y \left( \Lambda^R(\pi(y))  -  \Lambda^L(\pi(y)) \right)  g(y) dy \right| \\
&\stackrel{(1)}{=}& \frac{\partial }{\partial \eta} \left( \int_{y \le 0} (-y) \left( \Lambda^L (\pi(y))  - \Lambda^R (\pi(y)) \right) g(y) dy +  \int_{y >0}  y (  \Lambda^R(\pi(y))  -  \Lambda^L(\pi(y)) )  g(y) dy \right) \\
&\stackrel{(2)}{=}& \frac{\partial }{\partial \eta} \left( 2 \int_{y >0}  y \left( \Lambda^R(\pi(y))  -  \Lambda^L(\pi(y)) \right)  g(y) dy \right) \\
&=&  2 \int_{y >0}  y \left(  \frac{\partial \Lambda^R(\pi(y)) }{\partial \eta} -  \frac{ \partial \Lambda^L(\pi(y))  }{\partial \eta}  \right) g(y) dy  \\ 
&\stackrel{(3)}{=}&  2 \int_{y >0}  y \left(   \Lambda^R_+ \pi(y) \frac{\partial \pi(y) }{\partial \eta} -  \Lambda^L_+  \pi(y) \frac{\partial \pi(y) }{\partial \eta} \right) g(y) dy  \\ 
&=&  2 \int_{y >0}  y \left(   \Lambda^R_+  -  \Lambda^L_+   \right) \pi(y) \frac{\partial \pi(y) }{\partial \eta} g(y) dy  \\ 
&=&  2 \int_{y >0}  y \left(   \Lambda^R_+  -  \Lambda^L_+   \right) (1+ \eta \mu_H (y) ) \frac{(M-1)(\mu_H(y)-\bar{\mu}_H)}{\left(M(1+ \eta \bar{\mu}_H) + \eta ( \mu_H(y) - \bar{\mu}_H ) \right)^3} g(y) dy  \\ 
&\stackrel{(4)}{>}& 0,
\end{eqnarray*}
where (1) follows because of $\R_{-}$ we have $\Lambda^L(\pi(y)) > \Lambda^R(\pi(y))$, and on $\R_{+}$ we have $\Lambda^R(\pi(y)) > \Lambda^L(\pi(y))$, (2) follows by symmetry of the limit clicking distribution, (3) follows since $\Lambda^R, \Lambda^L$ are linear and hence,  for $y$ on $\R_{+}$, ${\Lambda^R}' (y) \equiv \Lambda_+^R$ and ${\Lambda^L}' (y) \equiv \Lambda_+^L$ are positive constants with $\Lambda^R_+  >  \Lambda^L_+$,  and finally (4) follows for the same reasons as with the previous cases ($ENG$ and $MIS$) since $\mu_H(y)$ and $y$ are increasing in $y$ and $\Lambda^R_+  >  \Lambda^L_+$ on $\R_+$, ensuring $\int_{y > 0}  \frac{(1 + \eta \mu_H(y))(\mu_H(y)-\bar{\mu}_H)}{\left(M(1+\bar{\mu}_H) + \eta ( \mu_H(y) - \bar{\mu}_H ) \right)^3} g(y)dy > 0$. Note also that due to symmetry, $ \int_{y >0} (\mu_H(y)-\bar{\mu}_H) g(y) dy = 0$.

\bigskip
Suppose now that  highlighting behavior is flat. The above expressions continue to hold:
\begin{eqnarray*}
\frac{\partial ENG}{\partial \eta} 
&=& \Lambda' \cdot \int \left(1 + \mu_H(y) \right)  \frac{(M-1)(\mu_H(y)-\bar{\mu}_H)}{\left(M(1+ \eta \bar{\mu}_H) + \eta ( \mu_H(y) - \bar{\mu}_H ) \right)^3}g(y)dy  \\
&>& 0.
\end{eqnarray*}
However, in the flat case, as $|y|$ increases $\mu_H(y)$ decreases, or equivalently, $\mu_H(y)$ increases as $|y|$ decreases, but $\mu_H(y) - \bar{\mu}_H$ is positive for smaller values of $y$ and hence the above integral is positive.

In the case of misinformation, the argument is reversed, since it is now $\frac{|y|}{\left(M(1+ \eta \bar{\mu}_H) + \eta ( \mu_H(y) - \bar{\mu}_H ) \right)^2}$ that multiplies $\mu_H(y)-\bar{\mu}_H$ and hence the fact that as $|y|$ increases the whole fraction increases, while $\mu_H(y)$ decreases, which means that the overall integral is now negative.
\begin{eqnarray*}
\frac{\partial MIS}{\partial \eta} &=& \Lambda'  \cdot \int |y| \frac{(M-1)(\mu_H(y)-\bar{\mu}_H)}{\left(M(1+ \eta \bar{\mu}_H) + \eta ( \mu_H(y) - \bar{\mu}_H ) \right)^2}g(y)dy \\
&<& 0.
\end{eqnarray*}
A similar argument applies for polarization:
\begin{eqnarray*}
\frac{\partial POL}{\partial \eta} 
&=& 2 \left(   \Lambda^R_+  -  \Lambda^L_+   \right) \int_{y >0}  y  (1+ \eta \mu_H (y) )  \frac{ (M-1)(\mu_H(y)-\bar{\mu}_H)}{\left(M(1+ \eta \bar{\mu}_H) + \eta ( \mu_H(y) - \bar{\mu}_H ) \right)^3} g(y) dy \\
&<& 0.
\end{eqnarray*}
Here the inequality follows since the integral is on $\R_{+}$ so that it is $\frac{y (1+ \eta \mu_H (y) )}{\left(M(1+ \eta \bar{\mu}_H) + \eta ( \mu_H(y) - \bar{\mu}_H ) \right)^3}$ that multiplies the expression $\mu_H(y)-\bar{\mu}_H $, where it can be checked that the former is increasing in $y$, while $\mu_H(y)$ is decreasing in $y$, making the overall integral negative. Note that as before, due to symmetry, $ \int_{y >0} (\mu_H(y)-\bar{\mu}_H) g(y) dy = 0$.
\hfill $\Box$

\bigskip
\noindent
{\bf Proof of Proposition \ref{prop:personalization}.}
Set again $\theta=\widehat{\theta}=0$ and fix $\beta> 1$, and write $\Lambda$ for the function $\Lambda_{\beta}$, thus dropping the subscript $\beta$.
Applying Eq.~(\ref{eq:exprank}) to the personalized algorithm Eq.~(\ref{eq:clickpers}), we obtain for the expected rank:
\begin{equation} \label{eq:exprankpers}
    r^g(y) \approx \zeta_0 - \zeta_1 \cdot \frac{ \pi_g^g(y_m) + \lambda \pi_g^{\neg g}(y_m)}{1+\lambda}, \, \, g \in \{ L, R \} , 
\end{equation}
where $\zeta_0, \zeta_1 >0$ are constants and $\neg g$ denotes the group in $\{ L, R \}$ other than $g$. 
Here the expressions $\pi_g^g(y)$ and $\pi_g^{\neg g}(y)$ denote respectively the popularity from individuals in $g$ and in $\neg g$ in the ranking seen by group $g$:\footnote{The distinction is necessary because of the normalizations that get applied to the two different rankings and that therefore change the denominators in the two cases.}
\[ \pi_g^g(y) = \frac{1+ \eta \cdot \mu_H^g(y)}{M( 1+ \eta \cdot \bar{\mu}_H^g)+ \eta  (\mu_H^g(y)- \bar{\mu}_H^g ) + \lambda \left( M( 1+ \eta \cdot \bar{\mu}_H^{\neg g})+ \eta  (\mu_H^{\neg g}(y)- \bar{\mu}_H^{\neg g} ) \right)}  \]
and 
\[ \pi_g^{\neg g}(y) = \frac{\lambda ( 1+ \eta \cdot \mu_H^{\neg g}(y) )  }{M( 1+ \eta \cdot \bar{\mu}_H^g)+ \eta  (\mu_H^g(y)- \bar{\mu}_H^g ) + \lambda \left( M( 1+ \eta \cdot \bar{\mu}_H^{\neg g})+ \eta  (\mu_H^{\neg g}(y)- \bar{\mu}_H^{\neg g} ) \right)} , \]
where $\mu_H^g(y)$ is the propensity to highlight by individuals in $g$:
\[   \mu_H^g(y) =  \int_{ x \in H^{-1}(y)} \Ind{{x \in g}} p_A(x) f(x) dx . \]

 We can apply Lemma~\ref{lemma:LCD} and write engagement as:
\begin{eqnarray*}
ENG &=& \sum_{g = L, R} \int \left( LCD^g(y) + LHD^g(y) \right) dy =  \sum_{g = L, R} \int \left(1 + \mu_H^g (y) \right) LCD^g(y) dy \\
&=& \sum_{g = L, R} \int \left(1 + \mu_H^g(y) \right) \Lambda^g \left( \frac{  \pi_g^g(y)+\lambda \pi_g^{\neg g}(y) }{1+\lambda} \right) g(y) dy , \\
\end{eqnarray*}
where as in the proof of Proposition~\ref{prop:main1}, $\Lambda^g \left( \frac{  \pi_g^g(y)+\lambda \pi_g^{\neg g}(y) }{1+\lambda} \right)$ is the probability of being clicked by an individual in $g$:
\begin{equation} 
\Lambda^g \left( \frac{  \pi_g^g(y)+\lambda \pi_g^{\neg g}(y) }{1+\lambda} \right)
=  \frac{M + \log{\beta} \cdot \left(M - \zeta_0 + \zeta_1 \frac{  \pi_g^g(y)+\lambda \pi_g^{\neg g}(y) }{1+\lambda} \right) }{M + \log{\beta} \cdot \sum_{m' \in M} (M-m')} , 
\end{equation}

We can compute 
\begin{eqnarray*}
\frac{\partial ENG}{\partial \lambda} 
&=& \sum_{g = L, R}   \frac{\partial }{\partial \lambda}  \int \left(1 + \mu_H^g(y) \right)\Lambda^g \left( \frac{  \pi_g^g(y)+\lambda \pi_g^{\neg g}(y) }{1+\lambda} \right) g(y) dy \\
&=& \sum_{g = L, R}   \int \left(1 + \mu_H^g(y) \right) \frac{\partial \Lambda^g \left( \frac{  \pi_g^g(y)+\lambda \pi_g^{\neg g}(y) }{1+\lambda} \right)}{\partial \lambda} g(y) dy  \\
&=& \sum_{g = L, R}  \left( \int_{y \le 0} \left(1 + \mu_H^g(y) \right)  {\Lambda_{-}^{g}}' \frac{\partial  \left( \frac{  \pi_g^g(y)+\lambda \pi_g^{\neg g}(y) }{1+\lambda} \right)}{\partial \lambda} g(y) dy  \right. \\
&& \hspace{1.3in} + \left.  \int_{y > 0} \left(1 + \mu_H^g(y) \right)  {\Lambda_{+}^{g}}' \frac{\partial  \left( \frac{  \pi_g^g(y)+\lambda \pi_g^{\neg g}(y) }{1+\lambda} \right)}{\partial \lambda} g(y) dy \right)     \\
&=&  2 \sum_{g = L, R}  \int_{y > 0} \left(1 + \mu_H^g(y) \right)  {\Lambda_{+}^{g}}' \frac{\partial  \left( \frac{  \pi_g^g(y)+\lambda \pi_g^{\neg g}(y) }{1+\lambda} \right)}{\partial \lambda} g(y) dy \\
&=& 2 \sum_{g = L, R}  \int_{y > 0}   \frac{\left(1 + \mu_H^g(y) \right){\Lambda_{+}^{g}}'}{1+\lambda} \left( \frac{\partial  \pi_g^g(y)}{\partial \lambda} + \lambda \frac{\partial \pi_g^{\neg g} (y) }{\partial \lambda}  - \frac{\pi_g^g(y) - \pi_g^{\neg g}(y)}{1+\lambda} \right) g(y) dy
\end{eqnarray*}
Now, 
\[  
\frac{\partial \pi_g^g(y)}{\partial \lambda}  =  \frac{-( 1+ \eta \cdot \mu_H^{g}(y) ) \left( M( 1+ \eta \cdot \bar{\mu}_H^{\neg g})+ \eta  (\mu_H^{\neg g}(y)- \bar{\mu}_H^{\neg g} ) \right)  }{\left( M( 1+ \eta \cdot \bar{\mu}_H^g)+ \eta  (\mu_H^g(y)- \bar{\mu}_H^g ) + \lambda \left( M( 1+ \eta \cdot \bar{\mu}_H^{\neg g})+ \eta  (\mu_H^{\neg g}(y)- \bar{\mu}_H^{\neg g} ) \right) \right)^2} , 
\]
and 
\[  
\frac{\partial \pi_g^{\neg g}(y)}{\partial \lambda}  =  \frac{ ( 1+ \eta \cdot \mu_H^{\neg g}(y) ) \left( M( 1+ \eta \cdot \bar{\mu}_H^g)+ \eta  (\mu_H^g(y)- \bar{\mu}_H^g ) \right) }{\left( M( 1+ \eta \cdot \bar{\mu}_H^g)+ \eta  (\mu_H^g(y)- \bar{\mu}_H^g ) + \lambda \left( M( 1+ \eta \cdot \bar{\mu}_H^{\neg g})+ \eta  (\mu_H^{\neg g}(y)- \bar{\mu}_H^{\neg g} ) \right) \right)^2}  ,
\]
where it can be checked that:
\[
\frac{\partial \left( \pi_g^g(y)+ \pi_g^{\neg g} (y) \right)}{\partial \lambda} = \frac{( 1+ \eta \cdot \mu_H^{\neg g}(y) )\eta  (\mu_H^g(y)- \bar{\mu}_H^g )  -   ( 1+ \eta \cdot \mu_H^{g}(y) ) \eta  (\mu_H^{\neg g}(y)- \bar{\mu}_H^{\neg g} )  }{\left( M( 1+ \eta \cdot \bar{\mu}_H^g)+ \eta  (\mu_H^g(y)- \bar{\mu}_H^g ) + \lambda \left( M( 1+ \eta \cdot \bar{\mu}_H^{\neg g})+ \eta  (\mu_H^{\neg g}(y)- \bar{\mu}_H^{\neg g} ) \right) \right)^2} ,
\]
which, integrated on $\R_{+}$, is strictly negative for $g=R$ and outweighs (in absolute value) the case for $g=L$, such that for any $\lambda \in [0,1]$ and whether or not $\mu_H^g(y)$ increases in $y$:
 \[
 \sum_{g = L, R}  \int_{y > 0} \frac{\left(1 + \mu_H^g(y) \right){\Lambda_{+}^{g}}'}{1+\lambda}   \frac{\partial \left( \pi^g(y)+\lambda \pi^{\neg g} (y) \right)}{\partial \lambda} g(y) dy < 0.
 \]
Moreover, a similar argument applies for $- \frac{\pi_g^{g}(y)-\pi_g^{\neg g}(y)}{1+\lambda}$, such that also:
 \[
 \sum_{g = L, R}  \int_{y > 0} \frac{\left(1 + \mu_H^g(y) \right){\Lambda_{+}^{g}}'}{1+\lambda} \frac{(-1)(\pi_g^{g}(y)-\pi_g^{\neg g}(y))}{1+\lambda} g(y) dy < 0.
 \]
This shows that $\frac{\partial ENG}{\partial \lambda} < 0$ so that less personalization (larger $\lambda$) decreases engagement both with flat and non-flat highlighting.

Finally,
\begin{eqnarray*}
POL &=& \left| \int y LCD^R(y)  dy -   \int y  LCD^L(y) dy \right| ,
\end{eqnarray*}
so that, using the same reasoning as in the proof of Proposition~\ref{prop:main1}, we can write:
\begin{eqnarray*}
\frac{\partial POL}{\partial \lambda}
&=& \frac{\partial }{\partial \lambda} \left( 2 \int_{y >0}  y \left( \Lambda^R \left( \frac{  \pi_R^R(y)+\lambda \pi_R^{L}(y) }{1+\lambda} \right)  -  \Lambda^L \left( \frac{  \pi_L^L(y)+\lambda \pi_L^{R}(y) }{1+\lambda} \right) \right)  g(y) dy \right) \\
&=&  2 \int_{y >0}  y {\Lambda^R_+}' \left( \frac{\partial  \pi_R^R(y)}{\partial \lambda} + \lambda \frac{\partial \pi_R^{L} (y) }{\partial \lambda}  - \frac{\pi_R^R(y) - \pi_R^{L}(y)}{1+\lambda} \right) g(y) dy  \\ 
&& \hspace{1in} - 2 \int_{y >0}  y {\Lambda^L_+}' \left( \frac{\partial  \pi_L^L(y)}{\partial \lambda} + \lambda \frac{\partial \pi_L^{R} (y) }{\partial \lambda}  - \frac{\pi_L^L(y) - \pi_L^{R}(y)}{1+\lambda} \right) g(y) dy  . 
\end{eqnarray*}
Moreover, similar calculations as above show that the first integral is negative and dominates in absolute value the second one, showing that overall $\frac{\partial POL}{\partial \lambda} < 0$ so that again less personalization (larger $\lambda$) decreases polarization both with flat and non-flat highlighting.
\hfill $\Box$

\bigskip
\noindent
{\bf Proof of Proposition \ref{prop:welfare}.}
Recall from Eq.~(\ref{eq:welfare}):
\[
    W_{\psi}(\eta, \lambda) = \psi \cdot ENG ( \eta, \lambda) - (1-\psi) \cdot MIS( \eta, \lambda) \cdot POL( \eta, \lambda) .
\]
Hence, for $\psi=0$, we have $W_0 = MIS \cdot POL$, while, for $\psi = 1$, we have $W_1 = ENG$. The results then follow directly from Propositions~\ref{prop:main1} and ~\ref{prop:personalization}. 

Consider the non-flat case. It follows immediately that $W_0$ is maximized at a smallest possible value of $\eta$, since $- MIS \cdot POL$ is maximized when $MIS \cdot POL$ is minimized and $\frac{MIS}{\partial \eta} >0, \frac{POL}{\partial \eta}>0$. Also, $W_0$ is maxized at a largest possible value of $\lambda$ again since $\frac{POL}{\partial \lambda} <0$ (less personalization decreases $POL$) while $\frac{MIS}{\partial \lambda} \approx 0$. The contrary is true for $\psi = 1$.

By contrast, by analogous argument, in the flat case, $W_0$ is maximized at a largest possible value of $\eta$ and at a largest possible value of $\lambda$, while for $\psi = 1$, $W_1$ is maximized at a largest possible value of $\eta$ and a smallest possible value of $\lambda$.
\hfill $\Box$

\subsection{Empirical Evidence on Meaningful Social Interactions and Political Polarization: Robustness}
One possible concern regarding the causal interpretation of our results linking Facebook's MSI update and political polarization is due to the concurring general elections in Italy in March 2018. With respect to this issue we notice that, by including the date of interview fixed-effect, our empirical strategy takes into account and controls for any general trend in political polarization over time. At the same time, one might argue that the presence of elections might have led to a differential trend in political polarization between individuals that used internet to form an opinion and the ones who did not which was not due to the MSI algorithm \textit{per se} (e.g., increase in online fake news before elections). In response to this argument, we first point out that the MSI algorithm might have further amplified the diffusion of fake-news as predicted by our model. Second, we provide below evidence suggesting a polarization effect even when dropping the months immediately after the MSI update and before the elections (i.e., January-March 2018). Specifically, Tables \ref{tab:rob_extremist} and \ref{tab:rob_affective_pol} present results when comparing the period June-December 2017 (pre-MSI) with April-December 2018 (post-MSI and post-elections).


\begin{table}[h!]
\begin{center} 
{\fontsize{10}{11.5}\selectfont
\centering \caption{MSI and non-moderate ideological position: Robustness \label{tab:rob_extremist}}
\begin{tabular}{lcccc} \hline\hline
 & (1) & (2)& (3)& (4)  \\
 & Non-moderate & Non-moderate  &Non-moderate &Non-moderate  \\  
 &Ideology&Ideology&Ideology&Ideology\\\hline
 &   & &   &  \\
Opinion via internet websites $\times$ Post MSI & 0.053** & 0.047** & 0.048** & 0.048** \\
 & (0.018) & (0.018) & (0.020) & (0.021) \\
Opinion via internet websites & -0.016 & -0.011 & -0.013 & -0.013 \\
 & (0.018) & (0.018) & (0.022) & (0.022) \\
Post MSI & -0.005 &  &  &  \\
 & (0.010) &  &  &  \\
 &  &  &  &  \\
Observations & 29,570 & 29,570 & 29,570 & 29,570 \\
Mean outcome & 0.37 & 0.37 & 0.37 & 0.37 \\
SD outcome & 0.48 & 0.48 & 0.48 & 0.48 \\

&   & &   &  \\
Municipality FE & YES & YES & YES & YES \\
Date of interview FE & NO & YES & NO & NO \\
Province-Date of interview FE & NO & NO & YES & YES \\
&   & &   &  \\
 Cluster SE & Region & Region & Region & Province \\ \hline\hline
\end{tabular}
}
\begin{minipage}{17cm}
\vspace{.1cm}
\scriptsize{\textbf{Note:} Time horizon: June 2017-December 2017 and April-December 2018.  All estimates include the following control variables: age, age squared, gender, number of resident family members, level of education, type of occupation and religiosity of the respondent. Observations are weighted according to the sampling weights provided by Ipsos and thus the results are representative of the Italian voting age population. Robust Standard Errors in parenthesis. *** p$<$0.01, ** p$<$0.05, * p$<$0.1}
\end{minipage}
 \end{center}
\end{table}


\begin{table}[h!]
\begin{center} 
{\fontsize{10}{11.5}\selectfont
\centering \caption{MSI and Affective Polarization \label{tab:rob_affective_pol}} \vspace{.1in}
\begin{tabular}{lcccc} \hline\hline
 & (1) & (2)& (3)& (4)  \\
 & Affective & Affective &Affective &Affective  \\  
 &Polarization&Polarization&Polarization&Polarization\\\hline
 &   & &   &  \\
Opinion via internet websites $\times$ Post MSI & 0.057* & 0.051 & 0.064** & 0.064* \\
 & (0.032) & (0.030) & (0.031) & (0.034) \\
Opinion via internet websites & -0.008 & -0.007 & -0.008 & -0.008 \\
 & (0.022) & (0.021) & (0.023) & (0.022) \\
Post MSI & 0.239*** &  &  &  \\
 & (0.027) &  &  &  \\
 &  &  &  &  \\
Observations & 17,317 & 17,317 & 17,317 & 17,317 \\
Mean outcome & 1.38 & 1.38 & 1.38 & 1.38 \\
SD outcome & 0.67 & 0.67 & 0.67 & 0.67 \\

&   & &   &  \\
Municipality FE & YES & YES & YES & YES \\
Date of interview FE & NO & YES & NO & NO \\
Province-Date of interview FE & NO & NO & YES & YES \\
&   & &   &  \\
 Cluster SE & Region & Region & Region & Province \\ \hline\hline
\end{tabular}
}
\begin{minipage}{15cm}
\vspace{.1cm}
\scriptsize{\textbf{Note:} Time horizon: June 2017-December 2017 and April-December 2018.  All estimates include the following control variables: age, age squared, gender, number of resident family members, level of education, type of occupation and religiosity of the respondent. Observations are weighted according to the sampling weights provided by Ipsos and thus the results are representative of the Italian voting age population. Robust Standard Errors in parenthesis. *** p$<$0.01, ** p$<$0.05, * p$<$0.1}
\end{minipage}
\end{center}
\end{table}

\subsection{Additional Results}

\subsubsection{Heterogeneous Benchmarks} \label{section:hetaverageop}

Throughout the paper we assume that all individuals share the same benchmark $\widehat{\theta}$. Allowing individuals to have idiosyncratic benchmarks $\widehat{\theta}_n$ does not change the results qualitatively. 
In fact, the simulations suggest that,
for $\widehat{\theta}_n$'s centered around $\theta$ and not too dispersed  ($\widehat{\theta}_n \sim N(\theta, \sigma^2_{\hat{\theta}})$ with $\sigma_{\widehat{\theta}} \le \min\{ \frac{\sigma_{x}}{4}, \frac{\sigma_{y}}{4} \}$), our main results on engagement, popularity and misinformation are rather close to the cases where individuals have a common benchmark, $\widehat{\theta}_n = \widehat{\theta}$ for all $n$. This is illustrated in Figure~\ref{fig:sim_IDIOSYN} that shows the effect of $\eta$ on the variables $ENG, POL, MIS$ for the non-flat case.


\begin{figure*}[!t]
  \centering
  \includegraphics[width=.95\linewidth]{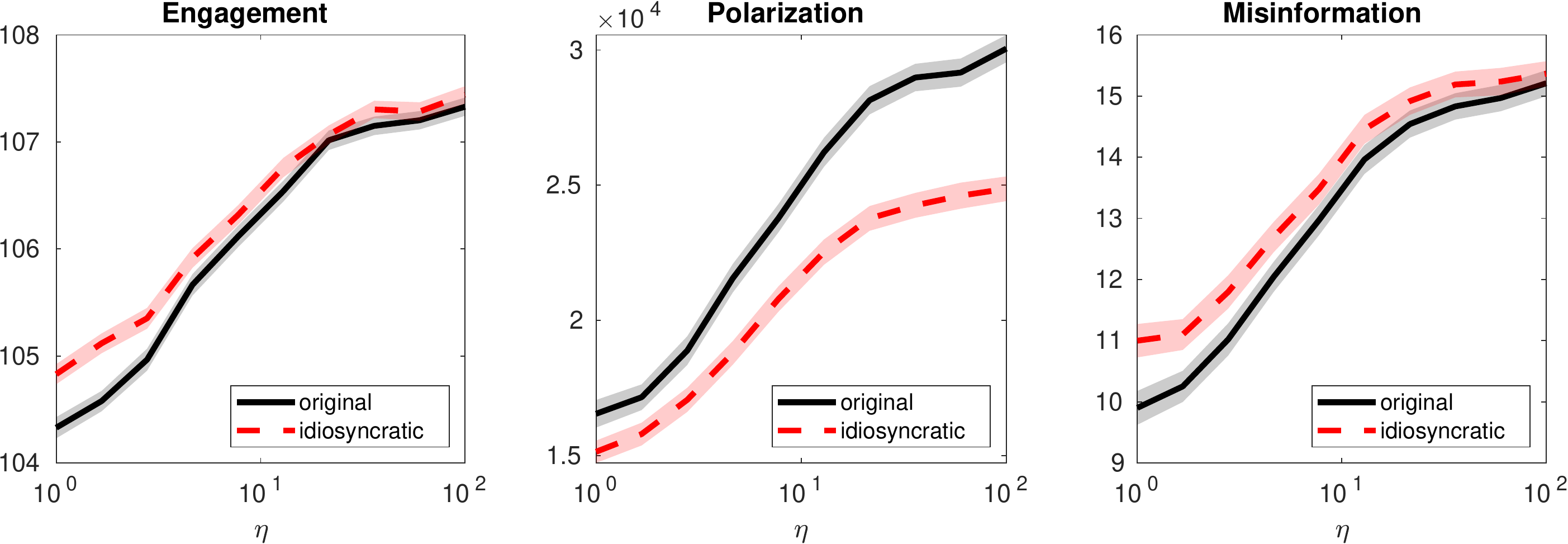}
     \caption{
     Engagement, polarization, and misinformation as a function of the highlighting parameter $\eta$ (non-flat case) with a common benchmark $\widehat{\theta}$ (solid line) and heterogeneous benchmarks $\widehat{\theta}_n$ (dotted line). The shaded areas represent the $95\%$ confidence intervals.
     }
    \label{fig:sim_IDIOSYN}
\end{figure*}

\subsubsection{Non-Centered Benchmark} \label{section:NCaverageop}

While it is natural to assume that the benchmark $\widehat{\theta}$ splits the signals roughly in half in symmetric environments, so that $\widehat{\theta} \approx \theta$, it may occur occasionally that the two are far apart. In such a situation, individuals' and news items' signals are shifted away from the benchmark $\widehat{\theta}$. This means that a potentially large mass of individuals have a prior belief far from $\widehat{\theta}$ and are hence likely to highlight news items far from it but potentially close to $\theta$.
In such a case, an increase in $\eta$ can contribute to both higher engagement and at the same time lower misinformation in the non-flat case. To see this consider Figure~\ref{fig_highlightdistr_NC} that illustrates a situation where clearly  $\widehat{\theta} \ne \theta$. Here $x^*\approx \theta$ so that a large mass of individuals with a signal close to the truth has a large highlighting propensity. An increase in $\eta$ leads to a more prominent ranking for items around $x^* \approx \theta$, which in turn, through the effect on the clicking distribution, leads to a lower level of misinformation as measured by $MIS$. Increasing the weight on highlights here actually accelerates individuals clicking on news items carrying truthful signals.


\begin{figure}[!t]
  \centering
  \includegraphics[width=15.5cm]{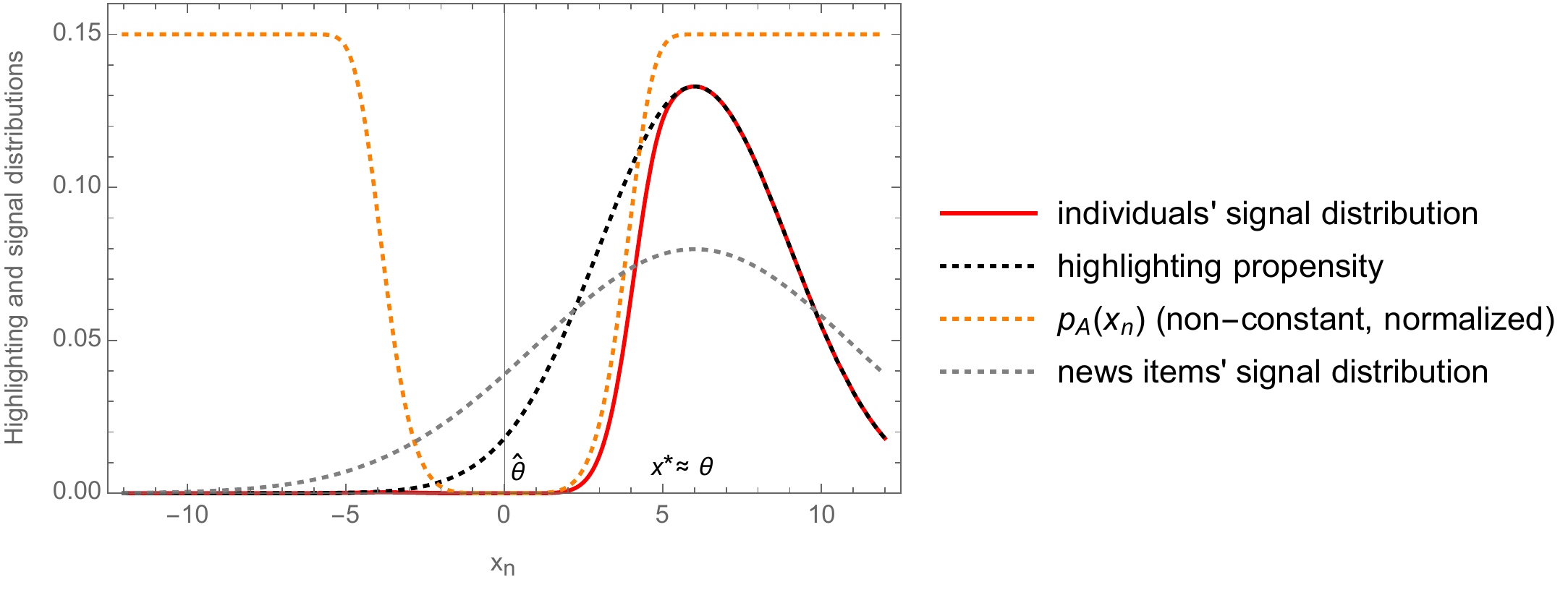}
    \caption{Individuals’ signal distribution and highlighting propensity in the non-flat case with non-centered $\widehat{\theta}$,  (with $\theta=6$ and $\widehat{\theta} = 0$); $x^*$ denotes the value of $x_n$ where the highlighting propensity is locally maximal.}
  \label{fig_highlightdistr_NC}
\end{figure}

\subsubsection{Website Concentration}

The effect of the highlighting parameter $\eta$ on the concentration of clicking traffic on the news items in $M$ is not clear-cut.
A widely used measure for the concentration within a given market is the Herfindahl index: $$HHI = \sum_{m \in M} \left(  \frac{100}{N} \cdot (\widehat{\kappa}^L_{N,m} + \widehat{\kappa}^R_{N,m}) \right)^2.$$


\begin{figure*}[!ht]
  \centering
  \includegraphics[width=.45\linewidth]{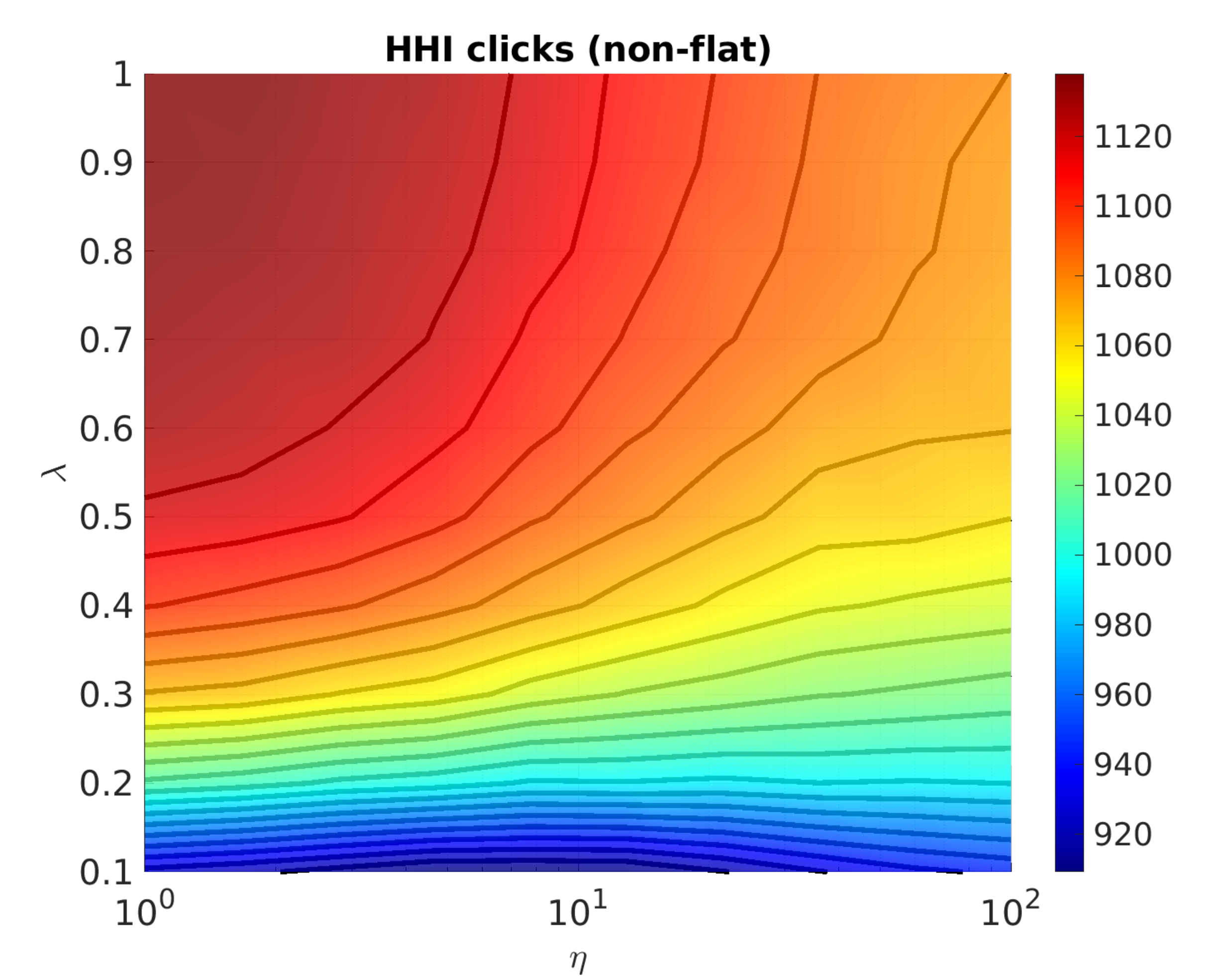}
  \includegraphics[width=.45\linewidth]{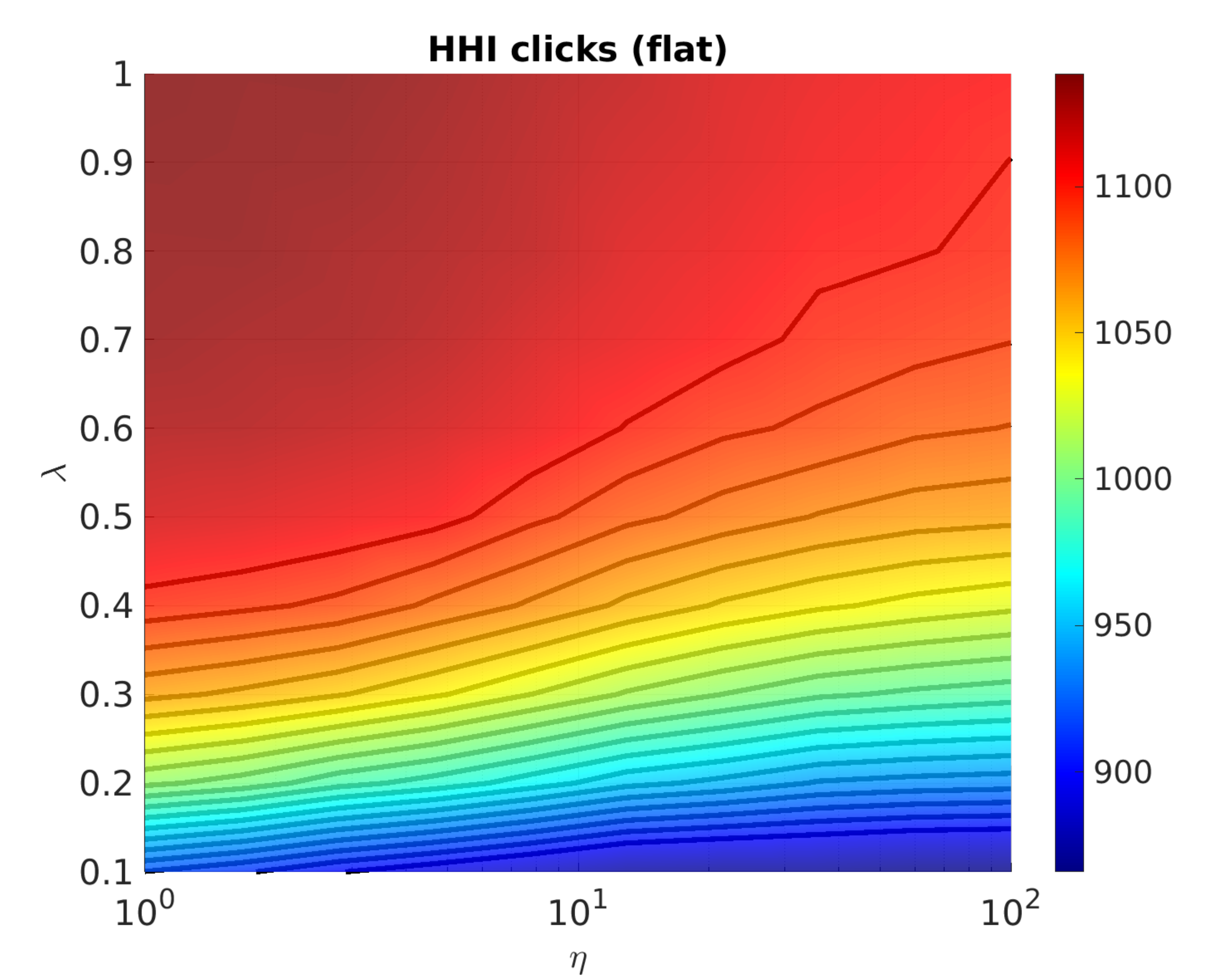}
    \caption{Herfindahl index ($HHI$) as a function of highlighting weight $\eta$ and personalization $\lambda$ for non-flat (left) and flat (right) individuals' highlighting behavior.}
    \label{fig:sim_hhi}
\end{figure*}

Figure~\ref{fig:sim_hhi} shows the Herfindahl index 
for simulations performed using the same setting as in Section~\ref{sec:empirical}. In the non-flat case, as $\eta$ increases, the clicking distribution becomes increasingly bimodal, which offsets the reinforced rich-get-richer effect due to the higher popularity weight. However, the flat case also does not exhibit an important tendency towards concentration due to a higher $\eta$. By contrast, more personalization tends to decrease concentration. Essentially, as $\lambda$ decreases individuals see increasingly uncorrelated rankings which increasingly tends to spread the users across two possibly different subsets of items, namely users in the Left group and ones in the Right group, resulting in a lower index $HHI$.

Another parameter which has an important effect on traffic concentration is the parameter $\beta$, which calibrates the attention bias. Quite generally, a larger $\beta$ tends to strengthen the rich-get-richer effect, thereby contributing to concentrating traffic on fewer items.

\end{document}